\documentclass[12pt]{article}

\usepackage{epsfig}

\newenvironment{appendletterA}
 {
  \setcounter{section}{0}
  \setcounter{equation}{0}
  
 }{
 }
\newenvironment{appendletterB}
 {
  \setcounter{section}{0}
  \setcounter{equation}{0}
  
 }{
 }

\newcommand{\s}{\smallskip}
\newcommand{\nn}{\noindent}
\newcommand{\lsim}{\raisebox{-0.13cm}{~\shortstack{$<$ \\[-0.07cm] 
$\sim$}}~}
\newcommand{\gsim}{\raisebox{-0.13cm}{~\shortstack{$>$ \\[-0.07cm] 
$\sim$}}~}

\def\be{\begin{equation}}
\def\ee{\end{equation}}
\def\bc{\begin{center}}
\def\ec{\end{center}}
\def\bea{\begin{eqnarray}}
\def\eea{\end{eqnarray}}

\def\nn{\nonumber}
\def\ov{\overline}

\def\at{\alpha_t}
\def\ab{\alpha_b}
\def\as{\alpha_s}
\def\atau{\alpha_{\tau}}
\def\oat{{\cal O}(\at)}
\def\oab{{\cal O}(\ab)}
\def\oatau{{\cal O}(\atau)}

\def\oatas{{\cal O}(\at\as)}
\def\oabas{{\cal O}(\ab\as)}
\def\oatababq{{\cal O}(\at\ab + \ab^2)}
\def\oatqatababq{{\cal O}(\at^2 + \at\ab + \ab^2)}

\def\oatasabas{{\cal O}(\at\as +\ab\as)}

\def\oatq{{\cal O}(\at^2)}

\def\oatauq{{\cal O}(\atau^2)}
\def\oabatau{{\cal O}(\ab \atau)}

\def\oatauqatab{{\cal O}(\atau^2 +\ab \atau )}

\def\mt{m_t}
\def\polemt{M_t}

\def\sqstop{\sqrt{m_{\tilde{t}_1}\,m_{\tilde{t}_2}}}

\def\mb{m_b}
\def\hb{h_b}
\def\msbu{m_{\tilde{b}_1}^2}
\def\msbd{m_{\tilde{b}_2}^2}
\def\diffb{\msbu-\msbd}
\def\Sdb{s_{2\theta_b}}
\def\Cdb{c_{2\theta_b}}

\def\sdb{s_{2\bar{\theta}_b}}

\def\ptb{\tilde{\varphi}_{b}}

\def\mtau{m_{\tau}}
\def\htau{h_{\tau}}
\def\mstauu{m_{\tilde{\tau}_1}^2}
\def\mstaud{m_{\tilde{\tau}_2}^2}
\def\mqstauu{m_{\tilde{\tau}_1}^4}
\def\mqstaud{m_{\tilde{\tau}_2}^4}
\def\difftau{\mstauu-\mstaud}

\def\sdtau{s_{2\bar{\theta}_{\tau}}}

\def\pttau{\tilde{\varphi}_{\tau}}
\def\Sdtau{s_{2\theta_{\tau}}}
\def\Cdtau{c_{2\theta_{\tau}}}

\def\mgut{M_{\rm GUT}}
\def\qewsb{M_{\rm EWSB}}
\def\m0{m_0}
\def\mhf{m_{1/2}}

\def\Veff{V_{\rm eff}}

\def\drbar{\overline{\rm DR}}
\def\msbar{\overline{\rm MS}}
\def\gev{{\rm GeV}}
\def\msqu{m_{\tilde{q}_1}^2}
\def\msqd{m_{\tilde{q}_2}^2}
\def\msqi{m_{\tilde{q}_i}^2}
\def\thsq{\bar{\theta}_{\tilde{q}}}
\def\thq{\theta_{\tilde{q}}}
\def\Sdq{s_{2\theta_q}}

\def\isajet{{\tt IsaJet}}
\def\spheno{{\tt SPheno}}
\def\suspect{{\tt SuSpect}}
\def\softsusy{{\tt SoftSusy}}
\def\sphenov{{\tt SPheno2.2.1}}
\def\suspectv{{\tt SuSpect2.3}}
\def\softsusyv{{\tt SoftSusy1.8.7}}
\def\feynhiggs{{\tt FeynHiggs}}

\def\gp{g^{\prime}}
\def\gpq{g^{\prime\,2}}
\def\tb{\tan\beta}

\def\cb{c_{\beta}}
\def\sb{s_{\beta}}

\def\mh{M_h}
\def\mH{M_{H}}

\def\mz{m_{Z}}

\def\polemz{M_{Z}}
\def\ma{M_{A}}
\def\marun{m_A}

\def\sq2{\sqrt{2}}
\def\hlf{\frac{1}{2}}

\def\dr{\ov{\rm DR}}

\catcode`@=11
\def\citer{\@ifnextchar
[{\@tempswatrue\@citexr}{\@tempswafalse\@citexr[]}}
\def\@citexr[#1]#2{\if@filesw\immediate\write\@auxout{\string\citation{#2}}\fi
  \def\@citea{}\@cite{\@for\@citeb:=#2\do
    {\@citea\def\@citea{--\penalty\@m}\@ifundefined
       {b@\@citeb}{{\bf ?}\@warning
       {Citation `\@citeb' on page \thepage \space undefined}}%
\hbox{\csname b@\@citeb\endcsname}}}{#1}}
\catcode`@=12

\long\def\symbolfootnote[#1]#2{\begingroup%
\def\thefootnote{\fnsymbol{footnote}}\footnote[#1]{#2}\endgroup} 

\begin{document}
\baselineskip=15.5pt
\thispagestyle{empty}

\bc
\hfill{MPP-2004-50, LAPTH-1046/04, {PM/04--13,}\\
\hfill ZU-TH-08/04, IFIC/04-21 ~~~~~~~~~~~~~}
\ec

\vspace{.2cm}
\bc
{\LARGE\bf Precise determination of the neutral Higgs}
\ec
\vspace{.1cm}
\bc
{\LARGE\bf boson masses in the MSSM} 
\ec

\vspace{.5cm}

\bc
{ \large \sc
B.C.~Allanach$^{\,a}$,
\symbolfootnote[0]{e--mail: {\tt benjamin.allanach@cern.ch,
abdelhak.djouadi@cern.ch, kneur@lpm.univ-montp2.fr,
porod@ific.uv.es, slavich@mppmu.mpg.de.}}
A.~Djouadi$^{\,b,c}$,
J.L.~Kneur$^{\,b}$,
W.~Porod$^{\,d,e}$}
\vspace{.2cm}

and  {\large \sc
P.~Slavich$^{\,f}$
}\\

\vspace{.5cm}
${}^a$
{\small \em LAPTH, 9 Chemin de Bellevue,
 F--74941 Annecy-le-vieux, France}

${}^b$
{\small  \em LPMT, Universit\'e de Montpellier II,
F--34095 Montpellier Cedex 5, France} 

${}^c$
{\small \em LPTHE, Universit\'es Paris 6\&7,  
4 Place Jussieu, F--75252 Paris Cedex 05, France }

${}^d$
{\small \em 
IFIC - Instituto de F\'\i sica Corpuscular,
E-46071 Valencia, Espa\~na}

${}^e$
{\small \em Physik Institut, Universit\"at Zurich,
Winterthurerstrasse 190, CH--8057 Zurich, Switzerland}

${}^f$
{\small \em Max Planck Institut f\"ur Physik,
F\"ohringer Ring 6, D--80805 M\"unchen, Germany}

\ec

\vspace{0.5cm}

\centerline{\bf Abstract}
\vspace{2 mm}
\begin{quote} \small

We present the implementation of the radiative corrections of the
Higgs sector in three public computer codes for the evaluation of the
particle spectrum in the Minimal Supersymmetric Standard Model,
\softsusy, \spheno\ and \suspect. We incorporate the full one--loop
corrections to the Higgs boson masses and the electroweak symmetry
breaking conditions, as well as the two--loop corrections controlled
by the strong gauge coupling and the Yukawa couplings of the third
generation fermions. We include also the corrections controlled by the
$\tau$ Yukawa coupling that we derived for completeness. The
computation is consistently performed in the $\drbar$ renormalisation
scheme. In a selected number of MSSM scenarios, we study the effect of
these corrections and analyse the impact of some higher order
effects. By considering the renormalisation scheme and scale
dependence, and the effect of the approximation of zero external
momentum in the two--loop corrections, we estimate the theoretical
uncertainty on the lighter Higgs boson mass to be 3 to 5 GeV. The
uncertainty on $M_h$ due to the experimental error in the measurement
of the SM input parameters is approximately of the same size.
Finally, we discuss the phenomenological consequences, using the
latest value of the top quark mass. We find, in particular, that the
most conservative upper bound on the lighter Higgs boson mass in the
general MSSM is $\mh \lsim 152$ GeV and that there is no lower bound
on the parameter $\tan\beta$ from non--observation of the MSSM
Higgs bosons at LEP2.
\end{quote}
\vfill
\newpage
\setcounter{equation}{0}
\setcounter{footnote}{0}
\vskip2truecm


\section{Introduction}

\vspace*{2mm}

The Minimal Supersymmetric Standard Model (MSSM) \cite{revs} provides
an attractive weak--scale extension of the Standard Model (SM). At the
price of introducing a bosonic (fermionic) super-partner for every
fermionic (bosonic) particle of the SM, plus an additional doublet of
Higgs fields and their fermionic partners, it allows one to solve the
hierarchy problem, leads to a consistent unification of the SM gauge
couplings and provides a natural candidate for the dark matter in the
universe. In the last two decades, a large effort has been devoted to
the detailed theoretical study of the properties of the supersymmetric
particles and the MSSM Higgs bosons, and to the experimental search
for these particles, either directly in high--energy collider
experiments or indirectly through high--precision measurements. \s

The task of providing a detailed phenomenological analysis of the
masses and couplings of the supersymmetric particles and the MSSM
Higgs bosons, and comparing it with the results or expectations of the
present and future experiments, proves to be an extremely difficult
one, due to the large number of new parameters ($>$100 in the most
general soft supersymmetry (SUSY) breaking Lagrangian and $>$20 when
some phenomenological constraints are imposed; see Ref.~\cite{GDR} for
a discussion).  However, there are well motivated theoretical models
where the soft SUSY--breaking parameters obey a number of universal
boundary conditions at some very high scale, such as the Grand
Unification Theory (GUT) scale $\mgut$, or the messenger scale $M_{\rm
mess}$, thus reducing the number of fundamental parameters to a
handful. This is for instance the case of the models of gravity
mediated (mSUGRA) \cite{msugra}, gauge mediated (GMSB) \cite{gmsb} or
anomaly mediated (AMSB) \cite{amsb} SUSY breaking. String--inspired
models of soft SUSY--breaking have also been proposed
\cite{string}. In all such models, the soft SUSY--breaking parameters
are given as input at the high--energy scale, then they are evolved by
means of the MSSM Renormalisation Group Equations (RGE) down to a
low--energy scale, $\qewsb$, where the conditions of electroweak
symmetry breaking (EWSB) are imposed and the spectrum of superparticle
masses and couplings is computed. \s

Several computer codes (that in what follows we will often denote as
{\em spectrum generators}) have been developed in the past years to
provide reliable determinations of the supersymmetric spectra in
models with high--energy boundary conditions, including the
calculations of the various radiative corrections. In particular, the
publicly available codes include \isajet~\cite{isajet},
\softsusy~\cite{softsusy}, \suspect~\cite{suspect} and
\spheno~\cite{spheno}. Detailed comparisons of the features and the
results of these codes can be found in Ref.~\cite{comparison} for
instance. \s

A crucial prediction of the MSSM is the existence of at least one
light Higgs boson \cite{haber} which, at the tree level, is bound to
be lighter than the $Z$ boson. If this upper bound was not
significantly raised by radiative corrections, the failure of
detecting this Higgs boson at LEP would have ruled out the MSSM as a
viable theory for physics at the weak scale. However, it was first
realised in Ref.~\cite{at} that the inclusion of the one--loop $\oat$
corrections, which rise quartically with the mass of the top quark and
logarithmically with the mass of its scalar superpartner, may push the
lighter Higgs boson mass well above the tree--level bound. \s

In the subsequent years, an impressive theoretical effort has been
devoted to the precise determination of the Higgs boson masses in the
MSSM. A first step was to provide the full one--loop computation,
performed in Refs.~\cite{1loop,dab,pbmz}. A second step was the
addition of the dominant two--loop corrections which involve the
strongest couplings of the theory: the QCD coupling constant and the
Yukawa couplings of the heavy third generation fermions (although the
masses of the bottom quark and the $\tau$ lepton are relatively tiny
compared to the top quark mass, the $b$ and $\tau$ Yukawa couplings
can be strongly enhanced for large values of $\tan \beta$, the ratio
of the vacuum expectation values of the two Higgs doublet fields).\s

The leading logarithmic effects at two loops have been subsequently
included via appropriate RGEs \cite{rge,HHH}, and the genuine
two--loop corrections of $\oatas$
\cite{hemphoang,Sven,Zhang,ezhat,dsz} and $\oabas$ \cite{bdsz2} have
been evaluated in the limit of zero external momentum. The two--loop
Yukawa corrections of $\oatq$ \cite{hemphoang,ezhat,bdsz} and
$\oatababq$ \cite{dds} have been also evaluated in the limit of zero
external momentum. To complete the calculation of the two--loop
corrections controlled by third--generation fermion couplings, only
the expectedly small corrections that are proportional to the
$\tau$--lepton Yukawa coupling are still to be determined. The
$\oatauq$ corrections can be straightforwardly derived from the
$\oatq$ corrections as explained in Ref.~\cite{dds}; the missing
$\oabatau$ contributions are provided in this paper.\s

The tadpole corrections needed to minimise the effective scalar
potential, $\Veff$, have also been calculated at the one--loop
\cite{pbmz,tad1l} and two--loop \cite{dds,dstad} levels for the strong
coupling and the top and bottom quark Yukawa couplings; the tadpole
corrections involving the $\tau$--lepton Yukawa coupling will also be
provided in this paper. Finally, the full two--loop corrections to the
MSSM effective potential have been calculated \cite{Martin1}, together
with a first study of the two--loop corrections to the lighter Higgs
boson mass controlled by the electroweak gauge couplings
\cite{Martin2}.\s

The calculation of the radiative corrections to physical observables
requires the choice of a renormalisation scheme. For example, one
might choose to express the corrections in terms of ``On--Shell'' (OS)
parameters, such as pole particle masses and suitably defined mixing
angles. This is the scheme adopted for the computation of the MSSM
Higgs boson masses by the computer code \feynhiggs~\cite{feynhiggs},
based on the results of Refs.~\cite{dab,Sven} and subsequently
expanded to include the results of Refs.~\cite{bdsz2,bdsz,dds,sven1L}
(see Ref.~\cite{dhhsw} for a discussion). \s

However, when the MSSM parameters at the weak scale are derived from a
set of unified parameters at the GUT scale via renormalisation group
(RG) evolution, they come naturally as unphysical ``running''
quantities expressed in a minimal subtraction scheme such as the
modified Dimensional Reduction scheme, $\drbar$, which is usually
adopted since it preserves SUSY, at least up to two--loop
order. Interfacing the output of the RG evolution of the MSSM
parameters with a two--loop calculation of the Higgs boson masses
based on the OS scheme requires translating the $\drbar$ input
parameters into physical quantities, at the two--loop level for the
parameters entering the tree--level mass matrix (such as the CP--odd
Higgs boson mass) and at the one--loop level for the parameters
entering the one--loop corrections (such as the squark masses and
mixing angles). This induces additional complications in the
procedure. \s

A more direct strategy would be to perform the computation of the
Higgs boson masses directly in the $\drbar$ renormalisation scheme,
using the MSSM parameters as they come from the RG evolution. The
results for the Higgs boson masses must be equivalent to those of the
OS calculation up to terms that are formally of higher order in the
perturbative expansion. The numerical differences between the results
of the two calculations can be taken as an estimate of the size of the
corrections that are still uncomputed. They can be viewed, together
with the differences due to changes in the renormalisation scale at
which the corrections are evaluated, as part of the theoretical
uncertainty in the calculation.\s

A first purpose of this paper is to present the implementation of a
purely $\drbar$ calculation of the neutral MSSM Higgs boson masses,
based on the one--loop results of Ref.~\cite{pbmz} and the two--loop
results of Refs.~\cite{dsz,bdsz,dds,dstad}, into the latest versions
of three public codes for the RG evolution of the MSSM parameters and
the computation of the superparticle and Higgs boson mass spectrum,
i.e.~\softsusyv, \sphenov\ and \suspectv.  We also derive the small
$\oabatau$ contributions to complete the calculation of the two--loop
corrections to the Higgs boson masses controlled by the
third--generation Yukawa couplings.\s

A second purpose of this paper is to estimate the various theoretical
uncertainties in the determination of the Higgs boson masses. Using
representative choices for the high--energy boundary conditions, known
as Snowmass Points and Slopes (SPS) \cite{snowmass}, we first compare
the outputs of the three codes, which are generally in very good
agreement. We outline the reasons for the small residual discrepancies
and briefly discuss the effect on the Higgs boson masses of various
two--loop refinements of the general calculation. We then compare the
results of the $\drbar$ computations of the Higgs masses with those of
the OS computation in the program \feynhiggs\ and obtain an estimate
of the residual scheme dependence.  We also discuss the
renormalisation scale dependence of our results for the physical
masses, which can also be taken as an estimate of the theoretical
uncertainty, and the effect of the approximation of zero external
momentum in the two--loop corrections.  A similar analysis is
performed in the context of the unconstrained MSSM.  Finally, we
estimate the uncertainties in the Higgs boson masses arising from the
experimental error in the determination of the top and bottom quark
masses and the strong and electromagnetic coupling constants.  \s

A final purpose of this paper is to study the impact of these results
on MSSM Higgs phenomenology, under different assumptions for the
mechanism of supersymmetry breaking. In particular, we estimate the
upper bounds on the lighter Higgs boson mass in the general MSSM and
in the mSUGRA, GMSB and AMSB scenarios once the latest values of the
SM parameters (in particular the new experimental value for the top
quark mass, $M_t= 178.0 \pm 4.3$ GeV \cite{toptev}) are used and the
theoretical and experimental uncertainties are included; the knowledge
of the precise value of this bound might be crucial to discriminate
various scenarios of SUSY--breaking.  We also discuss the implications
of the LEP2 constraints on the MSSM Higgs sector and derive a lower
bound on the value of $\tb$, once various sources of uncertainty are
taken into account. We then compare our results with those of previous
analyses \cite{scanadk,scansven1,scansven2}, which employ different
computations of the Higgs boson masses, different codes for the
determination of the (s)particle spectrum and which include only a
subset of the errors discussed in the present analysis.\s

The rest of the paper is organised as follows.  In section 2, we
describe the general structure of the spectrum generators, focusing on
the aspects that are more critical to the mechanism of EWSB and the
computation of the Higgs boson masses. In section 3 we compute the
MSSM Higgs boson masses, compare the results from the three codes, and
analyse the effect of various two--loop refinements, including the
small effect of the $\oabatau$ and $\oatauq$ corrections. We also
discuss the theoretical uncertainties associated with the scheme
dependence (comparing the $\drbar$ and OS computations), the
renormalisation scale dependence (variation with $\qewsb$), and the
possible effect of the non--zero external momenta in the two--loop
self--energies. In section 4 we study the impact of our results and
the effects of the uncertainties on the upper bound on the lighter
Higgs boson mass (and on its decay modes), as well as on the lower
bound on the parameter $\tb$, in various scenarios for the mechanism
of SUSY breaking.  We present our conclusions in section 5. Finally,
in appendix A we provide for completeness the general formulae for
the computation of the EWSB conditions and the Higgs masses, and in
appendix B we provide explicit analytic formulae for the
$\oabatau$ corrections to the MSSM Higgs boson masses and tadpoles.


\section{Determining the MSSM mass spectrum}
\label{sec:codes}

In this section, we describe the calculation of the MSSM mass spectrum
implemented in the latest versions of three publicly available
spectrum generators, i.e. \softsusyv, \sphenov\ and \suspectv. We
start with the general algorithm used, then discuss some of the
differences among the individual codes.  The spectrum generator codes'
manuals \cite{softsusy,suspect,spheno} describe in detail the
approximations used to perform their calculations. Here, we simply
discuss the aspects of the calculation that are more critical to the
Higgs boson masses.\s

Assuming CP and R--parity conservation, the MSSM RGEs consist of some
110 coupled non--linear first--order homogeneous ordinary differential
equations. The calculation of the MSSM mass spectrum consists of a
two--boundary problem: we must solve these differential equations
given boundary conditions at two values of the independent variable,
the renormalisation scale $Q$. Evolution of these RGEs is a standard
problem, and there exist many techniques to perform the
task~\cite{numericalrecipes}. The high--scale (theoretical) boundary
conditions are upon some of the parameters in the soft SUSY--breaking
Lagrangian, and are specified by the model assumed for SUSY breaking
(e.g., mSUGRA, AMSB, GMSB) and its parameters. They are often applied
at the scale $\mgut \sim {\mathcal O}(10^{16})$ GeV, at which the
running electroweak gauge couplings unify:
\begin{equation}
g_1(\mgut) = g_2(\mgut),
\end{equation}
where $g_1 \equiv \sqrt{5/3}\,\gp$ and $g_2 \equiv g$ with $g$ and
$\gp$ the couplings associated with the ${\rm SU(2)_L}$ and ${\rm
U(1)_Y}$ gauge groups of the MSSM, respectively. We do not enforce
that the ${\rm SU(3)_C}$ coupling constant $g_s \equiv g_3$ unifies
with $g_1$ and $g_2$ at the high scale; the few percent discrepancy
from real unification is assumed to be accounted for by threshold
corrections at $\mgut$. A case apart is the GMSB model, in which the
high--scale boundary conditions are set at the messenger scale $M_{\rm
mess}$.\s

The weak--scale boundary conditions set the gauge and Yukawa couplings
by matching the running MSSM parameters to the experimental data at
some renormalisation scale, usually taken as $Q=\polemz$. This step is
quite involved, and requires subtracting the radiative corrections
from the experimental data in order to arrive at the
$\drbar$--renormalised MSSM parameters. Using the formulae of
Ref.~\cite{pbmz} and references therein, the MSSM $\drbar$ gauge
couplings $g \,, \gp \,, g_3$ and the electroweak parameter $v$ can be
computed at $Q=\polemz$ from a set of four experimental input
parameters. \s

In the spirit of the {\em SUSY Les Houches Accord} (SLHA) for
interfacing the spectrum generators with other computer codes
\cite{slha}, these input parameters can be chosen as: $G_F$, the Fermi
constant determined from the muon decay; $\polemz$, the pole mass of
the $Z$ boson; $\alpha_{\rm em}(\polemz)^{\msbar}$, the five--flavour
SM electromagnetic coupling at the scale $\polemz$ in the $\msbar$
scheme; $\as(\polemz)^{\msbar}$, the five--flavour SM strong coupling
at the scale $\polemz$ in the $\msbar$ scheme. In particular, the
running parameters $g \,, \gp$ and $v$ are connected to the running
$Z$--boson mass $\mz$ by the relation:
\be
\label{mzrun}
\mz^2 = \polemz^2 + {\rm Re}\, \Pi_{ZZ}^T(\polemz^2) 
= \frac{1}{4}\,(g^2 + \gpq)\,v^2,
\ee
where $\Pi_{ZZ}^T(\polemz^2)$ is the transverse part of the $Z$ boson
self--energy computed at a squared external momentum equal to the
squared pole $Z$ boson mass. \s

The Yukawa couplings $h_u$ ($u = u,c,t$) for the up--type quarks,
$h_d$ ($d = d,s,b$) for the down--type quarks and $h_{\ell}$ ($\ell =
e,\mu,\tau$) for the leptons are determined from the corresponding
running fermion masses as
\be
\label{yukawas}
h_u = \frac{\sq2\,m_u}{v\,\sin\beta}\,,\;\;\;\;\;\;
h_d = \frac{\sq2\,m_d}{v\,\cos\beta}\,,\;\;\;\;\;\;
h_{\ell} = \frac{\sq2\,m_{\ell}}{v\,\cos\beta}\,,
\ee
where $\tb \equiv v_2/v_1$ is the ratio of the vacuum expectation
values (VEVs) of the two MSSM neutral Higgs fields, $H_1^0$ and
$H_2^0$, that also obey the relation $v_1^2 + v_2^2 = v^2$. \s

The running fermion masses $m_f$ (with $f=u,d,\ell$) in
eq.~(\ref{yukawas}) can be derived at the one--loop level from the
corresponding pole masses $M_f$, through the relation
\be
m_f = M_f + \Sigma_f(M_f)
\ee
where $\Sigma_f(M_f)$ is the one--loop fermion self--energy computed
at an external momentum equal to the pole mass. In the case of the top
quark, the self--energy includes also the leading two--loop standard
QCD corrections
\be
m_t = \polemt + \Sigma_t(\polemt) + (\Delta m_t)^{\rm 2-loop,\,QCD}
\label{corrmt}
\ee
where the precise expression for $(\Delta m_t)^{\rm 2-loop,\,QCD}$
depends on the renormalisation scheme in which the parameters entering
the one--loop self--energy $\Sigma_t$ are expressed. In the case of
the bottom quark, the SLHA prescribes to take as input the SM running
mass in the $\msbar$ scheme, $\mb(\mb)^{\msbar}$. Moreover, a
``resummation'' procedure is required (see e.g. Ref.~\cite{resum}) in
order to properly take into account the large QCD corrections, as well
as the $\tb$--enhanced SUSY corrections \cite{hrs}, to the relation
between the input bottom mass and the corresponding MSSM, $\drbar$
Yukawa coupling. We extract the latter, via eq.~(\ref{yukawas}), from
the MSSM, $\drbar$ bottom mass $\widehat{m}_b$, defined at the scale
$Q=\polemz$ by the following matching condition:
\be
\label{mbmssm}
\widehat{m}_b \equiv  m_b(\polemz)^{\drbar}_{\rm MSSM} =
\frac{\ov{m}_b}{1-\Delta_b}
\ee
where $\ov{m}_b \equiv m_b(\polemz)^{\drbar}_{\rm SM}$ is the SM,
$\drbar$ bottom mass, obtained by evolving $\mb(\mb)^{\msbar}$ up to
the scale $Q=\polemz$ with the appropriate RGE, in order to resum the
QCD corrections, and then converting it to the $\drbar$ scheme;
$\Delta_b \equiv \Sigma_b(\widehat{m}_b)/\widehat{m}_b$ accounts for
the remaining non--gluonic corrections, some of which are enhanced by
a factor $\tb$. It has been shown \cite{resum} that defining the
running MSSM bottom mass as in eq.~(\ref{mbmssm}) guarantees that the
large threshold corrections of ${\cal O}(\as \tb)^n$ are included in
$\widehat{m}_b$ to all orders in the perturbative expansion.  In the
case of the $\tau$ lepton, the only $\tb$--enhanced corrections to be
included are those controlled by the electroweak gauge couplings,
stemming from chargino--slepton loops.\s

The computation of the MSSM mass spectrum is also complicated by the
requirement that the electroweak symmetry is broken radiatively: at
some weak scale $\qewsb$, the Higgs VEVs $v_1$ and $v_2$ can be
computed by the minimisation of the MSSM effective potential,
$\Veff$. In principle, a set of high--energy boundary conditions is
acceptable if, after RG evolution of the parameters down to $\qewsb$,
it leads to the correct value of the squared running mass for the $Z$
boson, as given in eq.~(\ref{mzrun}).  In practice, however, it is
more convenient to assume that {\em there is} successful electroweak
symmetry breaking, and trade two of the high--energy input parameters
for $v_1$ and $v_2$ (or, equivalently, for $v$ and $\tb$). The
minimisation conditions of the effective potential then allow one to
determine the weak--scale values of two MSSM parameters, usually
chosen as the Higgs mass term in the superpotential, $\mu$, and its
soft SUSY--breaking counterpart, $B$, the latter being related to the
squared running mass for the CP--odd Higgs boson $A$, $\marun$,
through the relation
\be
\label{marun}
\marun^2 = 2\,B/\sin 2\beta\,.
\ee

The computation of $\mu$ and $B$ beyond tree level requires the
knowledge of the so--called tadpole corrections to the minimisation
conditions of the effective potential. We provide the corresponding
formulae in appendix A. At the one--loop level, \softsusy,
\spheno\ and \suspect\ use the complete calculation of the tadpoles
given in Ref.~\cite{pbmz}. Among the two--loop corrections, the most
relevant are those controlled by the top and bottom Yukawa couplings
and by the strong gauge coupling. The three codes use the results of
Ref.~\cite{dstad} for the $\oatasabas$ corrections, and those of
Ref.~\cite{dds} for the $\oatqatababq$ corrections. Besides being
required by consistency with the Higgs boson mass calculations, these
two--loop corrections are relevant to stabilise the scale dependence
of the MSSM parameters obtained from the minimisation conditions of
the effective potential (see the discussion in Ref.~\cite{dstad}).
Note that also for consistency reasons, the Higgs mass parameters
$m_{H_1}^2$ and $m_{H_2}^2$ which enter the minimisation conditions on
$\Veff$ have to be evolved using the two--loop RGEs. \s
 
After determining the full set of MSSM parameters, expressed in the
$\drbar$ renormalisation scheme at the scale $\qewsb$, the codes
compute the spectrum of physical masses of the supersymmetric
particles.  The approximations employed in the computation of
sfermion, gluino, chargino and neutralino physical masses differ from
code to code, and we refer readers interested in the details to the
respective manuals. \s

On the other hand, due to extreme phenomenological relevance of the
mass spectrum of the MSSM Higgs sector (as the failure of detecting a
light Higgs boson already rules out a considerable fraction of the
MSSM parameter space), it is mandatory to employ a precise calculation
of the neutral Higgs boson masses that encompasses all the presently
available radiative corrections. The general formulae for the
computation of the Higgs masses are provided for completeness in 
appendix A.  For the analysis presented in this paper, the spectrum
generators \softsusy, \spheno\ and \suspect\ have been upgraded to
include the state--of--the--art radiative corrections in the Higgs
sector. \s

In the newest versions \softsusyv, \sphenov\ and \suspectv\, the
complete one--loop formulae, including the tadpole corrections, are
taken from Ref.~\cite{pbmz}, which also provides formulae for the
one--loop corrections to the charged Higgs boson mass. Concerning the
two--loop part, the three codes employ the effective potential results
of Ref.~\cite{dsz} for the $\oatas$ corrections, and those of
Ref.~\cite{dds} for the $\oatqatababq$ corrections. The formulae for
the $\oabas$ corrections are obtained from those for the $\oatas$
corrections with appropriate replacements. The two--loop corrections
are computed in the $\drbar$ renormalisation scheme, i.e. they require
that the one--loop part of the corrections is expressed in terms of
running, $\drbar$--renormalised MSSM parameters. This is a
particularly convenient scheme in the situation in which the MSSM
parameters are computed via RG evolution from a set of unified
high--energy boundary conditions.  \s

The two--loop corrections to the Higgs boson masses controlled by the
third--family Yukawa couplings can be completed by including the
contribution of the $\tau$ Yukawa coupling. The inclusion of the
$\oatauq$ corrections was discussed in Ref.~\cite{dds}, and in 
appendix B we present a new computation of the remaining corrections
of $\oabatau$. However, we found that the numerical impact of the
$\tau$ radiative corrections is very small in all the scenarios that
we considered in our analysis. \s

Other efforts to improve the two--loop calculation of the Higgs boson
masses include an effective potential computation of the corrections
to the lighter Higgs boson mass controlled by the electroweak
couplings \cite{Martin2}, and a first attempt to go beyond the
effective potential approximation by including the effect of non zero
external momenta in the two--loop propagators
\cite{Martin3}. Two--loop corrections to the charged Higgs boson mass
should also be calculated. We plan to include these results, as soon
as they become available in a suitable form, in future updates of our
spectrum generators.\s

It appears from the discussion above that the problem of determining
the MSSM mass spectrum involves several mass scales, at which boundary
conditions and EWSB conditions are imposed. As the computation of the
radiative corrections at some scale requires knowledge about
parameters that are determined at a different scale, an iterative
procedure is adopted. The algorithm adopted by the codes can be
summarised as follows:
\begin{enumerate}
\item 
An initial guess for the MSSM parameters is taken at $Q=\polemz$.
\item
The parameters are then evolved to $\mgut$ (or $M_{\rm mess}$ in the
GMSB case), where the high scale boundary conditions are
imposed. \label{steptwo}
\item
The new set of MSSM parameters is then evolved to $\qewsb$, where the
EWSB constraints are imposed.
\item 
The MSSM pole mass spectrum is calculated.
\item
The parameters are then evolved to $Q=\polemz$, where the Yukawa and
gauge couplings are matched to empirical data.
\item 
One then proceeds to step \ref{steptwo}, continuing until the MSSM
parameters converge upon stable values.
\end{enumerate}

Although the three codes employ the same general procedure in the
determination of the MSSM mass spectrum, they differ in many details
when it comes to the implementation of the single steps of the
calculation. In particular, the differences that most affect the
results for the Higgs boson masses are:
\begin{itemize}
\item[--]
\spheno\ uses two--loop RGE for all the MSSM parameters, whereas
\softsusy\ and \suspect\ use one--loop RGE for the soft SUSY--breaking
masses and interaction terms of the sfermions (and two--loop RGE for
the remaining MSSM parameters);
\item[--]
\softsusy\ and \spheno\ include by default in the RGE the
contributions of the Yukawa couplings for the three generations of
fermions, whereas \suspect\ works under the approximation that the
Yukawa couplings of the first two generations are zero. However, we
found that the omission of the first two-family Yukawa couplings has
always a very tiny impact on the Higgs boson masses, which is limited to the
case of large $\tb$;
\item[--] 
\spheno\ defines the default EWSB scale $\qewsb = \sqstop$
in terms of the physical stop masses, whereas \softsusy\ and \suspect\
define it in terms of the running stop masses (note, however, that all
the codes also allow for an arbitrary choice of $\qewsb$);
\item[--]
when computing the running top mass at the scale $Q=\polemz$ according
to eq.~(\ref{corrmt}), \spheno\ uses the running top mass itself in
the one--loop part of the top self--energy, whereas \softsusy\ and
\suspect\ use the pole top mass.  This induces a difference in the
formulae for the two--loop QCD corrections, which is properly taken
into account by the codes;
\item[--]
when computing the running bottom mass at the scale $Q=\polemz$
according to eq.~(\ref{mbmssm}), \spheno\ defines $\Delta_b$ in terms
of the full bottom self--energy, whereas \softsusy\ and \suspect\
include only the $\tb$--enhanced contributions;
\item[--]
when computing the threshold corrections to the Yukawa couplings at
the scale $Q=\polemz$, \spheno\ uses the pole top mass in the $Z$ boson
self energy (required to obtain the running EWSB parameter $v$, see
eq.~(\ref{mzrun})) whereas \softsusy\ and \suspect\ use the running
top mass;
\item[--]
in the computation of the various threshold corrections at the scale
$Q=\polemz$, \softsusy\ and \spheno\ use for all the sparticle masses
and mixing angles the $\drbar$ running values computed at $Q=\polemz$,
whereas \suspect\ uses the values computed at $\qewsb$ for the
neutralino and chargino masses and mixing angles;
\item[--]
in the computation of the various radiative corrections at the scale
$\qewsb$, \spheno\ and \suspect\ compute $v(\qewsb)$ through
eq.~(\ref{mzrun}), whereas \softsusy\ takes the value computed at the
scale $Q=\polemz$ and evolves it up to $\qewsb$ with the appropriate
RGE.

\end{itemize}

All of the differences listed above correspond to effects that are of
higher order with respect to the accuracy required by the calculation
of the Higgs boson masses. However, as will be discussed in the next
section, they induce non negligible variations in the results for the
Higgs boson masses, amounting to something less than a GeV for the
lighter Higgs boson and a few GeV for the heavier Higgs bosons. For
the time being, such variations must be taken as contributing to the
uncertainties that still affect the calculation. Only when more
precise computations of the radiative corrections become available 
will it be possible to fix the procedural ambiguities and reduce the
uncertainty in the results.\s

Before presenting our numerical results in the next section, we point
out that the precise determination of the Higgs boson masses that we
are discussing here does not only apply in constrained models such as
mSUGRA, GMSB and AMSB. The neutral Higgs boson masses can also be
determined in a general (unconstrained) MSSM, where the soft
SUSY--breaking parameters are set by hand at the weak scale. The three
spectrum generators have options in which such a procedure can be
applied (although in the case of \softsusy\ the procedure is not yet
completely implemented). The programs can therefore also be viewed as
$\drbar$ Higgs boson mass calculators in a general MSSM framework,
analogous to the {\tt FeynHiggs} calculation in the On--Shell
scheme\footnote{In fact, some of us are planning to include the part
of the spectrum generators concerning the Higgs sector into the
program {\tt HDECAY} \cite{hdecay} which evaluates the Higgs boson
decay widths and branching ratios in the SM and in the MSSM.}.  At the
end of the next section, we will present some illustrations of the
calculation of the Higgs boson masses in the general or unconstrained
MSSM.


\newpage

\section{Precise results for the MSSM Higgs boson masses}

In this section we discuss in detail the results of the two--loop
$\drbar$ computation of the Higgs boson masses and of the EWSB
conditions in the constrained MSSM (cMSSM), focusing on the mSUGRA,
GMSB and AMSB models for SUSY breaking; we also display some results
in the unconstrained MSSM. In a first part, we specify the physical
scenarios in which we work.  We then compare the results of the three
codes, \softsusy, \spheno\ and \suspect, in the case of the cMSSM and
discuss the impact of the various radiative corrections that affect
the Higgs boson masses and the EWSB conditions.  We subsequently
discuss the various theoretical uncertainties which affect the Higgs
boson mass calculations: the scheme dependence, the renormalisation
scale dependence and the effect of the external momenta in the
two--loop self--energies.  We finally summarise the theoretical
uncertainty on the Higgs boson mass determinations and discuss the
impact of experimental errors on the SM input parameters on these
calculations.

\subsection{The physical scenarios}
\label{sec:scenarios}

We will first work in the framework of constrained MSSM scenarios and
illustrate our results in the case of the mSUGRA, GMSB and AMSB models
for SUSY--breaking. For what concerns the choice of the high--energy
boundary conditions, we restrict ourselves to six (out of ten)
Snowmass Points and Slopes (SPS) \cite{snowmass}. Four of these points
are mSUGRA ones, where the relevant input parameters are three
universal soft SUSY--breaking terms, $\m0\,,\mhf$ and $A_0$, the value
of $\tb$ (expressed in the $\drbar$ scheme at the renormalisation
scale $Q=M_Z$) and the scale--invariant sign of $\mu$:
\[
\begin{array}{lccccc}
{\rm SPS\; 1a}: &  
\m0 = 100\; \gev, &
\mhf = 250\; \gev, &
A_0 =   -100\;\gev, &
\tb = 10\,, & 
\mu > 0\,, \\
\vspace*{1mm} 
{\rm SPS\; 2}: &  
\m0 = 1450\; \gev, &
\mhf = 300\; \gev, &
A_0 =   0\,, &
\tb = 10\,, & 
\mu > 0\,, \\
\vspace*{1mm} 
{\rm SPS\; 4}: & 
\m0 = 400\; \gev, &
\mhf = 300\; \gev, &
A_0 = 0\,, &
\tb = 50\,, &
\mu > 0\,,\\ 
\vspace*{1mm} 
{\rm SPS\; 5}: & 
\m0 = 150\; \gev, &
\mhf = 300\; \gev, &
A_0 = -1\; {\rm TeV}\,, &
\tb = 5\,, &
\mu > 0\,.
\end{array}
\]

The first choice, SPS1a, corresponds to a ``standard'' mSUGRA
scenario; the second choice, SPS2, is called a ``focus point''
scenario and has interesting implications for dark matter abundance;
SPS4 is characterised by a large value of $\tb$ while SPS5 is
characterised by a large value of the stop mixing parameter and leads
to a light stop squark.  The focus point and the large $\tb$ scenario
are known to be the most sensitive to the various approximations made
by the spectrum generators \cite{comparison}. \s

We will also choose one point in the GMSB scenario, where the input
parameters are the SUSY--breaking scale $\Lambda$, the messenger scale
$M_{\rm mess}$, the messenger index $N_{\rm mess}$, $\tb$ and the sign
of $\mu$: the SPS8 point in which the lightest SUSY particle is the
lightest neutralino
\[
\begin{array}{lccccc}
\vspace*{1mm} 
{\rm SPS\; 8}: & 
\Lambda = 100\; {\rm TeV}, &
M_{\rm mess}= 200\; {\rm TeV}, &
N_{\rm mess} = 1\,, &
\tb = 15\,, &
\mu > 0\,.\\
\end{array}
\]

The last point, denoted SPS9, is for an AMSB scenario described by a
large gravitino mass $m_{3/2}$, a common mass term for the scalars
$m_0$, $\tb$ and the sign of $\mu$:
\[
{\rm SPS\; 9}:  \
m_{3/2}= 60\; {\rm TeV}, \
\m0 = 450\; \gev, \
\tb = 10\,, \
\mu > 0\,.
\]

We remark in passing that in Ref.~\cite{snowmass} the SPS scenarios
are defined in terms of the low--energy MSSM $\drbar$ parameters, as
computed by the code {\tt IsaJet7.58} at the weak scale
$\qewsb$. Thus, calling ``SPS scenarios'' the above choices of
high--energy boundary conditions always involves a slight abuse of
language.\s

In a next step, we will work in the case of the unconstrained MSSM.
For the latter scenario, we take the so--called ``phenomenological
MSSM" (pMSSM)~\cite{GDR}, in which some constraints have been imposed
(such as CP conservation, flavour diagonal sfermion mass and coupling
matrices and universality of the first and second generations). These
constraints ensure that an appreciable fraction of parameter space of
the pMSSM has viable phenomenology. The model involves 22 free
parameters in addition to those of the SM:

\begin{itemize}
\item[--]
\vspace*{-2mm}
the parameter $\tb$; 
\item[--]
\vspace*{-2mm}
the two soft SUSY--breaking Higgs mass parameters $m^2_{H_1}$ and
$m^2_{H_2}$, which can be traded against $\ma$ and $\mu$ by enforcing 
the EWSB conditions; 
\vspace*{-2mm}
\item[--]
the three gaugino mass parameters $M_1, M_2$ and $M_3$; 
\vspace*{-2mm}
\item[--] 
the diagonal sfermion mass parameters $m_{ {\tilde f}_{L,R}}$: five
for the third generation sfermions and five others for the
first/second generation sfermions;
\vspace*{-2mm}
\item[--]
the trilinear sfermion couplings $A_f$: three for the third generation 
sfermions and three others for the first/second generation sfermions.
\end{itemize}

Fortunately, most of these parameters have only a marginal impact on
the Higgs boson masses, a fact which considerably simplifies the
analysis. A very important parameter in the context of the radiative
corrections to the MSSM Higgs boson masses is the mixing parameter in
the stop sector, $X_t = A_t -\mu \cot\beta$. For our numerical
illustrations, we will consider three pMSSM scenarios, in which the
SUSY spectrum is rather heavy: the soft SUSY--breaking masses of the
sfermions (all set equal to a common mass $M_S$), the SU(3) gaugino
mass $M_3$ ($M_1$ and $M_2$ being related to the latter through the
usual GUT relation), the pseudoscalar Higgs boson mass $\ma$ and the
higgsino mass parameter $\mu$ are set to 1 TeV; for $\tb$ we choose a
moderate value, $\tb=10$; furthermore, we consider three cases for the
mixing $X_t$ in the stop sector: no mixing, $X_t=0$, typical mixing,
$X_t=M_S$, and large mixing, $X_t = \sqrt{6} M_S$ (the other sfermion
trilinear couplings will be also set to 1 TeV). All of the MSSM input
parameters are taken as $\drbar$ running quantities computed at the
scale $Q = 1$ TeV, apart from $\tb$ which is computed at $Q=\polemz$,
and $\ma$ which denotes the physical mass.  Thus we will have three
pMSSM points:
\bea
{\rm pMSSM1}: && \; M_S=A_{b,\tau}=M_3=M_A=\mu=1\; {\rm TeV}, \ \tb=10, \ 
X_t=0\,, \nonumber \\
{\rm pMSSM2}: &&\;  M_S=A_{b,\tau}=M_3=M_A=\mu=1\; {\rm TeV}, \ \tb=10, \ 
X_t=M_S\,, \nonumber \\
{\rm pMSSM3}: &&\; M_S=A_{b,\tau}=M_3=M_A=\mu=1\; {\rm TeV}, \ \tb=10, \ 
X_t=\sqrt{6}M_S\,. \nonumber 
\eea

The first and third of these scenarios are rather close to those which
have been used as benchmark points in the interpretation in the
unconstrained MSSM of the Higgs boson searches by the LEP2
collaborations \cite{benchmarks}. \s

Finally, the SM input parameters at the weak scale are fixed according to 
the SLHA prescription. In particular, we take for the electroweak and strong
parameters \cite{pdg}:
\[
G_F = 1.16639\, 10^{-5}\;{\rm GeV}^{-2},\;\;\;\;\;
\polemz = 91.1876\; {\rm GeV},
\]
\be
\label{inputew}
\alpha_{\rm em}^{-1}(\polemz)^{\msbar} = 127.934 \pm 0.027,\;\;\;\;\;
\as(\polemz)^{\msbar} = 0.1172 \pm 0.002, \
\ee
and for the third--generation fermion masses the values \cite{toptev,pdg}:
\be
\label{inputmass}
\polemt = 178.0 \pm 4.3\; {\rm GeV},\ \ \
m_b(m_b)^{\msbar} = 4.25 \pm 0.25\; {\rm GeV},\ \ \
M_{\tau} = 1.777 \; {\rm GeV}.
\ee

In most of our discussion we take the SM input parameters to be equal
to their central values (the errors on $G_F,\, \polemz$ and $M_{\tau}$
are indeed very small and we omit them in the equations
above). However, in section \ref{secexperr} we discuss the uncertainty
in the determination of the Higgs masses arising from the experimental
errors on some of these parameters.

\subsection{Determination of the neutral Higgs masses in the cMSSM}
\label{secSPS}

\subsubsection{Comparing the results of the three codes}
\label{comparecodes}

We start our discussion by comparing the values of the neutral Higgs
boson masses and of the superpotential parameter $\mu$ in the six SPS
scenarios discussed above, as they result from the computations of
\softsusy, \spheno\ and \suspect\ at the default renormalisation scale
$\qewsb = \sqstop$, once all of the two--loop radiative corrections
are implemented.  The corresponding values for the physical masses of
the CP--even Higgs bosons, $\mh$ and $\mH$, for the physical mass of
the CP--odd Higgs boson, $\ma$, and for the parameter $\mu$ (the
latter interpreted as a $\drbar$ running parameter computed at
$\qewsb$) are given in the tables \ref{tabmh}, \ref{tabmH},
\ref{tabma} and \ref{tabmu}, respectively. \s

Concerning the lighter CP--even Higgs boson mass, $\mh$, it can be
seen that the agreement between the three codes is very good, the
discrepancies being contained in a half GeV. In the cases of the
heavier Higgs boson masses and of $\mu$, which are more sensitive than
$\mh$ to small variations in the RG evolution of the soft
SUSY--breaking parameters, the discrepancies amount to at most few
GeV, generally below the 1\% level. We find this agreement very
satisfactory, taking into account the fact that the results are
obtained with three independent codes that -- although based on the
same set of formulae for the corrections to the Higgs boson masses and
the EWSB conditions -- differ in many details of the calculation, as
outlined in the previous section. \s

We have checked that, if we force the three codes to use the same
computation of the threshold corrections to the gauge and Yukawa
couplings and the same set of RGE, the residual discrepancies in the
results for the neutral Higgs boson masses and $\mu$ become
negligible. However, we stress again that the differences between the
three codes are a matter of choice, because they all correspond to
effects that are of higher order with respect to the accuracy required
by the calculation. The spread in the values of the Higgs boson masses
and $\mu$ resulting from tables \ref{tabmh}--\ref{tabmu} has to be
considered as part of the theoretical uncertainty, and will be reduced
only when more refined calculations of the radiative corrections
become available. \s

\begin{table}[!ht]
\renewcommand{\arraystretch}{1.2}
\begin{center}
\begin{tabular}{|l|c|c|c|c|c|c|c|c|c|c|}
\hline
Code & SPS1a & SPS2 & SPS4 & SPS5 & SPS8 & SPS9\\
\hline
\softsusy & 112.1 & 116.8 & 114.1 & 116.3 & 115.4 & 117.4 \\
\spheno & 112.2 & 117.1 & 114.3 & 116.5 & 115.8 & 117.8 \\
\suspect & 112.1 & 116.8 & 114.1 & 116.1 & 115.5 & 117.5 \\
\hline 
\end{tabular}
\caption{The lighter CP--even Higgs boson mass, $\mh$, in the six SPS
scenarios, as computed by \softsusyv, \sphenov\ and \suspectv. The SM
input parameters are chosen as in
eqs.~(\ref{inputew})--(\ref{inputmass}).}
\label{tabmh}
\end{center}
\vspace*{-.5cm}
\end{table}

\begin{table}[!ht]
\renewcommand{\arraystretch}{1.1}
\begin{center}
\begin{tabular}{|l|c|c|c|c|c|c|c|c|c|c|}
\hline
Code & SPS1a & SPS2 & SPS4 & SPS5 & SPS8 & SPS9\\
\hline
\softsusy & 406.5 & 1553.0 & 355.8 & 686.8 & 550.4 & 1056.9\\
\spheno & 406.0 & 1554.6 & 360.5 & 686.5 & 552.4 & 1051.1 \\
\suspect & 406.5 & 1552.1 & 355.3 & 686.9 & 550.6 & 1056.6\\
\hline
\end{tabular}
\caption{Same as table \ref{tabmh} for the heavier CP--even Higgs boson 
mass,
$\mH$.}
\label{tabmH}
\end{center}
\vspace*{-.5cm}
\end{table}

\begin{table}[!ht]
\renewcommand{\arraystretch}{1.1}
\begin{center}
\begin{tabular}{|l|c|c|c|c|c|c|c|c|c|c|}
\hline
Code & SPS1a & SPS2 & SPS4 & SPS5 & SPS8 & SPS9\\
\hline
\softsusy & 406.2 & 1552.9 & 355.8 & 687.0 & 550.1 & 1056.8 \\
\spheno & 405.7 & 1554.5 & 360.5 & 686.9 & 552.1 & 1051.0 \\
\suspect & 406.1 & 1552.0 & 355.3 & 687.2 & 550.3 & 1056.5\\
\hline
\end{tabular}
\caption{Same as table \ref{tabmh} for the CP--odd Higgs boson mass, $\ma$.}
\label{tabma}
\end{center}
\vspace*{-.5cm}
\end{table}

\begin{table}[!ht]
\renewcommand{\arraystretch}{1.1}
\begin{center}
\begin{tabular}{|l|c|c|c|c|c|c|c|c|c|c|}
\hline
Code & SPS1a & SPS2 & SPS4 & SPS5 & SPS8 & SPS9\\
\hline
\softsusy & 364.8 & 586.5 & 413.8 & 631.2 & 440.1 & 1011.8\\
\spheno & 364.3 & 588.2 & 414.7 & 631.2 & 442.2 & 1005.9 \\
\suspect & 364.7 & 583.6 & 413.6 & 631.3 & 440.3 & 1011.1 \\
\hline
\end{tabular}
\caption{Same as table \ref{tabmh} for the superpotential Higgs mass
parameter $\mu$, expressed in the $\drbar$ renormalisation scheme at
the scale $\qewsb$.}
\label{tabmu}
\end{center}
\vspace*{-.5cm}
\end{table}

\subsubsection{Impact of the radiative corrections}
\label{secradcorr}

We now discuss the importance of the various radiative corrections to
the Higgs boson masses and tadpoles that are taken into account by the
codes \softsusy, \spheno\ and \suspect. Tables \ref{twocorrh} and
\ref{twocorrH} show the variations of $\mh$ and $\mH$, respectively,
in the six SPS scenarios, as a result of different approximations for
the radiative corrections. The first three rows of each table contain
the values of the masses as obtained by \suspect\ employing,
respectively, the tree--level, one--loop and two--loop formulae for
the Higgs masses and EWSB conditions. The variations shown in the
second part of the tables result from progressively switching on the
various two--loop corrections to the Higgs boson masses and EWSB
conditions. Explicitly, we examine the two loop $\oatasabas$
corrections (first line), the $\oatqatababq$ corrections (second line)
and the newly calculated $\oatauqatab$ corrections (third line).

\begin{table}[h]
\renewcommand{\arraystretch}{1.1}
\vspace{.5cm}
\begin{center}
\begin{tabular}{|l|c|c|c|c|c|c|}
\hline
Approximation & SPS1a & SPS2 & SPS4 & SPS5 & SPS8 & SPS9 \\
\hline 
Tree--level &  88.6 &  87.8 &  90.2 &  82.9 &  89.2 &  88.2 \\
One--loop   & 109.7 & 112.3 & 111.1 & 113.0 & 111.3 & 112.6 \\
Two--loop   & 112.1 & 116.8 & 114.1 & 116.1 & 115.5 & 117.5 \\
\hline
$\oatasabas$   & +3.1 & +5.7 & +3.8 & +3.0 & +5.3 & +5.8 \\
$\oatqatababq$ & $-0.6$ & $-1.1$ & $-0.8$ & +0.2 & $-1.1$ & $-1.0$ \\
$\oatauqatab$  &$<10^{-3}$&$<10^{-3}$&$<10^{-3}$
&$<10^{-3}$&$<10^{-3}$&$<10^{-3}$ \\
\hline
\end{tabular}
\caption{The lighter CP--even Higgs boson mass, $\mh$, in the six SPS
scenarios, as computed by \suspect\ under different approximations for
the radiative corrections. The first three rows contain the mass (in
GeV) computed at tree--level, one-- and two--loop, respectively; the
last three rows contain the shifts (in GeV) due to the different
two--loop contributions.}
\label{twocorrh}
\vspace{.5cm}
\begin{tabular}{|l|c|c|c|c|c|c|}
\hline
Approximation & SPS1a & SPS2 & SPS4 & SPS5 & SPS8 & SPS9 \\
\hline 
Tree--level & 402.5 & 1542.7 & 356.3 & 689.6 & 525.7 & 1043.1 \\
One--loop   & 406.8 & 1551.3 & 356.0 & 688.2 & 549.0 & 1056.1 \\
Two--loop   & 406.5 & 1552.1 & 355.3 & 686.9 & 550.6 & 1056.6 \\
\hline
$\oatasabas$   & $-0.1$ & +1.1 & $-0.1 $ & $-0.9$ & $+2.5$ & +0.9 \\
$\oatqatababq$ & $-0.3$ & $-0.3$ & $-0.5$ & $-0.3$ & $-0.9$ & $-0.4$ \\
$\oatauqatab$  &$<10^{-3}$&$<10^{-3}$&0.01&$<10^{-3}$&$<10^{-3}$&$<10^{-3}$\\
\hline
\end{tabular}
\caption{Same as table~\ref{twocorrh} for the heavier CP--even Higgs boson
mass, $\mH$}
\label{twocorrH}
\vspace*{-5mm}
\end{center}
\end{table}

Comparing the first three rows of table \ref{twocorrh} it can be seen
that the one--loop corrections to $\mh$ are positive and of the order
of 20 GeV, while the two--loop corrections to $\mh$ are also positive
and of the order of 2--5 GeV. This is to be contrasted with the case
of the OS calculations (see e.g.~Ref.~\cite{bdsz}), where both the
one--loop and two--loop corrections are usually larger, and the
two--loop corrections are negative. However, as will be discussed in
section \ref{secosdrbar}, the two--loop results of the OS and $\drbar$
calculations agree within 2 GeV in general.  From the first three rows
of table \ref{twocorrH}, instead, it can be seen that the one--loop
corrections to $\mH$ amount in general to few GeV, and their impact is
usually of the order of 1\% (apart from the scenario SPS8, where they
reach 4\%). The two--loop corrections are also small, being of the
order of one GeV or less. This is due to the fact that, in the
considered scenarios, the heavier Higgs boson masses are large already
at the tree--level, so that the inclusion of radiative corrections can
have only a small effect. \s

Further information on the relative size of the various two--loop
corrections to the Higgs boson masses can be obtained from the lower
parts of tables \ref{twocorrh} and \ref{twocorrH}. In the case of
$\mh$, we see that the $\oatasabas$ corrections, controlled by the
strong gauge coupling, are dominant (amounting to 3--6 GeV) and
positive, whereas the $\oatqatababq$ corrections, involving only the
top and bottom Yukawa coupling, are smaller (of the order of one GeV)
and usually negative (apart from the scenario SPS5). This is again
to be contrasted with the case of the OS calculations, where the
$\oatas$ corrections, which are always leading at two--loop order, are
negative, and the subleading $\oatq$ corrections are positive (in both
OS and $\drbar$ calculations, the corrections controlled by the bottom
Yukawa coupling are negligible for small or moderate values of
$\tb$). In the case of $\mH$, instead, the trend is less definite: the
$\oatasabas$ corrections can be either positive or negative in the
different scenarios, and they are not always larger than the
$\oatqatababq$ ones.\s

Concerning the corrections controlled by the $\tau$ Yukawa coupling,
it can be seen that their effect is usually negligible, amounting to
less than a MeV for both $\mh$ and $\mH$. Only in the scenario
SPS4, in which the $\tau$ Yukawa coupling is enhanced by a large value
of $\tb$, a very tiny (10 MeV) effect is visible in the heavier Higgs
boson mass. This suppression of the $\tau$ corrections even in the
case of large $\tb$ is explained by the smallness of $\mtau$ (even in
comparison with $\mb$), by the absence of colour enhancements, and by
the fact that the only $\tb$--enhanced threshold corrections to the
relation between $\htau$ and $\mtau$ are those controlled by the
electroweak gauge couplings.\s

One has to bear in mind, however, that all of the above considerations on
the relevance of the various two--loop corrections to $\mh$ and $\mH$
depend to some extent on the choice of the electroweak symmetry
breaking scale, taken by default as $\qewsb = \sqstop$. For example,
as will be clear from the figures in section \ref{secscaledep}, with
the choice $\qewsb = \polemz$ the two--loop corrections to $\mh$
become smaller (1--2 GeV) and negative, while the two--loop
corrections to $\mH$ amount to several GeV and they are
positive. Also, in the scenario SPS4 the corrections to $\mH$
controlled by the $\tau$ Yukawa coupling get somewhat larger ($\sim
0.2$ GeV). It is thus clear that -- although it is always possible to
choose $\qewsb$ in such a way that some of the two--loop corrections
are small -- only the proper inclusion of all of the leading two--loop
corrections to the Higgs boson masses allows us to obtain reliable
results that do not depend strongly on a preconceived choice of the
renormalisation scale.


\subsubsection{Impact of some higher--order effects}

To understand the importance of several higher--order effects that are
taken into account by \softsusy, \spheno\ and \suspect, it may be
useful to discuss their individual impact on the numerical values of
the Higgs masses. 

\vspace{.5cm}

\begin{table}[htbp]
\renewcommand{\arraystretch}{1.1}
\begin{center}
\begin{tabular}{|l|c|c|c|c|c|c|}
\hline
Approximation & SPS1a & SPS2 & SPS4 & SPS5 & SPS8 & SPS9 \\
\hline
Default  & 112.17 & 117.06 & 114.34 & 116.46 & 115.80 & 117.82 \\
\hline
2--loop sfermion RGE &  0.07 &  0.15 &  0.04 &  0.03 &  0.03 &  0.03 \\
No $\Delta m_t^{\rm 2loop}$ &  0.26 &  0.33 &  0.28 &  0.35 &  0.35 &  0.29 \\
No $\Delta m_t^{\rm EW}$ &  0.14 &  0.40 &  0.50 &  0.15 &  0.44 &  0.29 \\
No $\mb$ resummation &  $<10^{-3}$ &  $<10^{-3}$ &  0.06 &  $<10^{-3}$ &  $<10^{-3}$ &  $<10^{-3}$ \\
Pole Higgs masses in 1--loop &  0.02 &  0.02 &  0.01 &  0.01 &  0.02 &  0.02 \\
\hline
\end{tabular}
\caption{Lighter CP--even Higgs mass, $\mh$, in the six SPS
scenarios, as computed by \spheno\ under different approximations
for the higher--order effects. The first row contains the default
values in GeV, the other rows contain the shifts (in GeV) due to the
approximations.}
\label{apprh}
\vspace{.5cm}
\begin{tabular}{|l|c|c|c|c|c|c|}
\hline
Approximation & SPS1a & SPS2 & SPS4 & SPS5 & SPS8 & SPS9 \\
\hline 
Default  &  406.0 & 1554.6 &  360.5 &  686.5 &  552.4 & 1051.1 \\
\hline 
2--loop sfermion RGE &  1.1 &  6.8 &  0.3 & -1.3 &  0.7 &  1.5 \\
No $\Delta m_t^{\rm 2loop}$ &  1.1 &  6.7 &  4.5 & -1.4 &  2.6 & -2.1 \\
No $\Delta m_t^{\rm EW}$ &  0.6 &  8.2 &  8.0 & -0.6 &  3.2 & -2.2 \\
No $\mb$ resummation &  0.2 &  0.3 & 31.0 &  0.1 &  0.3 &  $<10^{-2}$\\
Pole Higgs masses in 1--loop &  0.1 &  0.1 &  0.5 &  0.1 &  0.3 &  0.1 \\
\hline
\end{tabular}
\caption{Same as table~\ref{apprh} for the heavier CP--even Higgs boson
mass, $\mH$.}
\label{apprH}
\end{center}
\end{table}

Tables \ref{apprh} and \ref{apprH} show the variations of $\mh$ and
$\mH$, respectively, in the six SPS scenarios, as a result of different
approximations for the higher--order corrections that affect the RGE
evolution of the MSSM parameters. The first row of each table contains
the default values of the masses, as obtained with the public version
of \sphenov. In the first row of the second part of the table, we show
the effect (in GeV) from switching off the two--loop terms in the RGEs
for the sfermion soft SUSY--breaking parameters; in the second row,
from switching off the two--loop QCD corrections in the computation of
the top Yukawa coupling $h_t$; in the third, from switching off the
one--loop electroweak corrections in the computation of $h_t$; in the
fourth, from defining the running bottom mass as $\widehat{m}_b =
\ov{m}_b\,(1+\Delta_b)$ instead of adopting the ``resummed'' formula
of eq.~(\ref{mbmssm}); in the fifth, from using the physical values of
the Higgs masses, instead of the running values, as input parameters
for the computation of the one--loop corrections. \s

We see that the effect of these approximations on $\mh$ is in general
moderate, amounting at most to some tenths of a GeV, whereas their
effect on $\mH$ can amount to many GeV. This is due to the fact that
for $\ma \gg \polemz$, as is always the case in the SPS scenarios, the
tree--level value of $\mh$ is essentially proportional to $\mz$, and
the dependence on the MSSM parameters enters only through the
radiative corrections. On the other hand, the heavier Higgs boson masses
depend already at tree--level on the soft SUSY--breaking parameters
$m^2_{H_1}$ and $m^2_{H_2}$, and the RG--evolution of the latter is
very sensitive to the precise value of all of the other parameters,
especially $h_t$. We also see that, as expected, the precise
definition of the bottom Yukawa coupling has a sizeable effect on the
Higgs boson masses only in the scenarios with large $\tb$. Finally, we
see that the choice of the renormalisation scheme for the values of
the Higgs boson masses entering the one--loop corrections also induces
a small variation in the final results for the physical Higgs boson
masses. We note, however, that this variation amounts to a two--loop
effect controlled by the electroweak gauge couplings, thus it is of
higher order with respect of the accuracy of our calculation.


\subsection{Estimate of the theoretical uncertainties in the cMSSM}

In this subsection, we will discuss the shifts on the MSSM neutral
Higgs boson masses induced by the choice of a different
renormalisation scheme and by the variation of the renormalisation
scale. We will also try to estimate the impact of the approximation of
zero external momentum in the two--loop corrections. These shifts
indicate the size of the higher order radiative corrections that are
left uncomputed, and thus give a rough estimate of the theoretical
uncertainties in the calculation of the masses.

\subsubsection{Comparing the $\drbar$ and OS calculations}
\label{secosdrbar}

The first interesting comparison to make is between the results of our
$\drbar$ calculation of the CP--even Higgs boson masses and those of
the OS calculation implemented in \feynhiggs. The latter is based on
the one--loop formulae of Refs.~\cite{dab,sven1L} and the two--loop
formulae of Ref.~\cite{Sven} (for the $\oatas$ part) and
Refs.~\cite{bdsz2,bdsz,dds} (for the $\oabas$, $\oatq$ and $\oatababq$
parts, respectively). In the computation of \feynhiggs, the one--loop
corrections are expressed in terms of physical masses and mixing
angles, and appropriate counterterm contributions are inserted in the
two--loop corrections for consistency. \s

As \feynhiggs\ does not perform the evolution of the MSSM parameters
from the high--energy input scale to $\qewsb$, in order to obtain the
Higgs boson masses in the SPS scenarios we have to provide it with the
input parameters at the weak scale, as they are computed by one of the
RGE codes. In particular, the top quark mass must be the physical one,
$\polemt$; the parameters $\mu$ and $\tb$ must be the running ones,
computed at the scale $\qewsb$; the squark masses and mixings can be
either physical or running (in the latter case they are converted by
\feynhiggs\ into the physical ones by applying the appropriate
one--loop corrections); the definition of the gaugino mass parameters
is irrelevant at the perturbative order we are interested in; finally,
the $A$ boson mass, which enters the tree-level mass matrix of the
CP--even Higgs bosons, must be the physical one, computed at the
two--loop accuracy.\s

Table \ref{fhtabh} shows the values of $\mh$ and $\mH$ as computed by
\feynhiggs\ (version 1.5.1) in the six SPS scenarios, with the weak
scale input parameters provided by \suspect. The corresponding results
from the pure \suspect\ calculation are also shown for comparison.  It
can be seen from table \ref{fhtabh} that the values of $\mh$ resulting
from the OS calculation of \feynhiggs\ are consistently a couple of
GeV larger than the ones provided by the $\drbar$ calculation. As the
$\drbar$ and OS computations are equivalent up to the two--loop ${\cal
O}(\at\as + \ab\as + \at^2 + \at\ab + \ab^2)$ accuracy, the
discrepancies must be due to terms that are formally of higher order,
i.e. two--loop terms controlled by the electroweak gauge couplings and
three--loop terms, among which the most important are controlled by
the top Yukawa and strong gauge couplings. Concerning the heavier
Higgs boson mass, $\mH$, it can be seen in table \ref{fhtabh} that the
results of \feynhiggs\ are in excellent agreement with those of
\suspect.  This fact is not surprising, because for $\ma \gg \polemz$,
as is always the case in the SPS scenarios, $\mH$ and $\ma$ are very
strictly correlated, and in our comparison \feynhiggs\ takes as input
the value of $\ma$ as computed by \suspect.\s

\vspace{.5cm}
\begin{table}[htbp]
\renewcommand{\arraystretch}{1.5}
\begin{center}
\begin{tabular}{|c|l|c|c|c|c|c|c|c|}
\hline
Mass & Code & SPS1a & SPS2 & SPS4 & SPS5 & SPS8 & SPS9\\
\hline
$\mh$ & \suspect & 112.1 & 116.8 & 114.1 & 116.1 & 115.5 & 117.5 \\
&\feynhiggs & 113.8 & 118.3 & 116.1 & 118.5 & 117.3 & 118.3 \\
\hline
$\mH$ & \suspect & 406.5 & 1552.1 & 355.3 & 686.9 & 550.6 & 1056.6\\
& \feynhiggs & 406.5 & 1552.0 & 354.8 & 686.5 & 550.6 & 1056.7\\
\hline
\end{tabular}
\vspace{3mm}
\caption{The lighter and heavier CP--even Higgs boson masses, $\mh$
and $\mH$, in the six SPS scenarios, as computed by \suspect\ and
\feynhiggs. The weak scale input parameters, including the physical
mass $\ma$, are taken from the output of \suspect.}
\label{fhtabh}
\end{center}
\end{table}

\subsubsection{Renormalisation scale dependence}
\label{secscaledep}

Another measure of the effect of the higher orders consists in
studying the numerical dependence of the results for the physical
Higgs masses on the renormalisation scale $\qewsb$ at which the
effective potential is minimised and the radiatively corrected masses
are computed. In the ideal case of an all--orders calculation, the
physical observables should not depend on the choice of the scale. The
residual scale dependence still present in the real case can be taken
as a rough estimate of the magnitude of the corrections that are left
uncomputed. \s

\begin{figure}[p]
\begin{center}
\mbox{
\hspace{-1cm}
\epsfig{figure=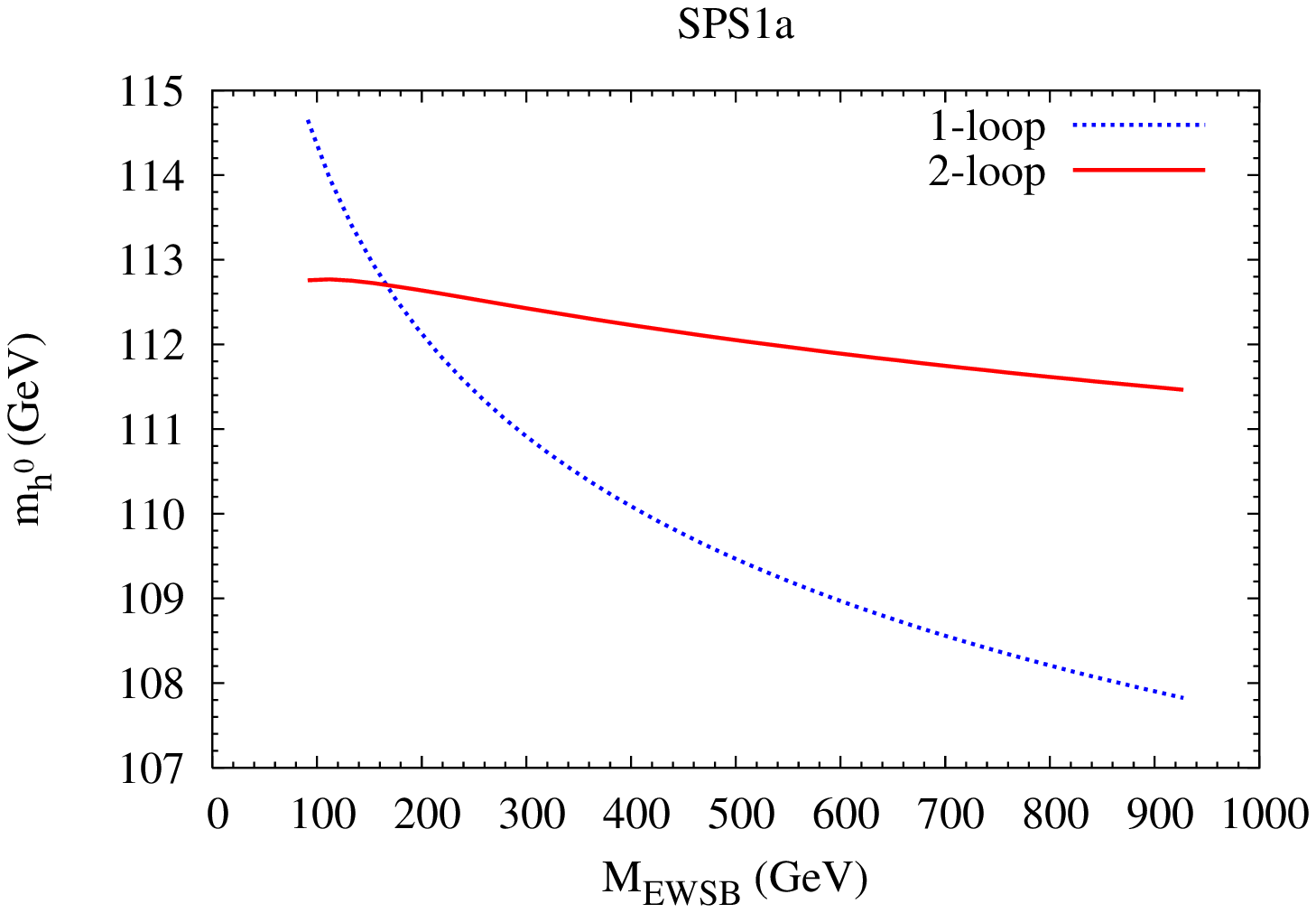,angle=-90,width=9cm}
\hspace{-0.8cm}
\epsfig{figure=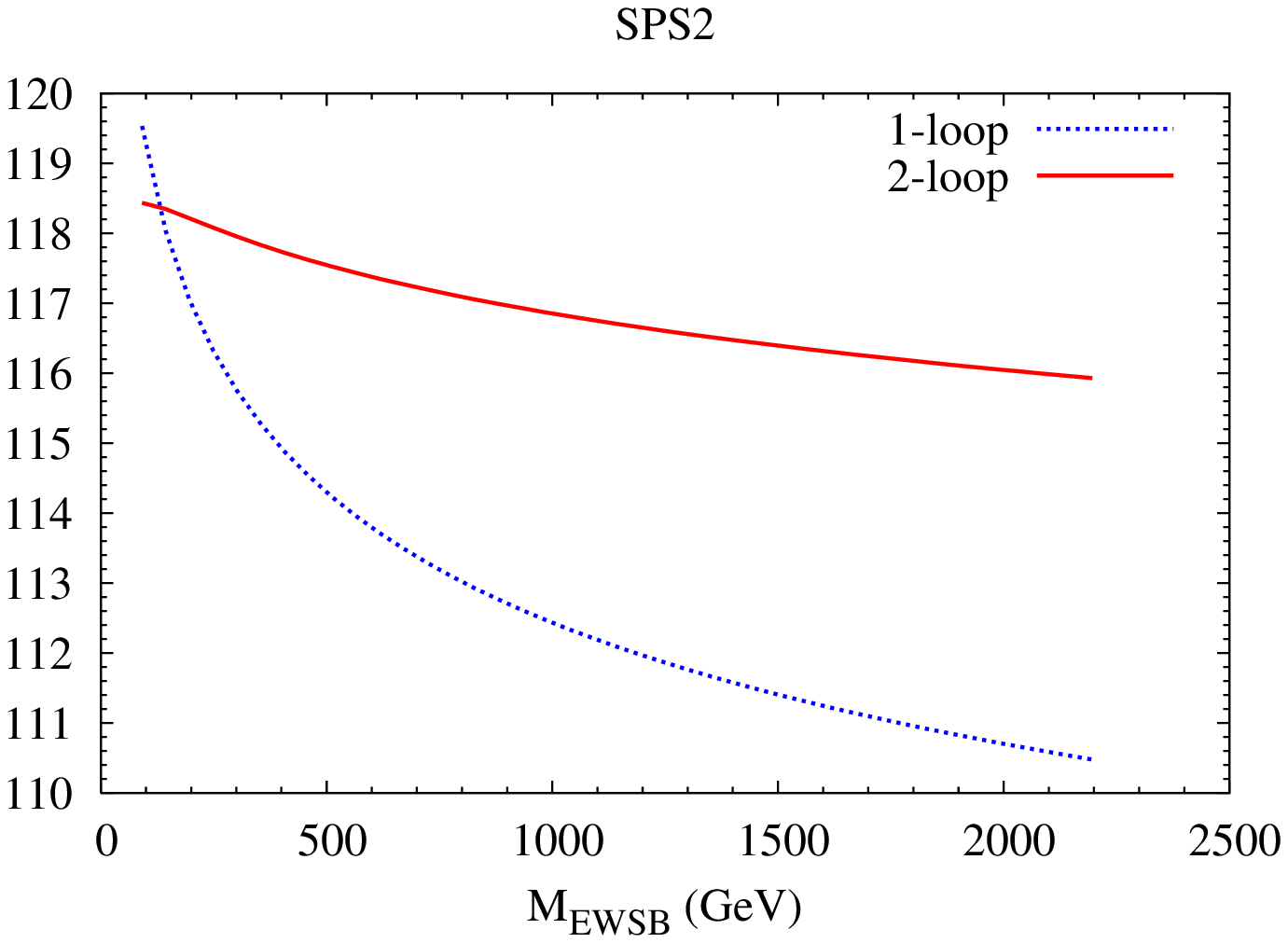,angle=-90,width=9cm}}
\vspace{3mm}
\mbox{
\hspace{-1cm}
\epsfig{figure=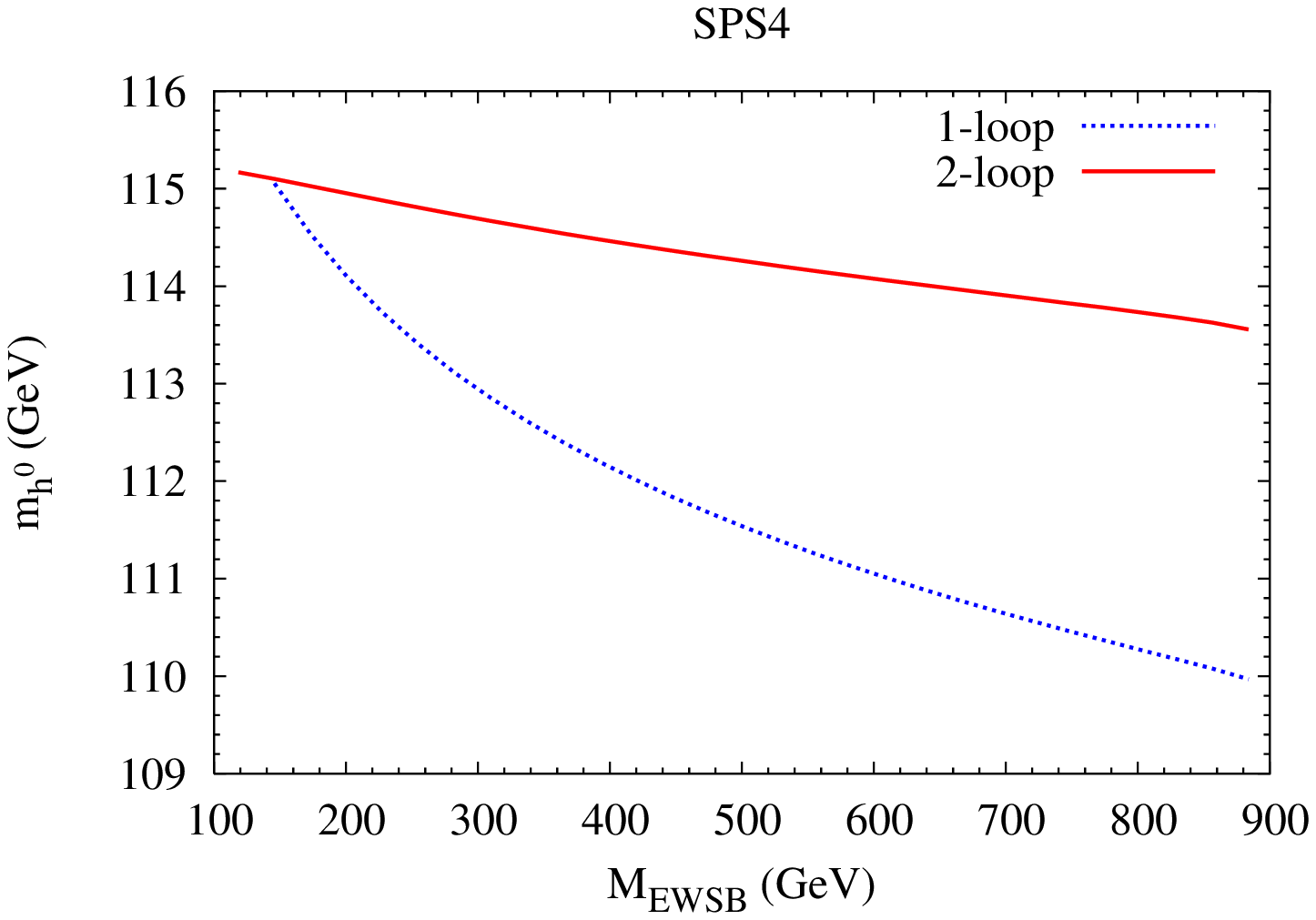,angle=-90,width=9cm}
\hspace{-0.8cm}
\epsfig{figure=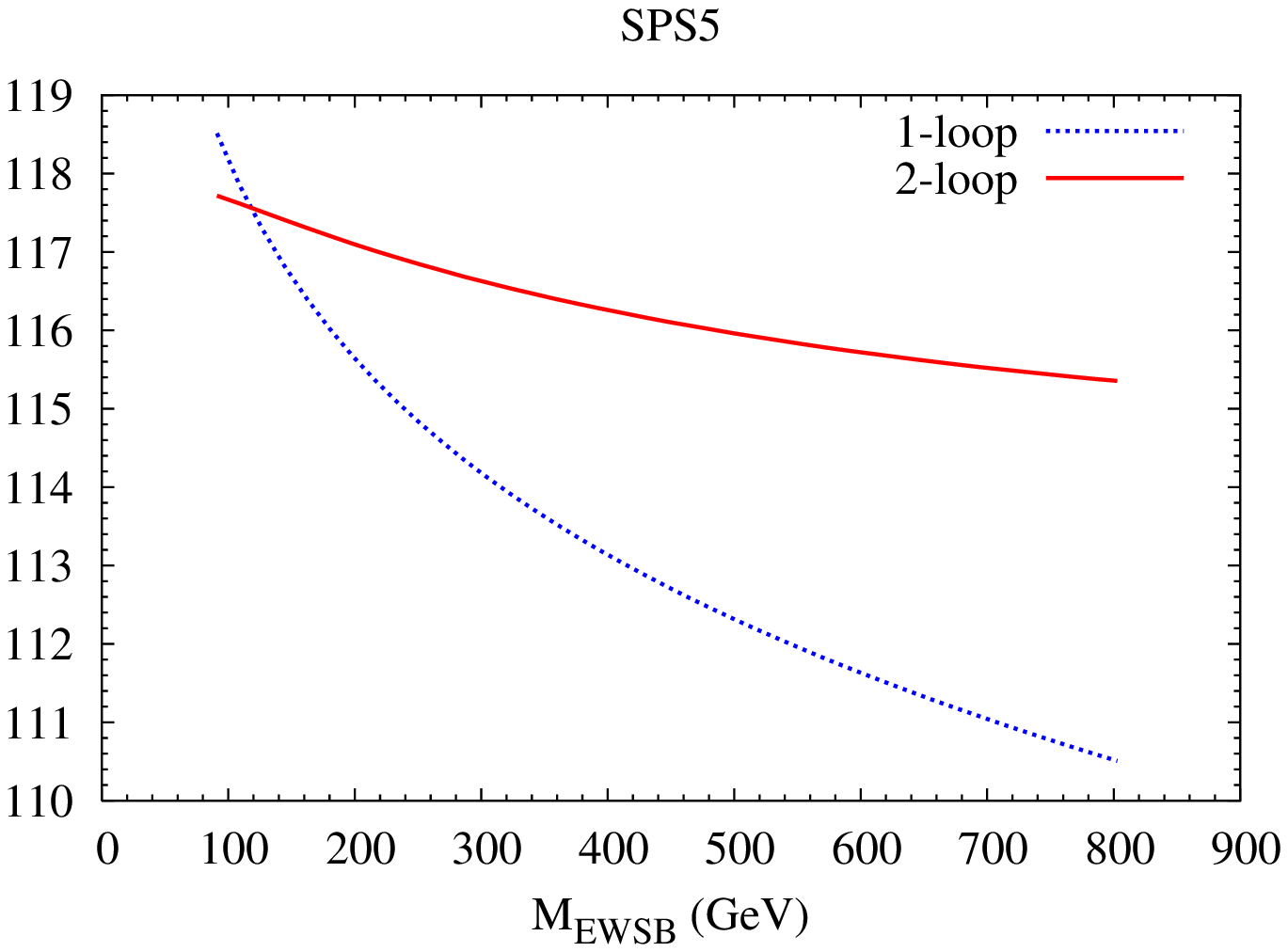,angle=-90,width=9cm}}
\vspace{3mm}
\mbox{
\hspace{-1cm}
\epsfig{figure=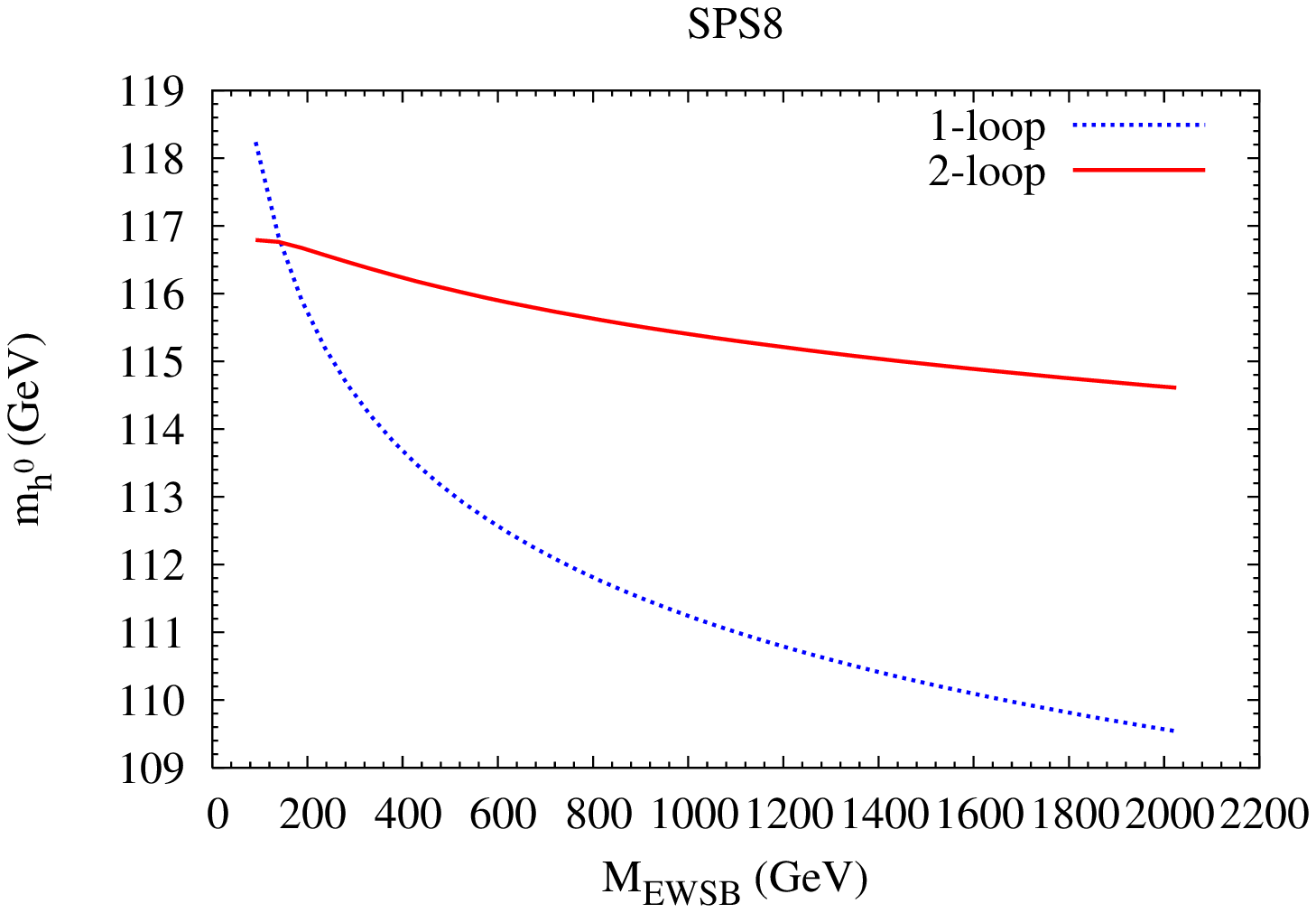,angle=-90,width=9cm}
\hspace{-0.8cm}
\epsfig{figure=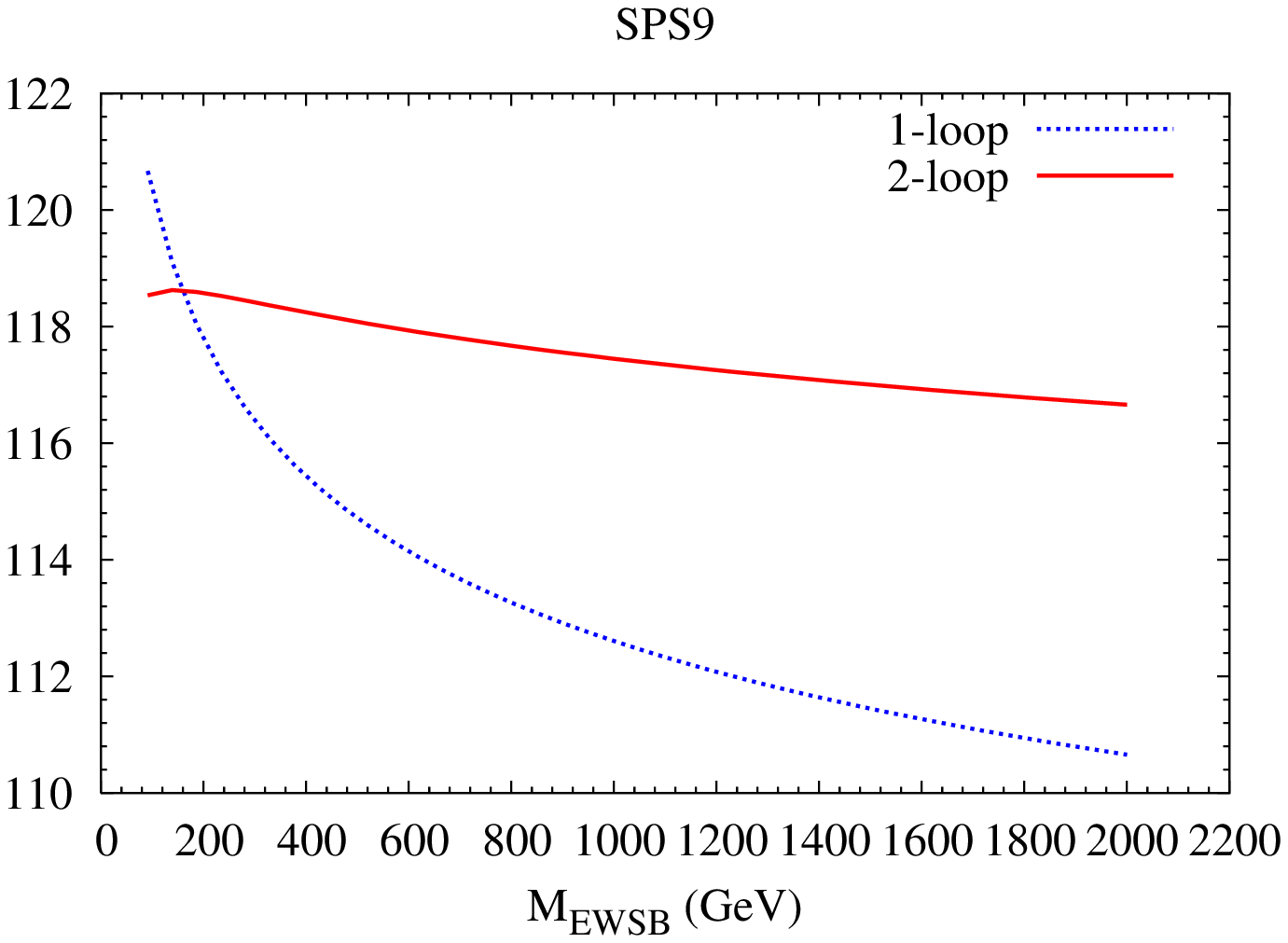,angle=-90,width=9cm}}
\end{center}
\vspace{-0.2cm}
\caption{The lighter CP--even Higgs boson mass, $\mh$, as computed by
\softsusy\ in the six SPS scenarios, as a function of the minimisation
scale of the effective potential, $\qewsb$. The dotted lines
correspond to the one--loop computation and the solid lines to the
two--loop computation.}
\label{plotmh}
\end{figure}
\begin{figure}[p]
\begin{center}
\mbox{
\hspace{-1cm}
\epsfig{figure=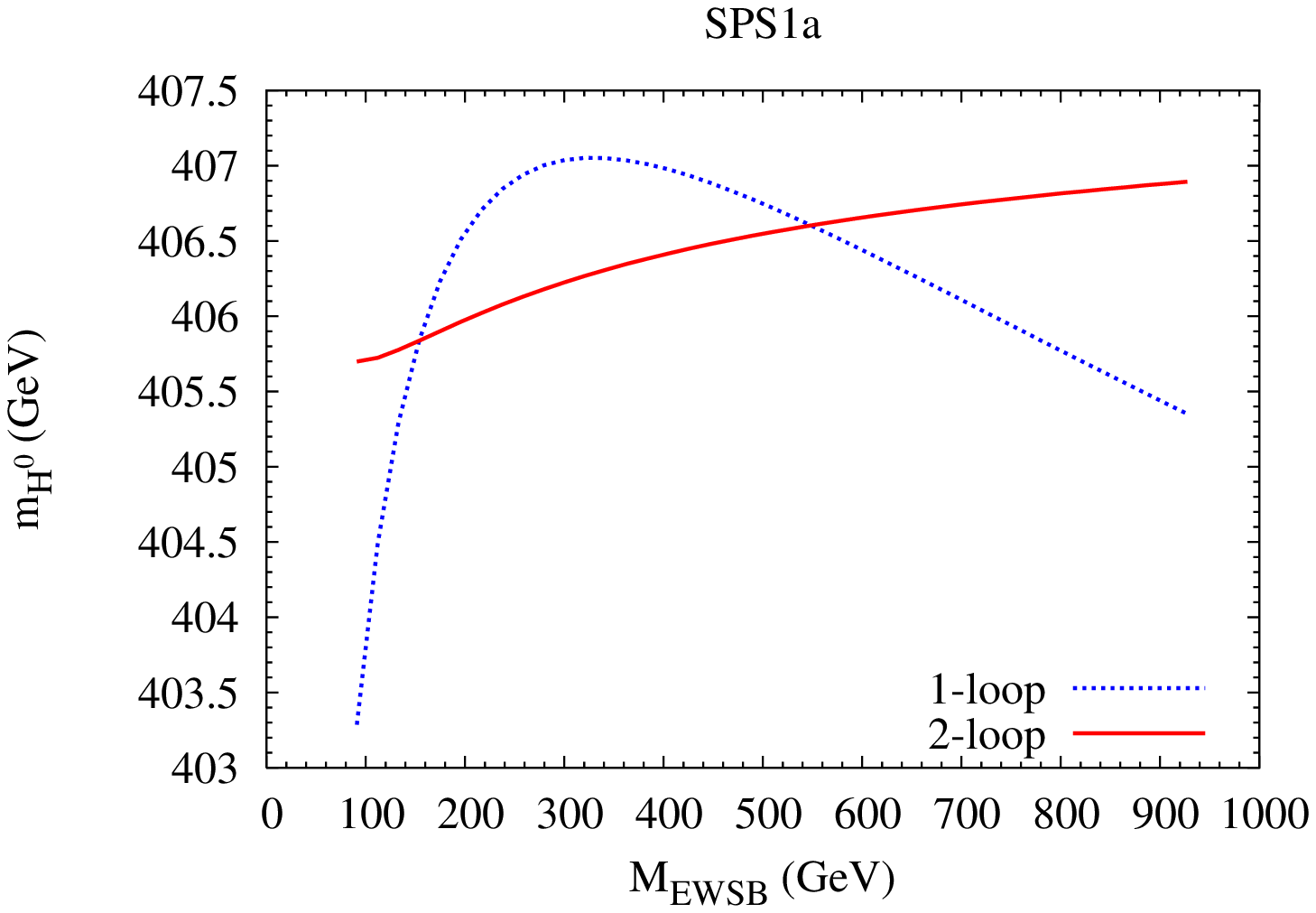,angle=-90,width=9cm}
\hspace{-0.8cm}
\epsfig{figure=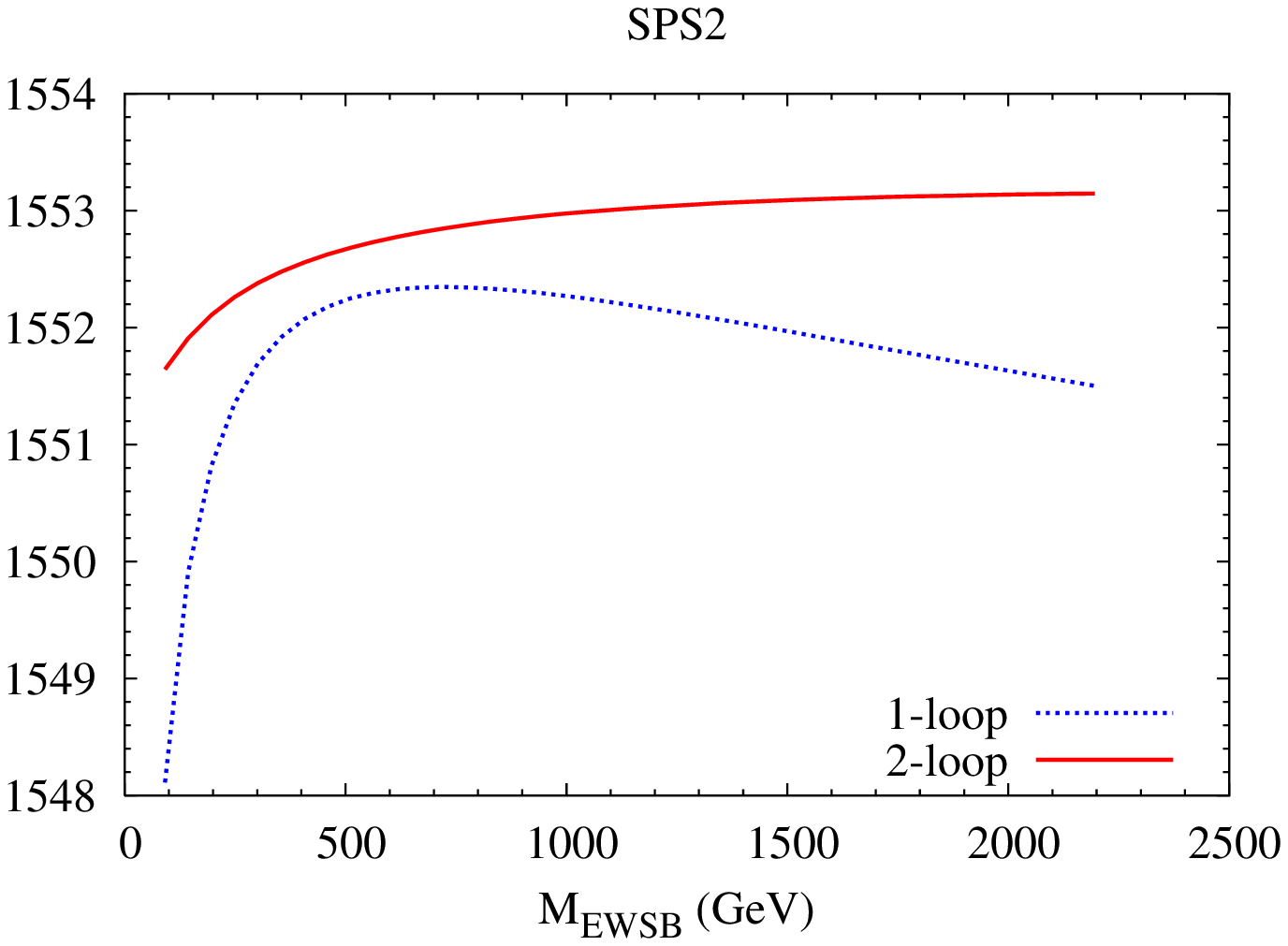,angle=-90,width=9cm}}
\vspace*{2mm}
\mbox{
\hspace{-1cm}
\epsfig{figure=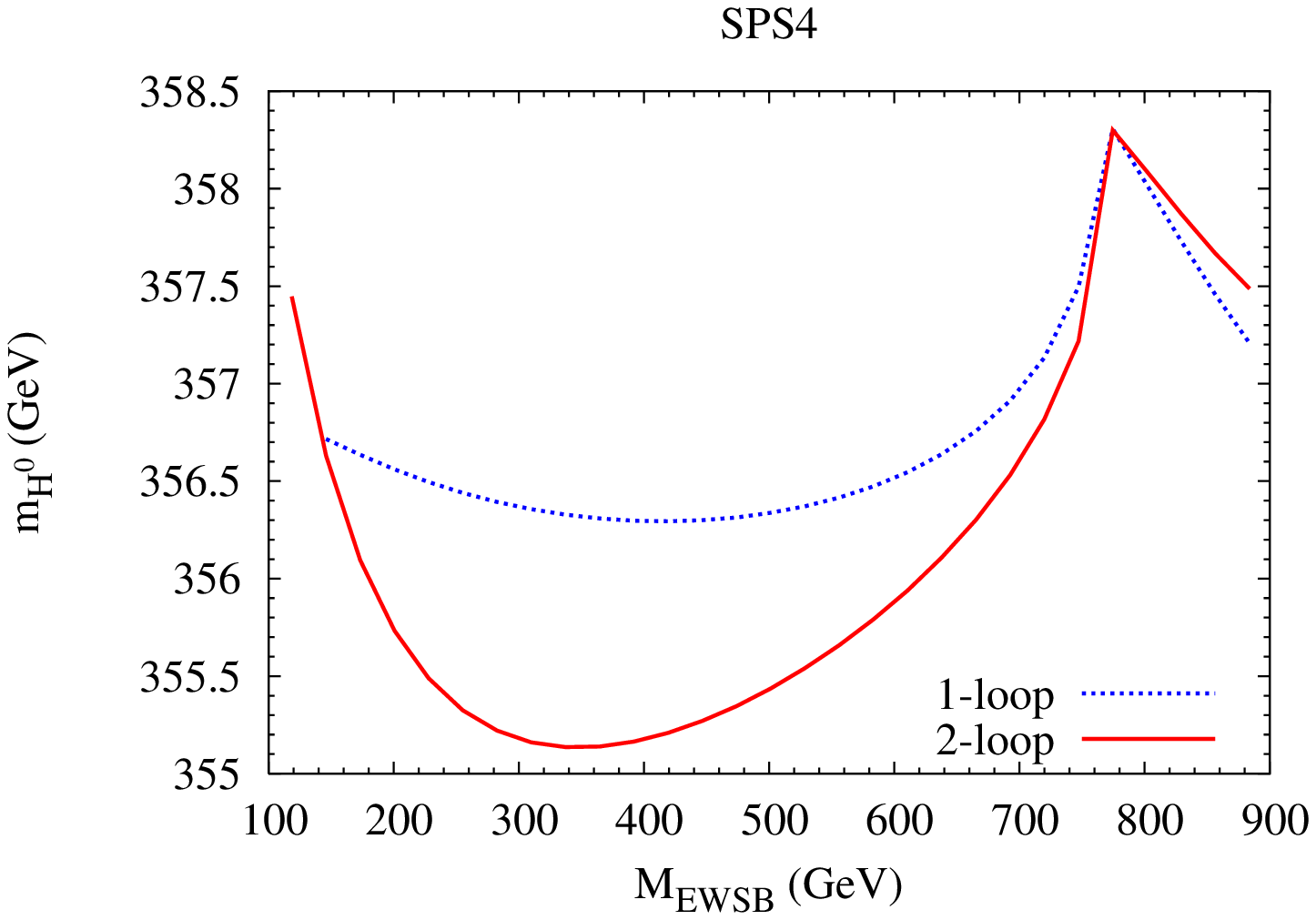,angle=-90,width=9cm}
\hspace{-0.8cm}
\epsfig{figure=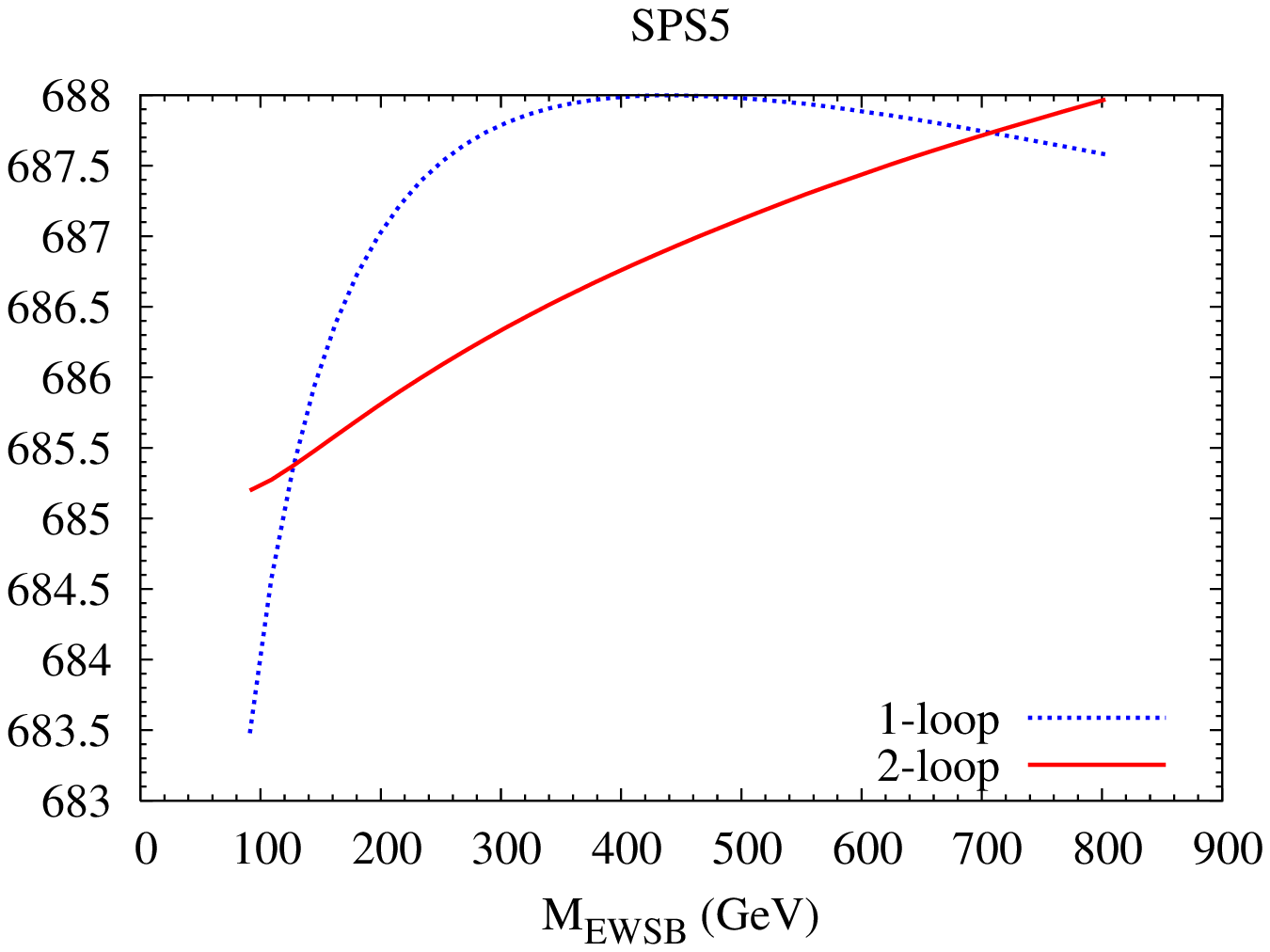,angle=-90,width=9cm}}
\vspace*{2mm}
\mbox{
\hspace{-1cm}
\epsfig{figure=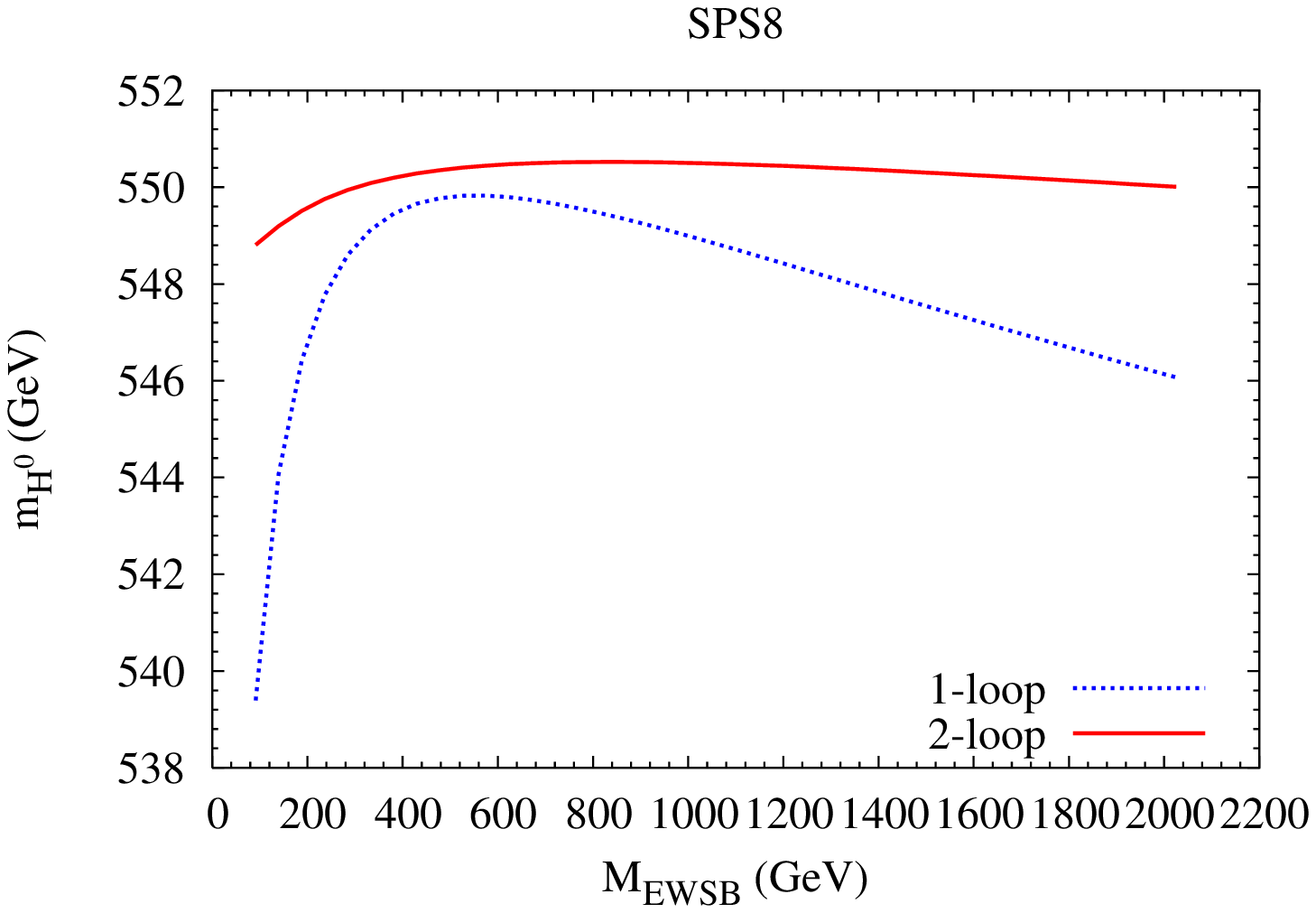,angle=-90,width=9cm}
\hspace{-0.8cm}
\epsfig{figure=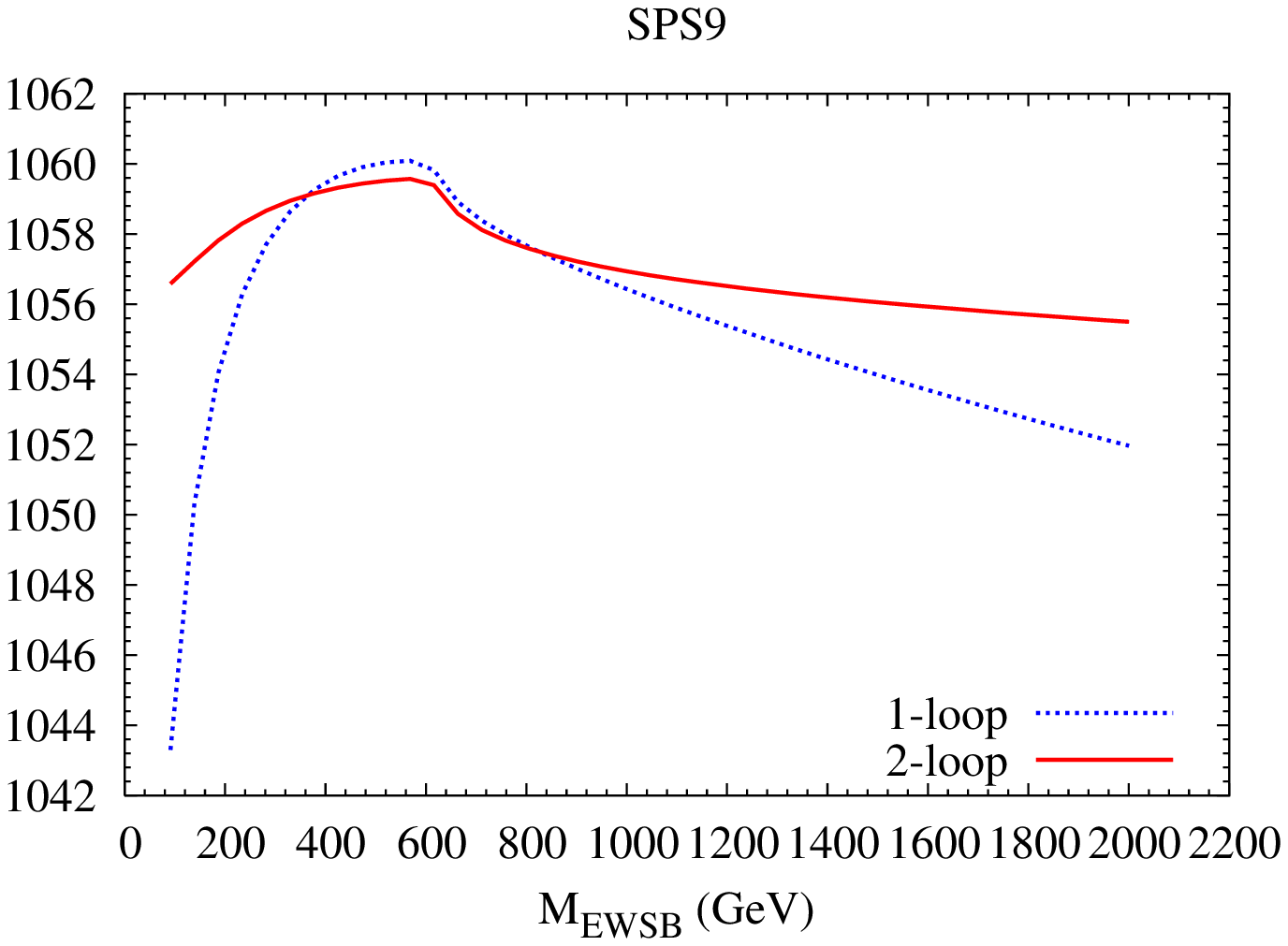,angle=-90,width=9cm}}
\end{center}
\vspace{-0.2cm}
\caption{Same as figure \ref{plotmh} for the heavier CP--even Higgs
boson mass, $\mH$.}
\label{plotmH}
\end{figure}

Figs.~\ref{plotmh} and \ref{plotmH} show the renormalisation scale
dependence of the CP--even Higgs boson masses $\mh$ and $\mH$,
respectively, as computed by \softsusy\ in the six SPS scenarios
discussed in section \ref{sec:scenarios}. The scale dependence of
$\ma$ is very similar to that of $\mH$ and is thus omitted 
for brevity. We have also checked that the same features can be
reproduced with \spheno\ and \suspect. The dotted line in each plot
corresponds to the one--loop computation of the relevant mass, whereas
the solid line corresponds to the two--loop computation. The scale
$\qewsb$ is varied from a minimum value of $\polemz$ to a maximum
corresponding to twice the stop mass scale $\sqstop$.\s

It can be seen from fig.~\ref{plotmh} that the one--loop results for
the lighter Higgs boson mass $\mh$ show a sizeable scale dependence,
varying by nearly 10 GeV in the considered range of $\qewsb$. This
dependence is essentially driven by the variation of the running top
mass, which decreases as the renormalisation scale increases (we
recall that the leading one--loop corrections to $\mh^2$ are
proportional to $\mt^4/v^2$). On the other hand, it appears that the
inclusion of the two--loop corrections significantly improves the
scale dependence of $\mh$, compensating for the dependence on $\mt$
and leaving a residual variation of 2--3 GeV in the considered range
of $\qewsb$.\s

Fig.~\ref{plotmH} shows that a similar situation occurs in the case of
the heavier Higgs boson mass $\mH$, with the inclusion of the two--loop
corrections clearly improving the renormalisation scale dependence of
the results. One exception is given by the plot corresponding to the
scenario SPS4, where the curves show a spike and the two--loop
corrections do not improve the scale dependence. This anomalous
behaviour is due to the fact that the codes use the running values of
the Higgs boson masses in the computation of the one--loop
corrections. For large values of $\tb$ (e.g. $\tb=50$ in SPS4) the
running masses for the heavier Higgs bosons vary by hundreds of GeV as
the renormalisation scale varies in the considered range. Thus, for
some values of the renormalisation scale, the loop integrals involving
the heavier Higgs boson masses encounter real--particle thresholds which
give rise to the characteristic spike behaviour.  We have checked that,
if we use in the one--loop corrections the physical values of the
Higgs boson masses, which vary only of a few GeV in the considered
range of $\qewsb$, the spike disappears and the plot for the SPS4
scenario looks qualitatively very similar to the ones for the other
scenarios (the same happens for another large--$\tb$ scenario, SPS1b,
which we did not include in our analysis). We recall, however, that
the choice of whether to use the running or physical Higgs masses in
the one--loop corrections amounts to a higher--order effect with
respect to the accuracy of our calculation.

\subsubsection{The external momentum dependence}
\label{secmom}

To simplify the calculation of the two--loop radiative corrections to
the Higgs boson masses, the squared external momentum $p^2$ in the
two--loop part of the self energies has been set to zero, rather than
to the squared mass of the corresponding Higgs boson. This
approximation can be considered as a source of theoretical
uncertainty. The actual size of the shift on the Higgs boson masses
induced by this approximation can only be obtained when the exact
two--loop computation, retaining the external momentum dependence, is
performed (see Ref.~\cite{Martin3} for a first attempt). However, it
can be roughly estimated by considering the error that the
zero--momentum approximation induces in the one--loop corrections, and
assuming that the error induced in the two--loop corrections will have
the same relative size (a similar analysis was performed in
Ref.~\cite{dhhsw}).\s

\begin{table}[htbp]
\renewcommand{\arraystretch}{1.2}
\begin{center}
\begin{tabular}{|l|c|c|c|c|c|c|c|}
\hline
Mass & Approximation & SPS1a & SPS2 & SPS4 & SPS5 & SPS8 & SPS9 \\
\hline 
$\mh$ & $p^2=0$ in 1--loop & +0.9 & +2.0 & +1.3 & +1.0 & +2.0 &+1.7 \\ 
& Estimated 2--loop effect & 0.1 & 0.4 & 0.2 & 0.1 & 0.3 & 0.3\\ 
\hline
$\mH$ & $p^2=0$ in 1--loop & $-0.3$ & +0.8 & +4.4 & +0.02 & $-0.1$ & $-0.9$\\ 
& Estimated 2--loop effect & 0.02 & 0.1 & 0.7 & 0.02 & $<$0.01& 0.04
\\ \hline
\end{tabular}

\caption{Effect of the zero--momentum approximation in the one--loop
corrections to the CP--even Higgs boson masses, $\mh$ and $\mH$, in
the six SPS scenarios, and an estimate of the corresponding two--loop
effect. All the numbers are in GeV.}
\label{tabmom}
\end{center}
\end{table}
In table \ref{tabmom}, we display the effect of the approximation of
zero external momentum on the lighter and heavier CP--even Higgs boson
masses, as computed by \suspect\ in the six SPS scenarios.  For each
mass, the first row contains the shift (in GeV) resulting from the
approximation $p^2=0$ in the one--loop part of the corrections, and
the second row contains the estimate for the size of the same effect
in the two--loop case. The latter is obtained by scaling the absolute
value of the shift induced in the one--loop corrections by the ratio
between the size of the two--loop zero--momentum corrections and that
of the one--loop zero--momentum corrections in the considered
scenario. As can be seen, the effect of the zero--momentum
approximation on the one--loop calculation of $\mh$ is of the order of
1--2 GeV. When rescaled to account for the relative size of the
two--loop to the one--loop corrections, this results in a shift of
less than half GeV in all of the considered scenarios. \s

On the other hand, we see from table \ref{tabmom} that in all
scenarios but SPS4 the effect of the zero--momentum approximation on
the one--loop computation of $\mH$ is of the order of a few tenths of
GeV. As a consequence, one can predict that in those scenarios the
effect of the same approximation on the two--loop corrections is
negligible. This is due to the fact that, as mentioned in section
\ref{secradcorr}, the heavier Higgs boson mass is dominated by the
tree--level value, and the impact of the one--loop radiative
corrections is anyway small. In the case of the large--$\tb$ scenario
SPS4, however, there are non--negligible momentum--dependent
contributions from the one--loop diagrams with bottom quarks and tau
leptons, inducing a shift of more than 4 GeV in $\mH$, and the effect
of the zero--momentum approximation on the two--loop corrections can be
predicted to get close to one GeV.

\subsection{The case of the pMSSM}
\label{secpmssm}

We now discuss the case of the phenomenological MSSM defined in
section \ref{sec:scenarios}. In principle, since there are 22 free
parameters in the model, the phenomenological analyses should be
rather complicated to carry out. However, as mentioned previously, only a
small subset of parameters plays a significant role in the evaluation
of the Higgs boson masses. Before presenting our numerical results, we
briefly summarise the role and the main effects of the various MSSM
parameters. \s

At the tree level, the Higgs sector of the pMSSM can be described by
two input parameters in addition to the SM ones; these parameters are
in general taken to be the mass of the pseudoscalar Higgs boson $\ma$
and $\tb$. In this parameterisation, the tree--level CP--even Higgs
boson masses are given by
\be
M_{h/H}= \frac{1}{\sqrt{2}} \left[ \ma^2+\polemz^2 \mp 
\sqrt{(\ma^2+\polemz^2)^2 - 4 \cos^22\beta \ma^2 \polemz^2} \right]^{1/2}.
\label{mhtree}
\ee

The charged Higgs boson mass is given at tree--level by $M_{H^\pm}
=\sqrt{M_A^2+M_W^2}$, while the angle $\alpha$ in the CP--even Higgs
sector is given by $\tan 2\alpha = \tan 2\beta
\,(M_A^2+M_Z^2)/(M_A^2-M_Z^2)$, with the constraint $- \frac{\pi}{2}
\leq \alpha \leq 0$. \s

The maximal value of the lighter CP--even Higgs boson mass at the
tree level, $\mh = \cos2\beta \,\polemz$, is obtained in the
decoupling regime, $\ma \gg \polemz$ (note that, in this case, the
lighter Higgs boson of the MSSM will have exactly the same properties as
the SM Higgs boson), and is saturated for large values of $\tb$. In
this limit, the mass of the heavier CP--even Higgs boson is simply
$\mH = \ma$. \s

The leading one--loop radiative corrections to $\mh^2$ are controlled
by the top and bottom Yukawa couplings, and in the decoupling regime
read
\begin{equation}
\epsilon = \frac{3\,m_t^4}{2\,\pi^2 \,v^2} 
\left( \ln \frac{M_S^2}{m_t^2} + \frac{X_t^2}{M_S^2} 
- \frac{X_t^4}{12\,M_S^4} \right)
-\frac{3\,m_b^4}{2\,\pi^2 \,v^2}\, \frac{X_b^4}{12\,M_S^4}\;\;,
\label{higgscorr}
\end{equation}
where $M_S$ is a common soft SUSY--breaking mass term for the
third--generation squarks and $X_{t,b}$ are given in terms of the 
stop/sbottom trilinear couplings, $\mu$ and $\tb$, by: 
\begin{equation}
X_t= A_t - \mu \cot \beta \ \ {\rm and} \ \ X_b= A_b - \mu \tan \beta\,.
\end{equation}

For $\ma \gg \polemz$ and $\tb \gg 1$, one obtains $\mh^2= \polemz^2+
\epsilon\,$ for the lighter CP--even Higgs boson, while for the
heavier Higgs boson one still has $\mH=M_A$.  The choice of the
renormalisation scheme for the parameters entering
eq.~(\ref{higgscorr}) becomes relevant when the two--loop corrections
are included; we recall that in our calculation we express the
one--loop corrections in terms of $\drbar$ parameters. \s

Because of the quartic dependence on $m_t$, the corrections controlled
by the top Yukawa coupling are the leading ones. These corrections are
larger when the logarithm in the first term of eq.~(\ref{higgscorr})
is larger, i.e.~for large $M_S$ values (corresponding to large stop
masses). In addition, the top quark corrections are maximal in the
so--called $M_h^{\rm max}$ scenario \cite{benchmarks}, where the
trilinear stop coupling in the $\drbar$ scheme is such that $X_t \sim
\sqrt{6}\,M_S$. \s

The corrections controlled by the bottom Yukawa coupling are in
general strongly suppressed with respect to those controlled by the
top Yukawa coupling, due to the overall factor $\mb^4$. However, in
the last term of eq.~(\ref{higgscorr}), proportional to $X_b^4$, this
suppression can be compensated by a large value of the product $\mu
\tb$, providing a non--negligible negative correction to $\mh^2$.  The
choice of the values for the remaining soft SUSY--breaking parameters
does not have a very large impact on the one--loop corrections, and in
the $\drbar$ calculation the two--loop corrections, although
numerically significant in the determination of the precise value of
the lighter Higgs boson mass, do not substantially alter the picture. \s

\vspace*{0.7cm}

\begin{figure}[ht]
\begin{center}
\epsfig{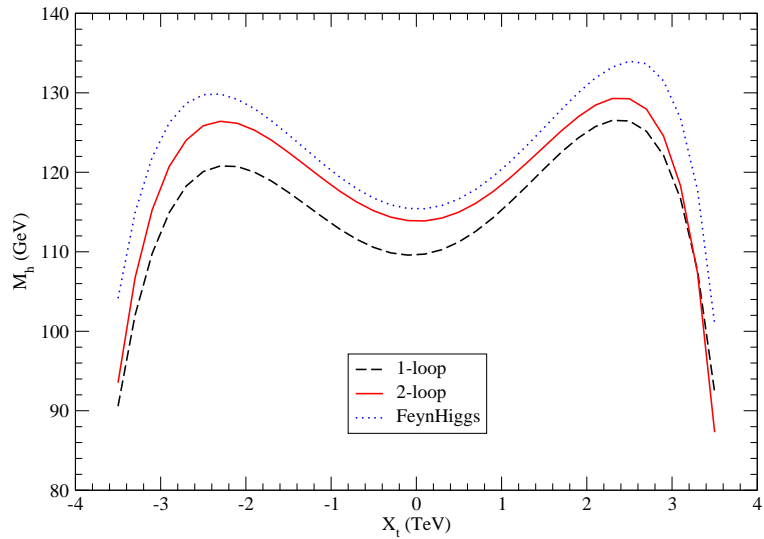}
\caption{The lighter MSSM Higgs boson mass as a function of $X_t$ in
the $\drbar$ scheme for $\tb=10$ and $M_S\!=\!M_A\!=\!1$ TeV with
$M_t=178$ GeV.  The full and dashed lines correspond, respectively, to
the two--loop and one--loop corrected masses as calculated with
\suspect, while the dotted line corresponds to the two--loop $\mh$
value obtained with \feynhiggs.}
\label{figmhvsxt}
\end{center}
\vspace{-0.5cm}
\end{figure}

The above features are exemplified in fig.~\ref{figmhvsxt}, where the
lighter Higgs boson mass is displayed as a function of the $\drbar$
parameter $X_t$, for $M_t=178$. In the figure, the MSSM parameters are
set to those of the three pMSSM points introduced in section
\ref{sec:scenarios}; in particular, the physical pseudoscalar mass
$M_A$ and the third--generation soft SUSY--breaking scalar masses
$M_S$ (the latter computed at the renormalisation scale $Q=1$ TeV) are
set to 1 TeV, while $\tb$ is fixed to $\tb=10$ at $Q=\polemz$. The
dashed curve for the one--loop corrections, and the full curve for the
two--loop corrections in the $\drbar$ scheme, have been obtained using
the program $\suspect$. As one can see, the lighter Higgs boson mass
$\mh$ has a local minimum for zero stop mixing, and it increases with
$|X_t|$ until it reaches a local maximum at the point $|X_t| =
\sqrt{6}\,M_S \sim 2.45$ TeV, where it starts to decrease again. Note
that the two--loop corrections amount to several GeV and, in this
scenario, decrease in size when $X_t$ takes on large and positive
values. \s

The dotted curve in fig.~\ref{figmhvsxt} is obtained with the program
\feynhiggs, using the MSSM input parameters as they come from
\suspect. In the OS renormalisation scheme adopted by \feynhiggs, the
maximal $\mh$ value is obtained for $X_t^{\rm OS} \sim 2\, M_S^{\rm
OS}$, where $X_t^{\rm OS}$ and $M_S^{\rm OS}$ are the unphysical
parameters obtained by rotating the diagonal matrix of the OS stop
masses by the OS mixing angle (see e.g.~Ref.~\cite{reconciling} for a
discussion). On the other hand, as we are plotting $\mh$ as a function
of the $\drbar$ parameter $X_t$ for both the \suspect\ and \feynhiggs\
computations, we find the maximum value of $\mh$ roughly at the same
place. Comparing the solid and dotted lines, it can be seen that for
small to moderate values of $X_t$ the results of the OS calculation of
\feynhiggs\ are a couple of GeV higher than those of the $\drbar$
calculation, as in the constrained MSSM scenarios. Around the maxima
of the curve, however, the difference between the two computations
reaches 4--5 GeV.  This indicates that for large mixing in the stop
sector, $X_t \sim \sqrt{6}M_S$, the corrections which are formally of
higher order (such as the three--loop terms controlled by the top
Yukawa coupling and the strong coupling) are larger.\s

We can now discuss the numerical results that we obtain with some of
the spectrum generators in the three pMSSM scenarios defined in
section \ref{sec:scenarios}. Table \ref{tabmhp} shows the results
obtained with the spectrum generators \spheno\ and \suspect\ for the
masses of the lighter and heavier CP--even Higgs bosons in the three
pMSSM scenarios (the option for the pMSSM mode is not yet completely
implemented in \softsusy). It can be seen from the table that the
results of the two codes agree within less than a GeV for $\mh$, as
was the case in the constrained MSSM scenarios (see table
\ref{tabmh}). The residual discrepancies are due, as discussed in
section \ref{sec:codes}, to the different ways in which the two codes
implement the various threshold corrections to gauge and Yukawa
couplings at the scale $Q=\polemz$~\footnote{A further difference
between the codes is that, in the pMSSM mode, \suspect\ uses the
physical Higgs boson masses in the computation of the various
threshold corrections at $Q=\polemz$, whereas \spheno\ switches to the
physical masses only if one of the squared running masses becomes
negative at $Q=\polemz$.}. On the other hand, it can be seen that the
two codes are in excellent agreement for the heavier CP--even Higgs
boson mass. This is due to the strict correlation between $\mH$ and
$\ma$: contrary to the case of the constrained MSSM scenarios, where
$\ma$ is determined through the EWSB conditions and is thus sensitive
to all of the details of the RG procedure, in the pMSSM scenario $\ma$
is given as an input and is thus the same for both codes.\s

\begin{table}[!ht]
\renewcommand{\arraystretch}{1.2}
\begin{center}
\begin{tabular}{|l|c|ccc|}
\hline
Mass & Code & pMSSM1 & pMSSM2 & pMSSM3 \\
\hline
$\mh$ & \spheno & 114.3 & 118.8 & 130.0 \\
& \suspect & 113.8 & 118.4 & 129.4 \\
\hline
$\mH$ & \spheno & 1000.2 & 1000.1 & 999.4 \\
& \suspect & 1000.2 & 1000.2 & 999.5 \\
\hline 
\end{tabular}
\caption{The lighter and heavier CP--even Higgs boson masses, $\mh$ and $\mH$,
in the pMSSM scenarios as computed by \sphenov\ and \suspectv.  The SM
and MSSM input parameters are chosen as in section
\ref{sec:scenarios}.}
\label{tabmhp}
\end{center}
\end{table}

The impact of the radiative corrections on the Higgs boson mass
calculations is similar to what has been discussed in the case of
the constrained scenarios and will not be repeated here. Furthermore,
for the heavier CP--even Higgs boson, all of the theoretical
uncertainties discussed in section 3.3 are very small, generating
shifts of less than 100 MeV. In the following, we will therefore only
concentrate on the uncertainties in the evaluation of the lighter
Higgs boson mass $M_h$ due to the renormalisation scheme, the
renormalisation scale and the dependence on the external momentum. The
results, obtained using \suspect, are summarised in table
\ref{tabpMSSMh} for the three pMSSM points.

\begin{table}[ht]
\renewcommand{\arraystretch}{1.3}
\begin{center}
\begin{tabular}{|c|c|c|c|}
\hline
Shift & pMSSM1 & pMSSM2 & pMSSM3 \\
\hline 
Default value     & 113.8 & 118.4 & 129.4 \\
\hline \hline
\feynhiggs\ shift   & $+1.6$ & $+1.9$ & $+4.4$ \\
\hline
$\qewsb =  150$ GeV 
& $+1.1$ & $+1.5$ & $+4.4$ \\
$\qewsb = 2$ TeV   
& $-0.7$ & $-0.8$ & $-0.8$ \\
\hline
$p^2=0$ in 1--loop       & +1.6 & +1.8 & +3.0 \\ 
Estimated 2--loop effect & +0.3 & +0.2  &+0.2 \\ 
\hline
\end{tabular}
\caption{Effect of renormalisation scheme, renormalisation scale and
estimate of the external momentum dependence in the determination of
$\mh$ for the three pMSSM scenarios (all the shifts are in GeV).}
\label{tabpMSSMh}
\vspace*{-.5cm}
\end{center}
\end{table}

The first line of the table, after the default values given by
\suspect, shows the shift in the value of $\mh$ when it is calculated
in the OS scheme by the program \feynhiggs, where the weak scale input
parameters are taken from \suspect. As mentioned previously, the
difference between the two calculations of $\mh$ amounts to a couple
of GeV in the first two scenarios, but reaches 4.4 GeV in the
large--mixing scenario pMSSM3.\s

To estimate the renormalisation scale dependence of the Higgs boson
masses in the pMSSM scenarios, we can vary the scale $\qewsb$ at which
the corrections are computed. By default, in the pMSSM scenarios this
scale is identified with the scale at which the input parameters are
given. In order to estimate the related uncertainty, we evolve the
parameters of the three pMSSM scenarios from the input scale $Q=1$ TeV
to some new scale, and compute again the Higgs boson masses in terms
of the new values of the parameters (there is an option in \suspect\
allowing for this study). The resulting shifts in the lighter Higgs
boson mass for two alternative values of the scale, $\qewsb = 150$ GeV
and $\qewsb = 2$ TeV, are shown in the middle part of table
\ref{tabpMSSMh}. As can be seen, the shifts in $\mh$ are usually
of the order of 1 GeV, but again reach 4.4 GeV in the case of the pMSSM3 
scenario.\s

Finally, we see from the lower part of table \ref{tabpMSSMh} that
the external momentum dependence of $\mh$ at the two--loop level, as
estimated from the one--loop momentum dependence calculated in
\suspect, is expected to generate a shift of 300 MeV or less, as was
approximately the case in the cMSSM scenarios discussed in the
previous sections.

\subsection{Summary of the theoretical uncertainties}
\label{secestimate}

From the results presented in the previous sections, it is possible to
attempt an estimate of the theoretical uncertainty that still affects
the calculation of the lighter MSSM Higgs boson mass. Such an estimate is
relevant because it affects the information on the MSSM parameter
space that can be extracted from the comparison between the
theoretical prediction on $\mh$ and the present experimental lower
bounds. The phenomenological implications of this theoretical
uncertainty will be further discussed in section \ref{secpheno}.\s

It has been shown in sections \ref{secosdrbar} and 3.4 that the
discrepancies between the OS and $\drbar$ calculations of the lighter
Higgs boson mass $\mh$, which formally amount to a combination of
two--loop effects controlled by the electroweak gauge couplings and
three--loop effects, are of the order of 2--3 GeV in most of the
considered cMSSM and pMSSM scenarios, but they can reach 4--5 GeV in
the pMSSM scenario with large stop  mixing.  Another measure of the size
of the uncomputed corrections is given by the residual renormalisation
scale dependence, and in sections \ref{secscaledep} and \ref{secpmssm}
it has been shown that the results of the $\drbar$ calculation of
$\mh$ vary by 2--3 GeV when the renormalisation scale is varied in a
reasonable range. The discrepancies between the results of the
different spectrum generators in the constrained MSSM scenarios
discussed in section \ref{comparecodes} correspond generally to
three--loop effects (i.e., two--loop differences in the determination
of parameters that enter the Higgs mass corrections at one loop), and
in the case of $\mh$ are contained within a half GeV. Finally,
another source of theoretical uncertainty is the approximation of zero
external momentum employed in the two--loop corrections. In sections
\ref{secmom} and 3.4 it has been shown that this uncertainty in $\mh$
is less than half GeV in the considered scenarios. \s

The way these different results should be combined into a figure for
the residual theoretical uncertainty on $\mh$ is debatable, but we
think that a reasonably conservative estimate of the global uncertainty 
can be taken as 3--5 GeV. The lower value is expected in most of the 
cMSSM and pMSSM scenarios, while the upper value is expected in some 
special scenarios such as in the pMSSM with large mixing in the stop 
sector. \s

In Ref.~\cite{dhhsw} the combined effect of the corrections yet to be
computed was estimated to induce an uncertainty in $\mh$ of the order
of 3 GeV.  We find that our results are in good agreement with that
earlier estimate for most of the scenarios that we have discussed in
this paper.  A recent analysis with the program \feynhiggs\
\cite{sventalk}, gives for the uncertainty on $\mh$ a range that is
similar to the one that we quote here, i.e. $\Delta \mh \sim 3$--5
GeV, depending on the considered MSSM scenario. \s

The fact that the predicted experimental accuracy in the determination
of $\mh$ at a future linear collider is 50 MeV \cite{tesla} indicates
that a huge effort is still necessary in order to improve the
theoretical uncertainty on the Higgs boson masses up to the level
required to compare with the forthcoming experimental results. An
experimental effort is also needed to measure the SM input parameters
more precisely, since they generate an error on $M_h$ which is of the
same order as the theoretical uncertainty, as will be discussed now.

\subsection{The impact of the experimental uncertainties}
\label{secexperr}

It is important to discuss how the predictions for the MSSM
Higgs boson masses are affected by the experimental errors in the
determination of the input SM parameters. This is a source of  
uncertainty independent of the ones discussed above, and it can be 
reduced by more precise experimental information. \s

It is well known, for example, that the precise values of the Higgs
boson masses in the MSSM depend strongly on the exact value of the
input top quark mass. Indeed, because the dominant one--loop
corrections to the Higgs masses grow as $M_t^4$, a shift of several
GeV on $M_t$ will produce a significant variation of $\mh$.  Since the
experimental error on the top quark pole mass, $ \Delta \polemt = \pm
4.3$ GeV, is rather large, it will have a large impact on the
numerical values of the Higgs masses.  The benefits arising from a
more precise measurement of $\polemt$ for the determination of the MSSM
parameters have been discussed in Ref.~\cite{topsw}. \s

Tables \ref{experrmh} and \ref{experrmH} show the variation of $\mh$
and $\mH$, respectively, in the six SPS scenarios, resulting from a
$\pm 1\sigma$ variation in the top quark mass $\polemt$ and in other
SM input parameters, i.e. the bottom quark mass $m_b(m_b)^{\msbar}$,
and the strong and electromagnetic coupling constants
$\as(\polemz)^{\msbar}$ and $\alpha_{\rm em}^{-1}(\polemz)^{\msbar}$,
which are given in eqs.~(\ref{inputew})--(\ref{inputmass}); the other
SM input parameters, such as $G_F, M_Z$ and $M_\tau$, are so precisely
measured that the impact of their experimental errors on the Higgs
boson masses is negligible. \s

As expected, the effect of the error on the top quark mass is rather
large. In the case of the lighter CP--even Higgs boson, a 4.3 GeV
variation in $\polemt$ results in a shift in $\mh$ of approximately 2
GeV in most SPS scenarios,\footnote{~There is a well known rule of
thumb for the so--called $M_h^{\rm max}$ scenario, according to which
a variation of one GeV in $\polemt$ results into a variation of one
GeV in $\mh$. We notice that this rule does not generally hold in the
constrained MSSM scenarios.} with larger (lower) values of $\polemt$
increasing (decreasing) the obtained value for $\mh$; an exception is
given by the scenario SPS5, where, due to a large stop mixing and a
low $\tan \beta$ value, the top quark mass plays a more prominent
role, and the shift can reach 3 GeV. In the case of the heavier
CP--even Higgs boson, the variations are at the level of a few
percent, except for the SPS4 scenario, where they reach the level of
$10\%$. Differently from the case of the lighter Higgs boson mass, $\mH$
can either decrease or increase with increasing (or decreasing)
$\polemt$. This is due to the fact that, as discussed previously, the
heavier Higgs boson masses are mostly sensitive to the values of the
soft SUSY--breaking Higgs mass parameters, whose RGE evolution has a
complicated dependence on the top Yukawa coupling. \s

The impact on the value of $\mh$ of the experimental error on the
bottom quark mass, $m_b(m_b)^{\msbar} = 4.25 \pm 0.25$ GeV, is
significant only in the SPS4 scenario with high $\tb$, where it can
lead to a shift of about 100 MeV; otherwise, it is negligible.  The
effect, however, can reach the level of 10\% or so for the
determination of $\mH$ in this high $\tb$ scenario, since, in this
case, the bottom Yukawa coupling plays an important role in the
evolution of the soft SUSY--breaking Higgs parameters upon which $\mH$
depends. Note that lower (higher) values of the bottom quark mass in
general increase (decrease) the values of $\mH$. \s

The main effect of the error on the strong coupling constant is to
generate an uncertainty in the top and bottom quark running masses,
which propagates on the soft SUSY--breaking parameters and eventually
on the Higgs mass spectrum. Varying $\as(\polemz)^{\msbar} = 0.1172
\pm 0.002$ within its $1\sigma$ range leads in general to an
uncertainty of less than one per mille for $\mh$ and one percent for
$\mH$ (an exception being the scenario SPS4 where the uncertainty for
$\mH$ is two percent).\s
    
Finally, as one might have expected, the uncertainty arising from the
error on the electromagnetic fine structure constant,
$\alpha^{-1}(\polemz)^{\msbar} = 127.934 \pm 0.027$ (due to the
hadronic contribution uncertainties), is essentially negligible: in
all cases, it barely reaches the per mille level for both $\mh$ and
$\mH$.\s

\begin{table}[p]
\renewcommand{\arraystretch}{1.65}
\begin{center}
\begin{tabular}{|l|c|c|c|c|c|c|}
\hline
Approximation & SPS1a & SPS2 & SPS4 & SPS5 & SPS8 & SPS9 \\
\hline
Default & 112.1 & 116.8 & 114.1 & 116.2 & 115.4 & 117.5 \\
\hline
$\Delta \polemt = \pm 4.3$ GeV & $^{+1.59}_{-1.55}$
                         & $^{+2.12}_{-2.04}$
                         & $^{+1.73}_{-1.69}$
                         & $^{+2.09}_{-3.05}$
                         & $^{+2.26}_{-2.14}$
                         & $^{+1.89}_{-1.84}$ \\
$\Delta m_b = \pm 0.25$ GeV & $^{<0.01}_{+0.02}$
                         & $^{<0.01}_{<0.01}$
                         & $^{-0.07}_{+0.09}$
                         & $^{<0.01}_{<0.01}$
                         & $^{<0.01}_{<0.01}$
                         & $^{-0.01}_{+0.01}$ \\
$\Delta \alpha_s =\pm 0.002$ & $^{+0.09}_{-0.06}$
                         & $^{-0.06}_{+0.07}$
                         & $^{+0.06}_{-0.05}$
                         & $^{-0.09}_{+0.03}$
                         & $^{-0.05}_{+0.05}$
                         & $^{+0.09}_{-0.10}$ \\
$\Delta \alpha^{-1} = \pm 0.027$ & $^{+0.02}_{<0.01}$
                         & $^{+0.01}_{<0.01}$
                         & $^{+0.01}_{<0.01}$
                         & $^{-0.01}_{<0.01}$ 
                         & $^{<0.01}_{<0.01}$
                         & $^{-0.01}_{<0.01}$ \\
\hline
\end{tabular}
\vspace{.2cm}
\caption{The lighter CP--even Higgs boson mass, $\mh$, in the six SPS
scenarios, as computed by \suspect\ using different input values
for the SM parameters. The first row contains the default
values in GeV, the other rows contain the shifts (in GeV) due to the
different inputs.}
\label{experrmh}
\vspace{1.cm}
\begin{tabular}{|l|c|c|c|c|c|c|}
\hline
Approximation & SPS1a & SPS2 & SPS4 & SPS5 & SPS8 & SPS9 \\
\hline
Default & 405.9 & 1551.2 & 354.4 & 686.6 & 550.3 & 1056.6 \\
\hline
$\Delta \polemt = \pm 4.3$ GeV & $^{+6.5}_{-6.7}$
                         & $^{+42.6}_{-42.5}$
                         & $^{+27.2}_{-29.4}$
                         & $^{-11.1}_{+7.5}$
                         & $^{+16.0}_{-15.8}$
                         & $^{-15.6}_{+10.2}$ \\
$\Delta m_b = \pm 0.25$ GeV & $^{-1.2}_{+1.2}$
                         & $^{-3.0}_{+2.8}$
                         & $^{-39.5}_{+34.2}$
                         & $^{-0.8}_{+0.7}$
                         & $^{-2.5}_{+2.3}$
                         & $^{-5.3}_{+4.9}$ \\
$\Delta \alpha_s =\pm 0.002$ & $^{+2.2}_{-1.9}$
                         & $^{-7.1}_{+7.4}$
                         & $^{+6.3}_{-6.2}$
                         & $^{+5.7}_{-6.1}$
                         & $^{+3.4}_{-3.5}$
                         & $^{+13.9}_{-14.4}$ \\
$\Delta \alpha^{-1} = \pm 0.027$ & $^{+0.3}_{-0.2}$ 
                         & $^{+0.4}_{-0.2}$
                         & $^{+0.2}_{-0.1}$ 
                         & $^{+0.2}_{-0.2}$
                         & $^{-0.1}_{+0.1}$ 
                         & $^{-0.1}_{<0.1}$ \\
\hline
\end{tabular}
\vspace{.2cm}
\caption{Same as table \ref{experrmh} for the heavier CP--even Higgs
boson mass, $\mH$.}
\label{experrmH}
\end{center}
\vspace*{-0.5cm}
\end{table}
\begin{table}[p]
\renewcommand{\arraystretch}{1.3}
\begin{center}
\begin{tabular}{|l|c|c|c|}
\hline
Approximation & pMSSM1 & pMSSM2 & pMSSM3 \\
\hline 
Default & 113.83 & 118.39 & 129.35 \\
\hline
$\Delta \polemt = \pm 4.3$ GeV & $^{+2.31}_{-2.21}$ 
                         & $^{+2.72}_{-2.60}$ 
                         & $^{+3.83}_{-3.66}$  \\
$\Delta \alpha_s = \pm 0.002$ & $\mp 0.17$ 
                             & $\mp 0.21$ 
                             & $\mp 0.30$  \\
\hline
\end{tabular}
\caption{The lighter CP--even Higgs boson mass, $\mh$, in the three pMSSM
scenarios, as computed by \suspect\ using different input values
for the SM parameters. The first row contains the default
values in GeV, the other rows contain the shifts (in GeV) due to the
different $M_t$ and $\as$ inputs; the errors due to $\Delta m_b$ and
$\Delta \alpha$ are negligible.}
\label{experrmhpMSSM}
\vspace*{-0.5cm}
\end{center}
\end{table}

We now turn to the case of the pMSSM. First of all, in the considered
scenarios the errors due to the variation of all SM parameters within
their $1\sigma$ range are completely negligible in the case of the
heavier CP--even Higgs boson: the maximal shift is less than 100 MeV
(i.e. $ \lsim 0.01\%$) and occurs when the top mass is varied within
its 4.3 GeV uncertainty. As discussed previously, this is due to the
fact that in the pMSSM the value $\mH$ is essentially controlled by
the one of $\ma$, which is fixed to 1 TeV in these scenarios. Second,
even in the case of the lighter Higgs boson, the errors due to the
variation of $\alpha$ and $m_b$ are very small and can be safely
neglected (the insensitivity to the precise value of $\mb$ is due to
the moderate value chosen for $\tb$ in the considered scenarios). \s

The only SM parameters which have an impact on the value of $\mh$ in
the considered scenarios are $\polemt$ and $\as$. The shifts due to
their experimental errors in the three scenarios are shown in table
\ref{experrmhpMSSM}. The main effect of the variation of $\as$ is to
shift the $\drbar$ top mass $\mt$ which enters the loop corrections to
the lighter Higgs boson mass; the shift in $\mh$ increases with the
magnitude of the stop mixing and reaches the level of 300 MeV in the
large mixing scenario. The largest uncertainty is, as expected, due to
the experimental error on the top quark mass. The effect follows the
same trend as in the SPS scenarios: for small mixing (pMSSM1) the
induced error is of about 2 GeV, and it reaches the level of 4 GeV in
the large mixing scenario (pMSSM3). To obtain an accuracy of 50 MeV on
the lighter Higgs boson mass (as will be measured experimentally at
the future colliders) one needs an experimental determination of $M_t$
to better than 100 MeV. Our results are in full agreement with
Ref.~\cite{topsw}, where the uncertainty in $\mh$ induced by the
errors on $M_t,\, \as,\,m_b$ and $M_W$ is discussed. \s

In conclusion, the experimental uncertainties on the SM parameters
lead to an error of about 3--4 GeV on the lighter Higgs boson mass
$M_h$. This error is thus of the same magnitude as the total
theoretical uncertainty which has been discussed in the previous
section, but will be decreased when new measurements of the top mass
become available. \s

Another independent source of uncertainty are the experimental errors in 
the determination of the masses and couplings of the supersymmetric particles.
A detailed study of this issue goes beyond the scope of our paper.

%


\section{Phenomenological consequences}
\label{secpheno}

In this section, we discuss the phenomenological impact of the
radiative corrections in the MSSM Higgs sector and of the related
theoretical and experimental uncertainties.  In particular, we study
the maximal values for the lighter Higgs mass as well as the
constraints on $\tb$ coming from the negative searches of the Higgs
bosons at LEP2.  The analyses will be performed in the context of the
constrained MSSM models (mSUGRA, AMSB and GMSB), but also in the
context of the unconstrained model (pMSSM).

\subsection{The upper bound on the lighter Higgs boson mass}
\label{sec:mhmax}

At the tree-level, the mass of the lighter MSSM Higgs boson $\mh$ is
bounded from above by $\polemz$. Loop corrections increase this bound
to $\sim 135$ GeV as discussed previously. The knowledge of the
precise value of this upper bound is crucial for several reasons.  

\begin{itemize}
\vspace*{-2mm}

\item[--] It is entirely conceivable that the lighter Higgs boson be
discovered before the SUSY particles. In this case, it is important to
accurately know the maximal possible value of $\mh$ in order to
discriminate between the SM and the MSSM. Furthermore, even if SUSY
particles have been observed, the knowledge of this value, in the
absence of extra information, could allow one to distinguish between
the MSSM and some non minimal SUSY models where the Higgs sector is
extended.

\vspace*{-2mm}
\item[--] Even in the context of the MSSM itself, the knowledge of
the maximal $\mh$ value could allow one to discriminate between various
scenarios of SUSY breaking. This approach would be then complementary
to the study of the differences in the SUSY particle spectra induced by
the various breaking schemes.

\vspace*{-2mm}
\item[--] The value $\mh \sim 135$ GeV has a rather special status in
the decoupling regime $\ma \gg \polemz$ (which occurs very often in
the constrained scenarios) where all of the MSSM Higgs bosons except for
the lighter are too heavy to be observed experimentally at the next
generation of high--energy colliders.  Indeed, if $\mh$ is below this
critical value, the Higgs particle will dominantly decay into $b$
quark pairs, while above this value it will dominantly decay into $W$
bosons (for a detailed discussion of the decay modes, see
Ref.~\cite{decays} for instance). 
\vspace*{-2mm}
\end{itemize}

\subsubsection{The upper bound on $M_h$ in the pMSSM}
\label{sec:mhmaxp}

As discussed in section 3.4, in the pMSSM only a small subset of input
parameters have a significant impact on the MSSM Higgs sector. In
first approximation, to obtain the maximal value of lighter Higgs
boson mass one does not need to scan over the 22 free pMSSM
parameters, but, rather, to choose the relevant parameters in such a
way that the one--loop radiative correction $\epsilon$ in
eq.~(\ref{higgscorr}) is maximised. In particular, one can obtain a
reasonable approximation of the maximal $\mh$ when one has:

\begin{itemize}
\vspace*{-2mm}

\item[i)] large values of the parameter $\tb$, $\tb \gsim 20$ (but
still $\tb \lsim 60$ to keep the bottom Yukawa coupling in a 
perturbative regime);
\vspace*{-2mm}

\item[ii)] a decoupling regime with a heavy pseudoscalar Higgs boson, 
$\ma \sim \mathcal{O}$(TeV);
\vspace*{-2mm}

\item[iii)] heavy stops, i.e. large $M_S$ values; 
we note, however, that heavier stops correspond to more fine tuning of the 
parameters in order to achieve the correct minimum of the Higgs 
potential~\cite{fineTuning} and we choose $M_S=2$ TeV as a maximal value;
\vspace*{-2mm}

\item[iv)] a stop trilinear coupling such that $X_t$ is close to
$+\sqrt{6}\, M_S$, as exemplified by fig.~\ref{figmhvsxt}.
\vspace*{-2mm}
\end{itemize} 

As a starting point, we realise the assumptions above by adopting the
$M_h^{\rm max}$ scenario of Ref.~\cite{benchmarks}, which was used as
a benchmark point for the LEP2 Higgs analyses; we choose however to be
conservative, scaling the relevant soft SUSY--breaking parameters by a
factor of two and using the upper limit $\tb \sim 60$:
\begin{eqnarray}
{M_h}^{\rm max}_{\rm bench} \ : \ \begin{array}{c} 
\tb=60 \, , \ M_S=M_A=2~{\rm TeV} \, , \ A_t=A_b =\sqrt{6}\,M_S\,, \\
M_2  \simeq 2 \,M_1 = -\mu=400~{\rm GeV} \, , \ M_3=0.8 \,M_S\,.
\label{pbenchmark}
\end{array}
\end{eqnarray}

For the central value $\polemt=178$ GeV, and varying the values of
$\as$ and $m_b$ (recall that we are in the large $\tb$ regime where
the bottom Yukawa coupling is enhanced) in their $1\sigma$ allowed
range, one obtains the maximal value of the lighter MSSM Higgs boson,
\be
{M_h}^{\rm max}_{\rm bench}\simeq 138~{\rm GeV}\ {\rm for}\ M_t=178~{\rm GeV}.
\label{Mh-bench}
\ee  

This value is higher than the one which is often quoted in the
literature, $M_h \lsim 135$ GeV.  The tendency of our $\drbar$
calculation to give lower values of $\mh$ than the OS calculation
(using the same parameters as in Ref.~\cite{benchmarks} we would find
$\mh \lsim 130$ GeV) is compensated by the increase of the central
value of the top quark mass and by the fact that we conservatively set
the SUSY--breaking scale to 2 TeV. \s

However, the bound in eq.~(\ref{Mh-bench}) on the lighter Higgs boson
mass is not yet fully optimised.  In order to find the {\em maximum
maximorum}~\footnote{We thank Daniel Treille for suggesting this
expression.} $M_h$ value, one can still vary in a reasonable range the
SUSY parameters entering the radiative corrections and add the
estimated theoretical (and experimental) uncertainties.  We have
therefore studied how to tune the pMSSM parameters in such a way that
the corrections to $\mh$ are maximised. \s

For instance, the one--loop electroweak corrections involving Higgs,
gauge bosons and their fermionic superpartners tend to lower the value
of $\mh$ when the chargino and neutralino masses are large. On the
other hand, those corrections get small, allowing for a maximal $\mh$,
when the chargino and neutralino masses are of the same order of the
gauge boson masses (see e.g.~Ref.~\cite{HHH} for a discussion). For
$\mu \sim M_2 \sim \frac{3}{2}M_1 \sim 150$ GeV, the upper bound on
$M_h$ can be pushed upwards by about $2$ GeV, while the lightest
chargino mass is still larger than 104 GeV, as required by the
negative searches at the LEP2 \cite{pdg}.\s

The gluino mass enters the two--loop QCD corrections in the Higgs
sector and can have a non--negligible impact on the value of
$M_h$. For large $\tb$, lowering the value of the gluino mass
parameter to $M_3 \sim 0.4 \,M_S$ leads to an increase of $M_h$ of
about 2 GeV. \s

Another increase of the upper bound on $M_h$ can be obtained by
relaxing the equality of the stop mass parameters at the weak scale
(this turns out to affect $\mh$ through changing the threshold
correction to the top Yukawa coupling). For large $\tb$, assuming
$M_{\tilde{t}_L} \simeq 2 M_{\tilde{t}_R} \simeq 2.5$ TeV (which still
keeps the EWSB scale, defined as the geometric mean of the two stop
masses, in the vicinity of 2 TeV), one obtains an increase of $M_h$ of
about 2 GeV.\s

One can then optimise the soft SUSY--breaking parameters to maximise
all the effects mentioned above. Using the codes \spheno\ and
\suspect, we have performed a scan of the MSSM parameter space with
the following constraints: $(i)$ $M_{\rm EWSB} \leq 2$ TeV, even if
some SUSY particles (in particular the heavier stop) may have masses
above this value; $(ii)$ the trilinear couplings of the third
generation sfermions should not exceed the value $A_{t,b,\tau} \sim
3\, M_{\rm EWSB}$, from which problems with charge and colour breaking
(CCB) minima might appear \cite{CCB}; ($iii)$ all sparticles, in
particular the lightest chargino and the gluino, should have masses
that exceed their experimental bounds; ($iv$) the resulting spectrum
should lead to acceptable contributions to the $\rho$ parameter, to
the anomalous magnetic moment of the muon $(g-2)_\mu$ and to the
radiative $b\to s\gamma$ decay\footnote{The $(g-2)_\mu$ bound is
easily satisfied as the second--generation sfermion masses can be
taken to be large without affecting the Higgs mass calculation, while
the $\Delta \rho$ constraint is in general weaker than the related CCB
constraint. On the other hand, constraints coming from dark matter are
not studied, being beyond the scope of this paper.}.  The two codes
give rather close results, the differences being of the order of 1
GeV. To provide a conservative upper bound on $\mh$, we quote the
result of the code that gives the largest output value. \s

Using the central value of the top quark mass, we find the following
upper bound on the lighter Higgs boson mass in the unconstrained MSSM:
\be
\mh \lsim 143~{\rm GeV} \ \  {\rm for} \ \  M_t=178~{\rm GeV}.
\ee

Thus, tuning the SUSY parameters\footnote{For files detailing the
choices of parameters that lead to the maximal $\mh$ and the resulting
MSSM mass spectra, see the \suspect~\cite{suspect} and
\spheno~\cite{spheno} webpages.}  in the way discussed above is very
important, leading to an increase of $\sim 5$ GeV of the upper bound
on $M_h$. \s

If we vary the top quark mass within its $1\sigma$ experimental range 
we obtain the following upper bounds on $M_h$ for the lower and upper 
$M_t$ values
\begin{eqnarray}
\mh &\lsim & 138~{\rm GeV} \ \ {\rm for} \ \polemt=173.7~ {\rm GeV},
 \nonumber \\
\mh &\lsim & 148~{\rm GeV} \ \ {\rm for} \ \polemt=182.3~ {\rm GeV}. 
\end{eqnarray}

Finally, to obtain the most conservative maximal value of $\mh$, one
has to include the theoretical uncertainty. For the highest value of
the top quark mass, $M_t=182.3$ GeV, when we also include the
theoretical uncertainty that we estimate to be $\Delta \mh \approx 4$
GeV in this case, we obtain the following {\em maximum maximorum}
bound on the lighter Higgs boson mass in the unconstrained MSSM:
\be
\mh \lsim 152~{\rm GeV}.
\label{mhcons}
\ee 

Results for the maximal $\mh$ value in the unconstrained pMSSM, based
on the two--loop OS calculation implemented in \feynhiggs, have been
presented e.g.~in Ref.~\cite{dhhsw}, where the upper bound $\mh \lsim
140$ GeV can be inferred for $M_S=2$ TeV and $\polemt = 179.4$ GeV.  A
recent update \cite{sventalk}, using the latest upper value $\polemt =
182.3$ GeV, gives $\mh \lsim 143$ GeV. In both cases, the MSSM
parameters are chosen as in the $M_h^{\rm max}$ scenario of
Ref.~\cite{benchmarks}, and the theoretical uncertainty is not taken
into account.  Including the theoretical uncertainty, these results
become only a few GeV lower than the upper bounds discussed above.

\subsubsection{The upper bound on $M_h$ in the constrained MSSM}
\label{sec:mhmaxc}

In the mSUGRA, GMSB and AMSB scenarios, the various parameters which
enter the radiative corrections are not all independent, due to the
relations between SUSY breaking parameters that are set at the
high--energy scale. In addition, the radiative EWSB constraint must be
fulfilled for each set of input parameters (in the pMSSM, this is
automatic since $\ma$ and $\mu$ are used as input parameters). Thus,
in contrast to what occurs in the pMSSM, it is not possible to freely
tune all relevant weak-scale parameters in order to get a maximal
value of $\mh$.  This should in principle make the analysis more
complicated. However, since we have only a small set of input
parameters in the constrained case, one can easily perform scans of
the parameter space. \s

\begin{figure}[t]
\begin{center}
\vspace*{-1.5cm}
{\includegraphics[width=0.7\textwidth]{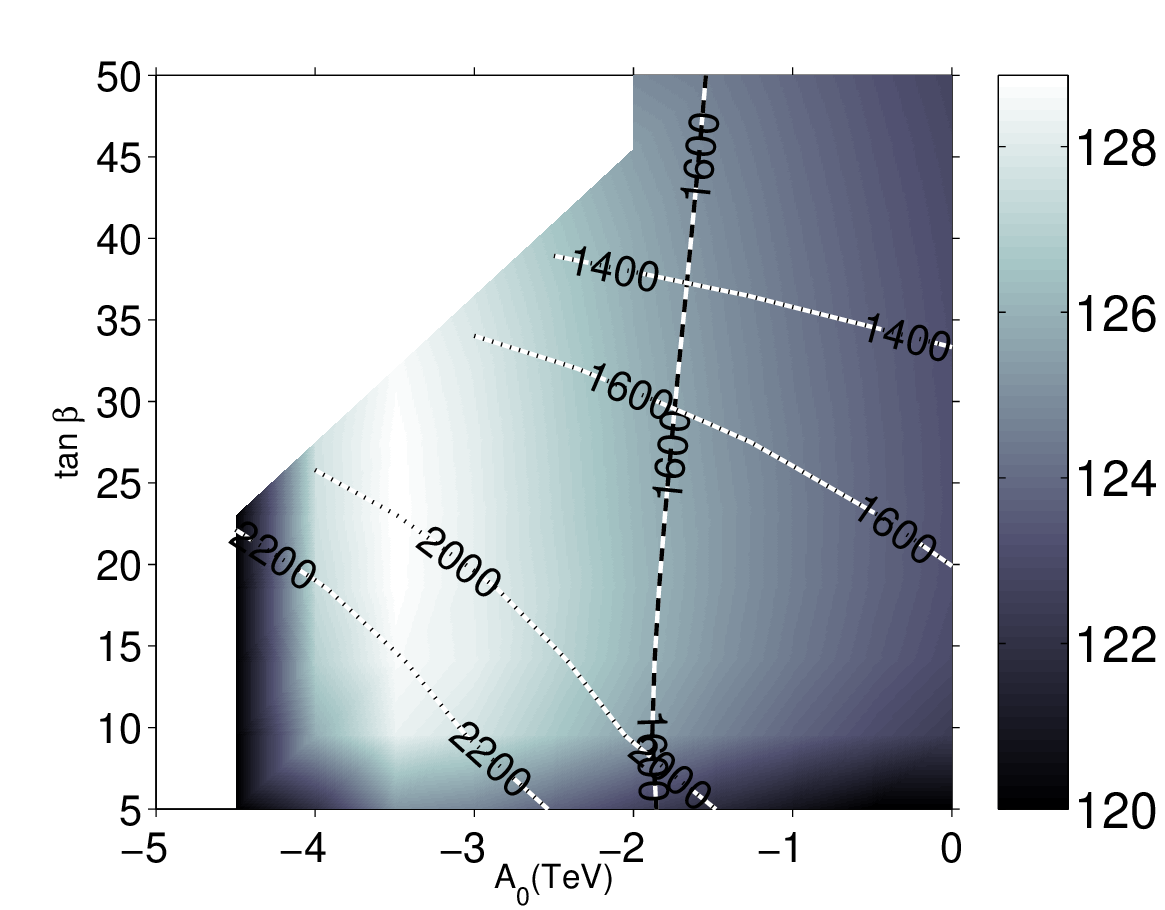}}
%
\caption{The lighter Higgs boson mass $\mh$ in the $\tb-A_0$ plane of the
mSUGRA scenario. Other input parameters are $\m0 = \mhf = 1$ TeV and
$\mu>0$. The SM input parameters are fixed to their default
values. The lighter Higgs boson mass is shown as background density, as
measured by the bar on the right.  Dashed contours are lines of equal
$M_S$, dotted contours are lines of equal $\mH$. The white region
contains scalars with negative squared mass.}
\label{fig:mhscans1}
\end{center}
\vspace*{-0.5cm}
\end{figure}
\begin{figure}[t] 
\begin{center}
\vspace*{-1cm}
{\includegraphics[width=0.7\textwidth]{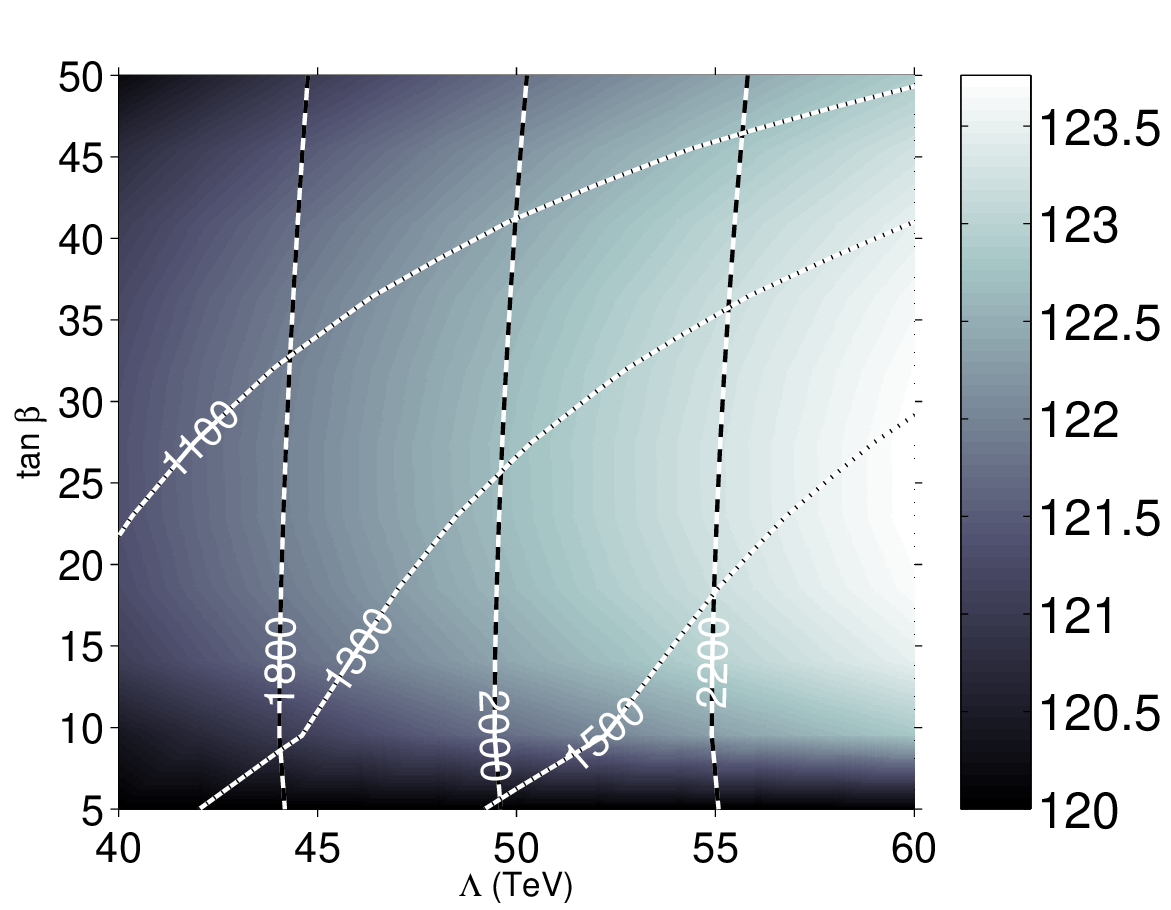}}
%
\caption{Same as fig.~\ref{fig:mhscans1} in the $\tb-\Lambda$ plane of the
GMSB scenario. Other input parameters are $M_{\rm mess}=10^5 \times \Lambda$,
$N_{\rm mess}=8$ and $\mu>0$.}
\label{fig:mhscans2}
\end{center}
\vspace*{-0.5cm}
\end{figure}
\begin{figure}[t]
\begin{center}
\vspace*{-1.5cm}
{\includegraphics[width=0.72\textwidth]{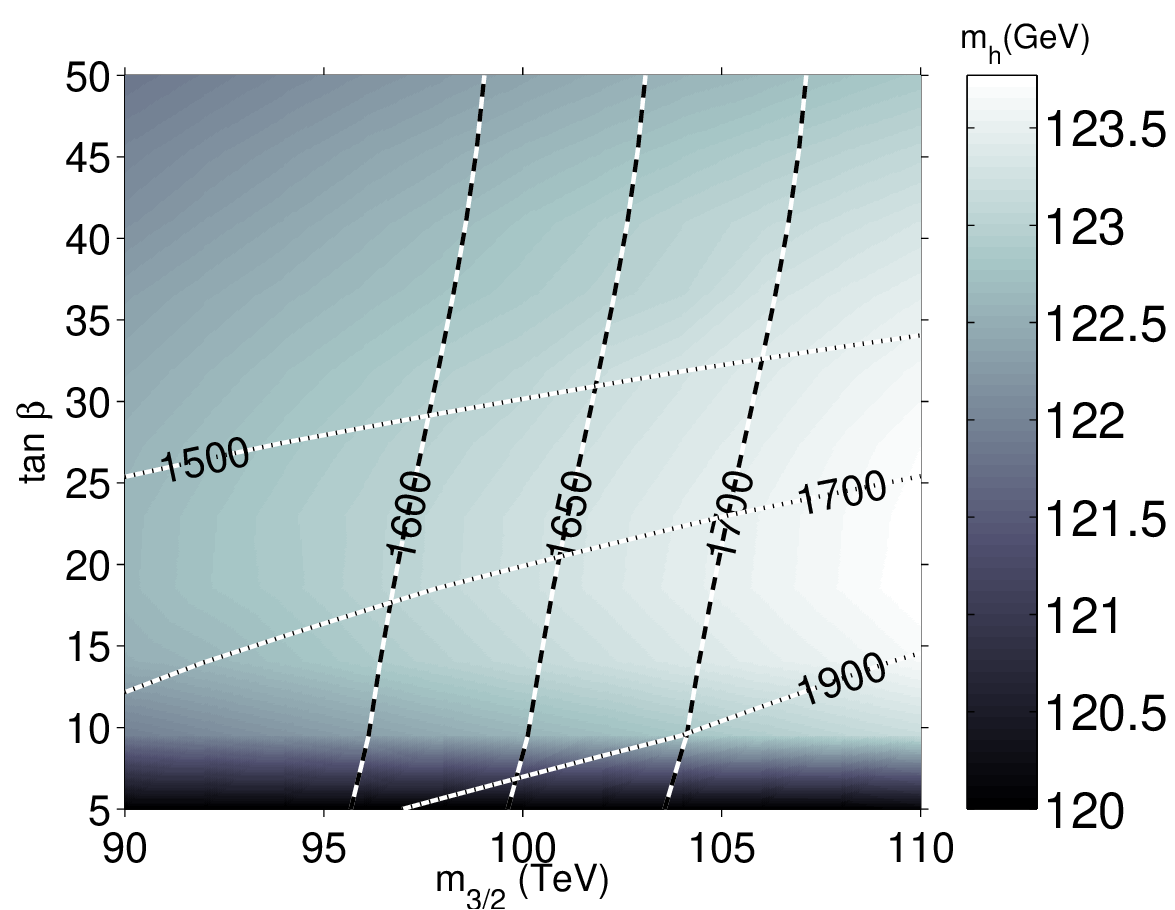}}
%
\caption{Same as fig.~\ref{fig:mhscans1} in the $\tb-\m0$ plane of the
AMSB scenario. Other input parameters are $m_{3/2}= 100$ TeV and
$\mu<0\,.$}
\label{fig:mhscans3}
\end{center}
\vspace*{-.5cm}
\end{figure}

For the purpose of illustration, we show in
figs.~\ref{fig:mhscans1}--\ref{fig:mhscans3} how $\mh$ depends on some
relevant parameters of the mSUGRA, GMSB and AMSB scenarios. The results 
shown in the figures are obtained with \softsusy. 
Fig.~\ref{fig:mhscans1} shows the value of $\mh$ as the background
density in the plane $A_0-\tb$ of the mSUGRA model. $\m0$ and $\mhf$
are set to 1 TeV in order to have an average stop mass $M_S$ of the
order of 2 TeV, and $\mu$ is taken positive. In the white region in
the upper--left corner of the plot a sparticle squared mass is
negative, resulting in an unacceptable scalar minimum. Contours of
$M_S$ and $\mH$ are also shown. Large stop mixing is achieved for
large and negative $A_0$, around $-4$ TeV, corresponding to the maximal
$\mh \approx 128.5$ GeV region in the figure. \s

Fig.~\ref{fig:mhscans2} shows the value of $\mh$ as the background
density in the plane $\Lambda-\tb$ of the GMSB model. The messenger
mass $M_{\rm mess}$ is taken equal to $10^5 \times \Lambda$. Moreover,
$N_{\rm mess}=8$ and $\mu$ is taken positive. It can be seen that the
maximal value for $\mh$, which is around 123.5 GeV in this example, is
obtained for large values of the SUSY--breaking scale $\Lambda$, which
correspond to large stop masses. \s

Finally, in fig.~\ref{fig:mhscans3} we show $\mh$ in the $m_{3/2}-\tb$
plane of the AMSB model. The gravitino mass $m_{3/2}$ is varied in a
range that allows for stop masses of the order of 2 TeV, the scalar
mass term $\m0$ is set to 1 TeV, and $\mu$ is taken to be negative. Even in
this case, it can be seen that $\mh$ increases for larger values of
$m_{3/2}$, which correspond to larger values of the stop masses. \s

Plots like the ones discussed above can give useful indications about
the choice of parameters that leads to the maximal $\mh$. However, in
order to get a reliable determination of the maximal $\mh$ in a given
SUSY--breaking scenario it is necessary to scan through the allowed
range of values for all of the relevant parameters.  Such scans were
previously performed, e.g., in
Refs.~\cite{scanadk,scansven1,scansven2}.  The most recent among these
analyses, Ref.~\cite{scansven2}, used the combination of {\tt
FeynHiggs} with different private codes \cite{privcodes} for the
computation of the MSSM parameters in the various SUSY--breaking
scenarios. For convenience, we take the same ranges of soft
SUSY--breaking input parameters as in Ref.~\cite{scansven2}:
\begin{center}
\begin{tabular}{rccc}
mSUGRA: & 50 GeV $\leq\m0\leq$ 1 TeV, & 50 GeV $\leq\mhf\leq$ 1 TeV,
& $|A_0|\leq3 $ TeV;\\
&&&\\
GMSB: & 10 TeV $\leq\Lambda\leq$ 200 TeV, &
1.01 $\leq M_{\rm mess} / \Lambda \leq 10^5$, & 1 $\leq N_{\rm mess} \leq$ 8;\\
&&&\\
AMSB: & 20 TeV $ \leq m_{3/2}\leq$ 100 TeV, & 50 GeV $\leq\m0\leq$ 2 TeV. & 
\end{tabular}
\end{center}

Moreover, we require 
$$
1\leq\tan \beta\leq 60
$$
and we allow for both signs of $\mu$.  The SM input parameters are fixed to 
their default values as in eqs.~(\ref{inputew})--(\ref{inputmass}). In 
particular, we take $\polemt = 178.0 \pm 4.3$ GeV \cite{toptev}, whereas
Ref.~\cite{scansven2} used the old world average $\polemt=175$ GeV. \s

To avoid the need for excessive fine--tuning in the EWSB conditions,
we also impose an additional bound on the weak--scale parameters, i.e. 
that the geometrical average of the stop masses is less than 2 TeV, 
\be
M_S = \qewsb = \sqrt{m_{\tilde{t}_1}m_{\tilde{t}_2}} < 2~{\rm TeV}\,.
\ee

However, it is important to bear in mind that, in the absence of a
compelling criterion to define the maximal acceptable amount of
fine--tuning, the choice of the upper bound on $M_S$ is somehow
subjective. This upper bound on $M_S$ affects the maximal attainable
value of $\mh$, as larger values of $M_S$ push the value of $\mh$
upward. \s
\begin{table}[t]
\renewcommand{\arraystretch}{1.4}
\begin{center}
\begin{tabular}{|l|c|c|c||c|}\hline
$M_h^{\rm max}$ & $M_t=173.7$ GeV & $M_t=178.0$ GeV & $M_t=182.3$ GeV 
& conservative \\ \hline
pMSSM & 138  & 143 & 148 & 152 \\ \hline
mSUGRA &  126.2 & 129.0 & 131.7 & 136 \\ 
GMSB  &  120.8 & 123.7  & 126.7 & 131 \\ 
AMSB &   122.0 & 124.6  & 127.1 & 131 \\ \hline
\end{tabular}
\end{center}
\vspace*{-0.5cm}
\caption{Maximal lighter Higgs boson mass (in GeV) in the pMSSM,
mSUGRA, GMSB and AMSB scenarios as obtained by the three spectrum
generators for three values of the top quark mass. The last column
gives the most conservative estimate, where $\polemt=182.3$ GeV and a
theoretical uncertainty around 4 GeV is added.}
\label{tab:mhmax}
\end{table}

Taking the range of SUSY--breaking parameters described above and
scanning with \softsusy, \spheno\ and \suspect\ leads to the results
for the maximal $\mh$ given in table \ref{tab:mhmax}, where $\polemt$
is set to 173.7, 178.0 and 182.3 GeV; we also display, for
completeness, the upper bounds in the case of the pMSSM. \s

We display in table \ref{tab:mhmax} only the result of the code that
gives larger upper bound on $\mh$; the differences between the results
of the three codes in the search for the maximal $\mh$ can amount to 1
GeV, somewhat larger than those found in section \ref{comparecodes}
for the SPS scenarios. Apparently, the higher order effects that
induce the residual differences among the codes are enhanced for the
extreme choices of the SUSY parameters that give rise to the maximal
values of $\mh$. In addition, in the last column of the table we show
our most conservative estimate for the upper bound on the lighter
Higgs boson mass in the mSUGRA, GMSB and AMSB scenarios. For this, we
take $\polemt =182.3$ GeV, the central values for the other SM input
parameters and add to the results an uncertainty of approximately 4
GeV (which includes both the theoretical uncertainty as estimated
previously, and the much smaller error due to the variation of the
remaining SM input parameters). \s

It can be seen that, for the central value of $\polemt$, the GMSB
model has the tightest bounds upon the maximal lighter Higgs mass,
$M_h \lsim 123.7$ GeV, followed closely by the AMSB model, where the
maximal value is obtained for $\mu<0$ (for the central value of the
top mass, setting $\mu>0$ would induce a 0.6 GeV decrease in the
maximal $\mh$). The mSUGRA model has a much looser bound, $\sim 5$ GeV
above the GMSB case, but this bound is more than 10 GeV smaller than
in the case of the pMSSM. This pattern can be qualitatively understood
by considering in each model the allowed weak scale values of $A_t$,
which essentially determines the stop mixing parameter $X_t$ and,
therefore, the value of $M_h$.
The GMSB model has $A_t=0$ at the relatively low scale $M_{\rm mess} <
2 \times 10^{7}$ TeV, thus its magnitude does not increase greatly in
the RG evolution down to $\qewsb$. A small value of $A_t$ implies a
smaller value for the stop mixing, therefore producing less high $\mh$
(see eq.~(\ref{higgscorr})). 
The AMSB model has a non-zero $A_t$ that is fully predicted at any
renormalisation scale in terms of the Yukawa and gauge couplings, and
of the overall SUSY breaking scale $m_{3/2}$ which determines also
the common squark mass $M_S$.  Thus, the ratio between $X_t$ and $M_S$
cannot be tuned at will in order to get the maximal value of
$\mh$. Nevertheless, $X_t/M_S$ turns out to be sizeable at the weak
scale, resulting in a larger value of $\mh$ than in the GMSB model. 
Finally, in the mSUGRA model the only restriction on $A_t(\qewsb)$
comes from $|A_0|<3$ TeV, allowing for close to maximal mixing, but
only for values of the SUSY scale lower than the value $M_S=2$ TeV
which is used as input in the pMSSM case; if we allow for larger and
negative values of $A_0$ we can get closer to, but not saturate, the
pMSSM bound. In fact, it can be seen from fig.~\ref{figmhvsxt} that,
in the unconstrained MSSM, the maximal $\mh$ is obtained for positive
values of $X_t$. On the other hand, in the constrained models the RG
evolution tends to drive $A_t$ (thus, $X_t$) towards negative
values. In order to get at the weak scale a positive $X_t$ large
enough to obtain the maximal $\mh$, we should start from an
unreasonably large positive value of $A_0$. \s

Note that table \ref{tab:mhmax} also shows that changing $\polemt$ by
one standard deviation adds or subtracts about 3 GeV to the maximal
$\mh$ value in each of the cMSSM scenarios, to be compared with the 5
GeV variation discussed previously for the pMSSM.\s

In Ref~\cite{scansven2}, the upper bounds $\mh<126.6, 123.2$ and 124.5
GeV are quoted for the mSUGRA, GMSB and AMSB models,
respectively. These results follow the same qualitative pattern as
those given in table \ref{tab:mhmax}.  However, the good agreement
found in the GMSB and AMSB scenarios using the central value of
$\polemt$ is to some extent accidental: as discussed above, our
$\drbar$ calculation gives results for $\mh$ that are a few GeV
smaller than the ones of the OS calculation in \feynhiggs, but this
effect is compensated by the higher central value of $\polemt$ taken
here ($\polemt=178$ GeV instead of $\polemt=175$ GeV).  Also, the
authors of Ref.~\cite{scansven2} did not impose an upper bound on
$M_S$, resulting for example in stop masses up to 10 TeV in GMSB (and
therefore an enhancement of $\mh$).  Finally, the case $\mu<0$ was not
considered in their analysis, which resulted in a reduced upper bound
on $\mh$ in the AMSB case. \s

Note that the bounds on $M_h$ discussed above will not be in conflict
with any phenomenological constraint. Since the obtained SUSY spectra
in the cMSSMs is rather heavy, one easily evades the collider
experiment bounds on the sparticle and Higgs boson masses and
satisfies the indirect constraints from from high--precision
measurements such as $(g-2)_\mu$, BR($ b \to s \gamma)$ and $\Delta
\rho$ (the dark matter constraint is not included).

\subsection{The lower bound on $\tb$}

The present LEP2 limit on the mass of the SM Higgs boson,
$M_{H^0}>114.4$ GeV \cite{lephiggs}, sets strong constraints on the
parameter space of the MSSM Higgs sector. 
In general, a detailed investigation of the Higgs boson production and
decay modes would be required to establish whether a point in the
MSSM parameter space is excluded by the experimental searches. However,
in the decoupling regime where $\ma \gg \polemz$, the lighter CP--even
Higgs boson of the MSSM has the same couplings to fermions and gauge
bosons as the SM Higgs boson (and hence the same production and decay
properties) thus the LEP2 bound can be extrapolated to $\mh>114.4$
GeV. The tree--level value of $\mh$ decreases for small
values~\footnote{Also, for $\tan \beta<4$, one is close to the
infrared quasi fixed point regime of the MSSM~\cite{fixed}, where the
lighter Higgs boson mass $\mh$ becomes smaller~\cite{fixed2}.} of
$\tb$, as can be seen from eq.~(\ref{mhtree}), and the lower
experimental bound on $\mh$ translates into a lower bound for
$\tb$. \s

In this section, we provide such a bound in the different SUSY
breaking scenarios discussed previously.  Such investigation is
interesting for several reasons:

\begin{itemize}

\item[--] An appealing possibility is that the Yukawa couplings of
the third generation fermions are unified [e.g., $h_b(M_{GUT})=h_\tau
(M_{GUT})$ in minimal SU(5)].  In such models, the value of $\tb$ is of 
critical importance and determines the viability of bottom--tau 
Yukawa coupling unification~\cite{Yukawaunif}.

\item[--] The experimental determination of the value of $\tb$ is an
important measurement to be performed once the supersymmetric and
Higgs particles are detected. Unfortunately, such a measurement can be
difficult (see Ref.~\cite{tanbeta} for instance): in most cases the
relations between the SUSY or Higgs parameters or observables which
can be used for this determination depend on $\cos(2\beta)$ rather
than on $\tb$ itself. This quantity becomes approximately equal to
unity as soon as $\tb \geq 3$, leaving little room for an accurate
measurement of $\tb$, if low values of $\tb$ are excluded by the LEP2
constraint\footnote{Note that, for high values, $\tb$ could be possibly
measured by considering associated production of the heavier MSSM
Higgs bosons with bottom quarks \cite{bottomfus}, tau fusion into
Higgs bosons \cite{choi} or stau decays into Higgs bosons
\cite{bartl}.}. It is thus interesting to see whether there are still
viably low values of $\tan \beta$ that allow for a measurement.

\item[--] A related reason is that, in the MSSM Higgs sector, the
parameter space of the decoupling regime expands for higher values of
$\tb$. For $\tb \gsim 5$ for instance, the lighter Higgs boson becomes
SM--like as soon as $\ma$ exceeds the maximal value of $\mh$. This is
not the case for low values of $\tb$: even for $A$ boson masses of
several hundred GeV, the lighter MSSM Higgs boson has different
couplings to fermions and gauge bosons (and thus different decay and
production properties) than the SM Higgs boson.
\end{itemize}

\subsubsection{The lower bound on $\tb$ in the pMSSM}
\label{sec:tbminp}

In the pMSSM, the absolute lower limit on $\tb$ can be obtained by
maximising the lighter Higgs mass in such a way that it exceeds the
value $\mh>114.4$ GeV~\cite{lephiggs} from direct searches at LEP2
(for earlier studies, see e.g.~Refs.~\cite{sventanb,
dhhsw}). Therefore, one can perform the same analysis as for the
determination of the maximal $M_h$ value discussed in the previous
subsection, except that here, one looks for the minimal value of $\tb$
which leads to $\mh>114.4$ GeV. Thus, one has to be in the regime
defined by the items (ii)--(iv) of section 4.1, i.e. being in the
decoupling regime with a large mixing scenario and using large stop
masses. In addition, as in the case of the maximal $M_h$, one still
has to tune the values of $\mu$ and of the soft SUSY--breaking
parameters in order to maximise the radiative corrections. \s

\begin{figure}[!ht]
\begin{center}
\epsfig{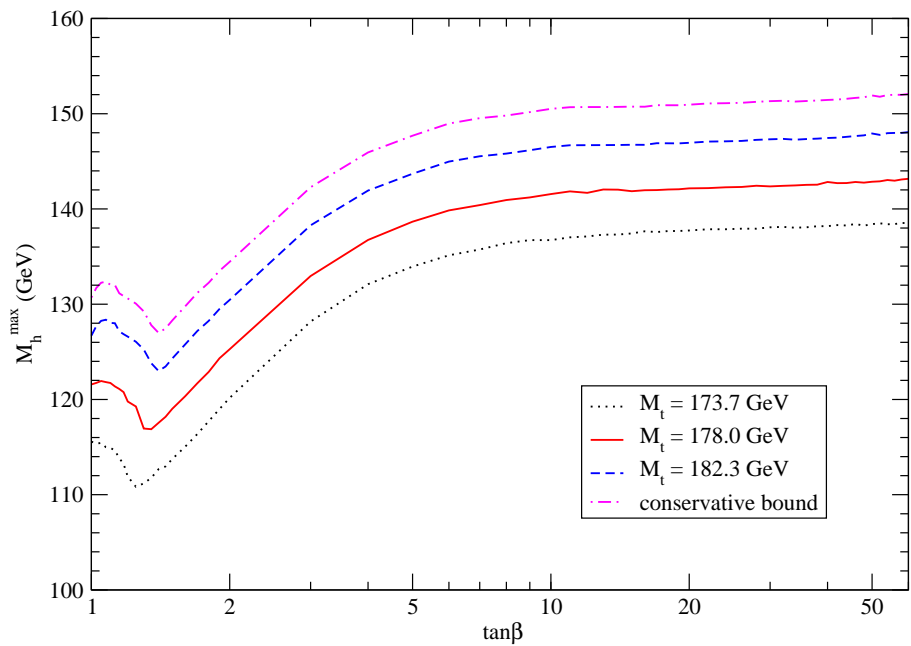}
\caption{Upper bound on the lighter Higgs boson mass $\mh$ in the
pMSSM as a function of $\tb$, as obtained from a full scan of the
parameter space.  The dotted, full and dashed lines correspond to the
top mass values $M_t=$ 173.7, 178.0 and 182.3 GeV, respectively, while
the dash--dotted line on the top is for the conservative case where
$M_t=182.3$ GeV is used and a 4 GeV theoretical uncertainty is added
to $M_h$.}
\label{fig:mhvstb}
\end{center}
\end{figure}

To obtain the lower bound on the parameter $\tb$, we have performed a
scan similar to the one discussed in section 4.1.1, looking for the
maximal value of $M_h$ corresponding to a given value of $\tb$. In
fig.~\ref{fig:mhvstb}, we display the variation of the maximal $\mh$
in the pMSSM as a function of $\tb$.  The dotted, full and dashed lines
show the values of $M_h$ for the top mass values $M_t=173.7, 178.0$
and 182.3 GeV, respectively, while the dash--dotted line on the top is
for the conservative case where $M_t=182.3$ GeV is used and a 4 GeV
theoretical uncertainty is added to $\mh$. \s

\vspace{1cm}

As can be seen, for the default value $M_t=178$ GeV the LEP2 bound of
114.4 GeV on $\mh$ is always satisfied and therefore, for this top
quark mass value, no empirical bound on $\tb$ (provided that it is
larger than unity) can be derived in the pMSSM.  This is of course
also the case for the larger value of $M_t=182.3$ GeV, and {\it a
  fortiori} for the conservative case where a theoretical uncertainty
is taken into account, where all values $1 \leq \tb \leq 60$ are
allowed by the LEP2 constraint.  Only in the case of a lighter top
quark, $M_t=173.7$ GeV, the range $1.1 \lsim \tb \lsim 1.6$ is
excluded if the LEP2 bound $M_h \geq 114.4$ GeV is to be satisfied.
However, if a theoretical uncertainty of 4 GeV on $M_h$ is included
(meaning, in practice, that the LEP2 Higgs mass bound translates to
the bound $M_h \geq 110.4$ GeV on the prediction obtained without
including the theoretical uncertainty) no bound on the parameter $\tb$
can be obtained from the LEP2 constraint. \s

The local maximum around $\tb=1.2$ for the maximal $\mh$ in
fig.~\ref{fig:mhvstb} is due to the fact that the top Yukawa coupling
is maximised by the threshold corrections in that point. On the other
hand, in the large $\tb$ regime one recovers the maximal $M_h$ values
discussed in the previous section, as shown in the right--hand side of
the figure.  In particular, one sees that the upper bound on the
lighter Higgs boson, in the conservative case where $M_t=182.3$ GeV
and a 4 GeV theoretical uncertainty on $\mh$ is added, is slightly
above 150 GeV.  \s

\subsubsection{The lower bound on $\tb$ in the constrained MSSM}
\label{sec:tbminc}

In the constrained MSSM scenarios, we parameterise the 95\%
C.L. bounds from LEP2 in the $\sin^2(\beta-\alpha)$--$\mh$ parameter
space~\cite{lephiggs}, rather than just using the lower bound on
$\mh$, in case the decoupling regime is not obtained. In practice, the
lowest values of $\tb$ that saturate the LEP2 limit are indeed
obtained in the decoupling regime. To reduce the volume of the space
that is scanned over, we use the parameter ranges as described in
section \ref{sec:mhmax} for the mSUGRA, GMSB and AMSB cases.  The
results are summarised for these three cases in table \ref{tab:tbmin}
where the values $\polemt=173.7, 178.0$ GeV and $182.3$ GeV are used,
while for the other SM inputs we used the default values given in
eqs.~(\ref{inputew})--(\ref{inputmass}).  The last column shows the
most conservative lower bounds, obtained by adding a theoretical uncertainty
of 4 GeV to the prediction of the lighter Higgs boson mass.  For
completeness, we also include in the first row the results in the
pMSSM case discussed previously. In each entry of the table, we quote
the result obtained with the code that gives the weakest lower bound
on $\tb$.\s

\vspace{0.3cm}
\begin{table}[ht]
\renewcommand{\arraystretch}{1.1}
\begin{center}
\begin{tabular}{|l|c|c|c||c|}\hline
$\tb^{\rm min}$ & $M_t=173.7$ GeV & $M_t=178.0$ GeV & $M_t=182.3$ GeV 
& conservative \\ \hline
pMSSM & 1.6  & -- & -- & -- \\ \hline
mSUGRA & 2.8  & 2.4 & 2.1 & 1.9  \\ 
GMSB  & 4.2  & 3.3 & 2.7 & 2.2  \\ 
AMSB & 3.7   & 3.1 & 2.7 & 2.3  \\ \hline
\end{tabular}
\end{center}
\vspace*{-0.3cm}
\caption{Lower bounds on the parameter $\tb$ in the pMSSM, mSUGRA,
GMSB and AMSB scenarios for three values of the pole top mass.  The
last column gives the most conservative estimate, where we use
$\polemt=182.3$ GeV and add a 4 GeV theoretical uncertainty on $\mh$.}
\label{tab:tbmin}
\end{table}

One can see from the table that, as expected, the minimal values of
$\tb$ follow the same pattern as in the evaluation of the maximal
$\mh$ value. The bounds are most stringent in the GMSB scenario,
closely followed by the AMSB scenario, and finally by the mSUGRA
scenario. In all cases the bounds are much stronger than in the
pMSSM. In fact, contrary to the pMSSM case, in the constrained
scenarios there is always a significant bound on the parameter $\tb$
even for the largest value of $M_t$, $\tb \gsim $2.1--2.7, and when a
4 GeV theoretical uncertainty on $\mh$ is included, $\tb
\gsim$1.9--2.3.  This is merely a consequence of the $\sim14$ GeV
difference between the maximal values of $M_h$ in the constrained and
unconstrained cases. \s
 
In Ref.~\cite{scansven2} the lower bounds $\tb > 2.9$, 3.1 and 3.8 for
the mSUGRA, GMSB and AMSB scenarios, respectively, are quoted. In
contrast to our findings, summarised in table \ref{tab:tbmin}, the
bound quoted in Ref.~\cite{scansven2} for the AMSB scenario is
stronger than the one quoted for the GMSB scenario, presumably due to
the fact that the authors did not consider the case $\mu <0$ in their
analysis~\footnote{~We have checked that in the considered scenarios the
SUSY contribution is within the $2\sigma$ range for $(g-2)_{\mu}$
(see Ref.~\cite{gminus2} for a recent analysis).}. On the other hand,
our results for the mSUGRA and GMSB scenarios, using the central value
of $\polemt$, are relatively close to the results of
Ref.~\cite{scansven2}. This is again due to a compensation between the
tendency of the $\drbar$ calculation to give lower values of $\mh$
than the OS calculation (thus, stronger bounds on $\tb$), and the
increased value of $\polemt$, resulting into higher values of $\mh$
(thus, weaker bounds on $\tb$).

\subsection{Implications for the decays of the Higgs bosons}

As discussed in the previous subsections, in the most conservative
case where the top quark mass is set to $M_t=182.3$ GeV and a
theoretical uncertainty of 4 GeV is added to $M_h$, the upper bound on
the lighter Higgs boson mass in the unconstrained MSSM can reach the
level of 150 GeV, and there is no lower bound on the parameter $\tb$
from the negative Higgs boson searches at LEP2. Such large values of
$\mh$ and small values of $\tb$ would have a decisive impact on the
decays of the MSSM Higgs bosons (and also on their production rates)
and thus on their experimental detection at future high--energy
colliders, in particular at the LHC, as will be briefly discussed
below.\s

\begin{figure}[!ht]
\begin{center}
\vspace*{-1cm}
\psfig{figure=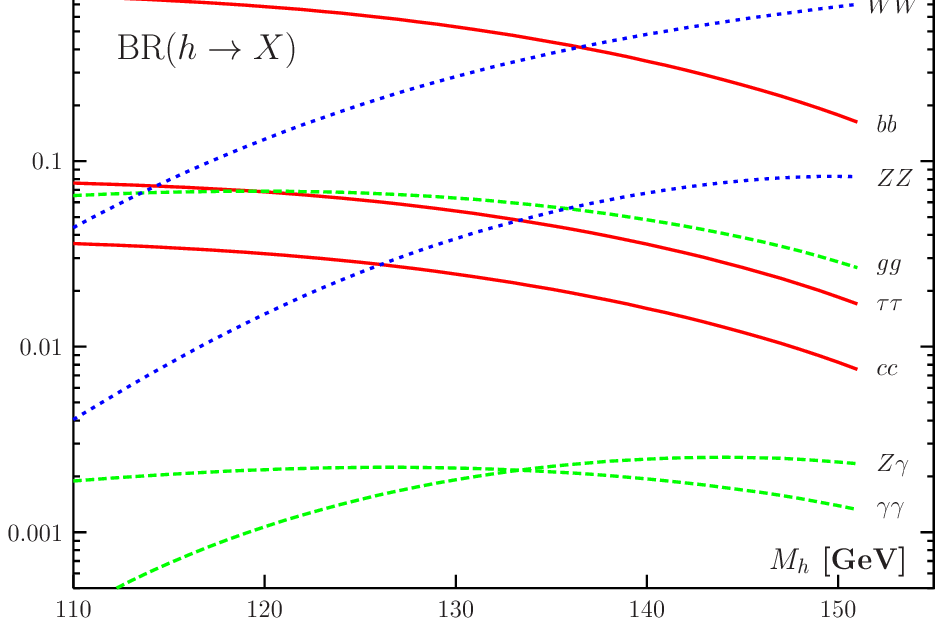, width=14.cm}
\vspace*{-11.5cm}
\caption{The branching fractions of the SM Higgs boson (equivalent to
those of the lighter MSSM Higgs boson $h$ in the decoupling regime) as a
function of its mass.}
\label{fig:hdecay}
\end{center}
\vspace*{-.5cm}
\end{figure}

In the decoupling regime, the lighter Higgs boson $h$ behaves
approximately like the SM Higgs particle. In particular, it has the
same couplings to fermions and gauge bosons, thus the same branching
fractions and production cross sections, provided that the SUSY
particles are heavy enough to make their contribution to the loop
induced decays safely negligible~\cite{loopdecays}. In
fig.~\ref{fig:hdecay}, obtained with the code {\tt HDECAY}, we display
the branching ratios of the SM Higgs boson (equivalent to those of $h$
in the decoupling regime) as a function of its mass. \s

As can be seen, while the $h \to b\bar{b}$ decay mode is dominant for
masses below $M_h \sim 135$ GeV, it is the decay $h \to WW$ (with one
of the gauge boson being off--shell) which becomes the most important
one above this Higgs mass value. In particular, for Higgs boson masses
close to the upper bound, $M_h \sim 150$ GeV, the decay into $W$ boson
pairs reaches a branching ratio of the order of 70\%. Contrary
to the $b\bar b$ decay mode, which suffers from a huge hadronic
background at the LHC, the $WW$ decay mode is much cleaner if one
considers the leptonic decays of one or two $W$ bosons, and therefore
becomes the most promising detection channel at the LHC
\cite{LHCsignals}. \s

Because the partial decay width of the lighter Higgs boson into $WW$
final states is enhanced by the gain in phase space, almost all of the
other decay modes will be suppressed. The relatively clean decays into
$\tau$--lepton pairs, which are a useful signal when the lighter Higgs
boson is produced in the fusion of two gauge bosons at the LHC $pp \to
qqh$ with $h \to \tau^+ \tau^-$ \cite{LHCsignals}, can be suppressed
to the level of 2\% and this detection channel might therefore be
disfavoured.  The rare but very clean decay of the lighter Higgs boson
into two photons is not strongly affected, being at the level of a few
per mille, and still constitutes a viable detection
channel\footnote{~The branching ratio for this decay is not suppressed
because it is dominantly mediated by $W$ boson loops, whose
contribution becomes maximal near the $WW$ threshold.}.\s

Another consequence of a large $\mh$ would be that, because of the
increase of the phase space, the branching ratio for the decay of the
lighter Higgs boson into $Z$ boson pairs can reach the level of
10\%. This would make the detection channel $h \to ZZ^* \to 4 \ell^\pm$
more important at the LHC, and the larger number of events could
eventually allow for the determination of the spin--parity quantum
numbers of the Higgs particle from the study of the angular
correlations between the final state leptons \cite{spin}. \s

Let us now turn to the consequences of the constraints on the
parameter $\tb$ obtained from the negative searches of the MSSM Higgs
bosons at LEP2.  First of all, we recall that the lower $\tb$ bounds
discussed in the previous subsection, have been derived in the
decoupling regime where the pseudoscalar Higgs boson is rather heavy,
$M_A=2$ TeV.  To derive this constraint outside the decoupling regime,
one can no longer use the LEP2 bound on the SM Higgs boson mass, but
rather, the $\sin^2(\beta-\alpha)$--$\mh$ parameter space excluded by
the LEP collaborations \cite{lephiggs} in the search of the MSSM Higgs
bosons in all production channels. A detailed analysis of the lower
bound on $\tb$ for any value of $M_A$ and its impact on the decays of
the MSSM Higgs bosons and on their production rates at the various
high--energy colliders is beyond the scope of this
paper. Nevertheless, a rough analysis shows that, in the conservative
case discussed above, values of $\tb \sim 1.8$ are still allowed for
$M_A$ values as low as 200 GeV, which leads to the following
qualitative conclusions:

\begin{itemize}
\item[--] 
The decays of the three neutral MSSM Higgs bosons into
$b\bar{b}$ and $\tau^+\tau^-$ final states may not be too strongly
enhanced. The lighter MSSM Higgs boson will then behave more like the
SM Higgs particle even outside the decoupling regime.
\item[--]
The decays of the heavier CP--even Higgs bosons into two
gauge bosons and into two lighter Higgs bosons would be less
suppressed. In particular, if the decay $H \to hh$ is allowed it
could lead to the possibility of measuring the trilinear $Hhh$
coupling. 
\item[--] 
The decays of the pseudoscalar and charged Higgs bosons into
a gauge boson and the lighter Higgs boson, $A \to hZ$ and $H^\pm \to
hW^\pm$, could provide interesting detection signals. 
\end{itemize}

A detailed analysis of these features will be presented in a forthcoming 
publication \cite{tocome}.

\section{Conclusions}

In this paper, we provided a detailed analysis of the Higgs sector of
the Minimal Supersymmetric extension of the Standard Model, both in
the constrained cases such as mSUGRA, GMSB and AMSB and in the
unconstrained case, where 22 parameters are allowed to be free at the
EWSB scale. We have provided a very precise determination of the
masses of the neutral Higgs bosons including all the presently
available radiative corrections and using the latest value of the top
quark mass, as recently measured at the Tevatron. \s

In three public RGE codes which evaluate the supersymmetric particle
spectrum, \softsusy, \spheno\ and \suspect, we have implemented all of
the available radiative corrections for the Higgs boson masses and for
the effective potential, in the $\drbar$ renormalisation scheme which
is more natural for these spectrum generators.  In particular, we have
incorporated the full set of one--loop radiative corrections to the
Higgs self--energies and to the tadpoles, including the contributions
of all SUSY particles and taking into account the external momentum
dependence. We have also included the dominant two--loop radiative
corrections at vanishing external momentum, i.e.  the corrections
which involve the strong gauge coupling and the Yukawa couplings of
the third generation fermions. To complete this picture, we have
derived the missing radiative corrections controlled by the $\tau$
lepton Yukawa coupling, which however turned out to be numerically
small. \s

We have performed a detailed comparison of the three programs, in six SPS
scenarios for the mSUGRA, GMSB and AMSB cases as well as in three
scenarios for the unconstrained MSSM. We have found a very good
agreement among the results of the three codes: to within half GeV for
the mass of the lighter CP--even Higgs boson and within 1\% for the
heavier CP--even and CP--odd Higgs bosons. We therefore conclude that
the three codes provide now a reliable and very precise determination
of the neutral Higgs boson masses, both in constrained and
unconstrained MSSM scenarios. \s

We have then studied the effects of the various radiative corrections
on the Higgs boson masses in constrained scenarios and in the $\drbar$
scheme.  The corrections due to the strong interactions, $\oatasabas$,
increase the lighter Higgs boson mass by up to 6 GeV, while the pure
Yukawa corrections due to the top and bottom quarks, $\oatqatababq$,
can decrease it by 1 GeV; the corrections due to the $\tau$--Yukawa
couplings, $\oatauqatab$, can be safely neglected. The impact of some
higher order effects, such as those originating from the inclusion of
the two--loop standard QCD and one--loop electroweak corrections to
the top quark mass, the resummation of the SUSY--QCD corrections in
the bottom quark mass, the use of the two--loop RGEs for the sfermion
masses and the use of the pole Higgs boson masses in the one--loop
corrections, has also been analysed. These higher order effects can
alter the results by up to half a GeV for $M_h$ and up to a few
percent in $M_H$. \s

A detailed study of the theoretical and experimental uncertainties in
the determination of the CP--even Higgs boson masses has been
performed in both the constrained and unconstrained MSSM
scenarios. The comparison between the three RGE codes and the program
\feynhiggs, where the MSSM Higgs boson masses are evaluated in the
On--Shell renormalisation scheme, gives an estimate of the scheme
dependence of the predictions. This dependence is in general of the
order of 2 GeV, but can reach the level of 4--5 GeV in some cases,
such as in the pMSSM scenario with large mixing in the stop sector.
The dependence on the scale at which the effective scalar potential is
minimised and the radiative corrections are computed is significantly
improved by the inclusion of the two--loop radiative corrections,
leaving a variation of less than 3 GeV on $M_h$ (apart from the case
of the pMSSM scenario with large stop mixing, where the variation is
larger). The effect of approximating the external momentum in the
two--loop Higgs boson self--energies to zero rather than to the mass
of the Higgs boson has been estimated to lead to a shift of less than
half a GeV on $M_h$ and to an even smaller shift on $M_H$. \s

All of these shifts indicate the size of the higher order corrections
that are left uncomputed, thus they are a measure of the theoretical
uncertainty which affects the calculation. As a summary, we conclude
that these effects induce a theoretical uncertainty on the mass of the
lighter Higgs boson of 3--4 GeV in the constrained MSSM and up to 5
GeV in the unconstrained MSSM, in particular when the mixing in the
stop sector is such that $M_h$ is maximised. The uncertainties on
$M_h$ due to the experimental errors in the measurement of the SM
input parameters are approximately of the same size. Indeed, a
variation of the top quark mass within its $1\sigma$ range, $M_t=178.0
\pm 4.3$ GeV, leads to a shift of about $\pm 2$ to $\pm 4$ GeV on
$M_h$, depending on the scenario.  The shifts on $M_h$ from the errors
in the measurement of $\mb$ and $\as$ are less than half a GeV, while
the error on $\alpha$ has a negligible impact. Thus, both a
theoretical and an experimental effort are needed to attain the level
where the lighter Higgs boson mass can be predicted with a precision
of less than 100 MeV, i.e. the level of precision at which this mass
can be measured at the future colliders. \s
  
Finally, using the RGE codes, we have performed a phenomenological
study of the impact of the radiative corrections, as well as the
theoretical and experimental uncertainties, on the Higgs sector of the
constrained and unconstrained MSSM. In particular, using the latest
Tevatron value of the top quark pole mass, $M_t=178.0 \pm 4.3$ GeV, we
have derived the maximal value of $M_h$ in the MSSM, as well as the
minimal value of the parameter $\tb$ that is allowed by the negative
search of the MSSM Higgs bosons at LEP2. In the unconstrained MSSM, we
get an upper bound $M_h \lsim 143$ GeV and we show that all values of
$\tb$ are allowed by LEP2 constraints. In the constrained MSSM
scenarios, as a result of the chosen ranges for the soft
SUSY--breaking parameters and of the relations among them, the bounds
on $M_h$ are stricter than in the pMSSM: 129.0, 123.7 and 124.6 GeV in
the mSUGRA, GMSB and AMSB cases respectively. Similarly, in the
constrained scenarios it is possible to derive lower bounds on $\tb$:
2.4, 3.3 and 3.1 in the mSUGRA, GMSB and AMSB cases respectively.  If
the upper value of the top quark mass, $M_t=182.3$ GeV, is used, and
an estimated 4 GeV theoretical uncertainty on $M_h$ is included, one
arrives at the conservative upper bound of $M_h \lsim 152$ GeV for the
value of the lighter Higgs boson in the unconstrained MSSM. Such a
bound has important implications for searches and phenomenology of the
MSSM Higgs bosons.
 

\section*{Note added}

During the final stages of the preparation of the present work, a
paper appeared~\cite{martinnew}, where a two--loop computation of the
leading contributions to the MSSM Higgs boson self--energies for
arbitrary external momentum is presented. In the numerical examples
presented in Ref.~\cite{martinnew}, the two--loop momentum--dependent
corrections to the lighter Higgs boson mass $\mh$ are of the order of
a few hundred MeV, while the corresponding corrections to the heavier
Higgs boson mass $\mH$ are considerably smaller.  These results are in
good agreement with our empirical estimates given in section
\ref{secmom}. We plan to include the results of Ref.~\cite{martinnew}
in our spectrum calculators as soon as they become available in a
suitable form.

\section*{Acknowledgements}

We would like to thank M.~Drees, S.~Heinemeyer, W.~Hollik and
G.~Weiglein for useful discussions, and K.~Desch for providing us with
LEP2 Higgs data.  This work was partially supported by the European
Community's Human Potential Programme HPRN-CT-2000-00149 (Collider
Physics). W.~P.~is supported by a Spanish MCyT Ramon y Cajal contract
and partly by the Swiss ``Nationalfonds''. P.~S.~thanks LAPTH and LPMT
for hospitality offered while part of this work was carried out.



\section*{Appendix A: ~EWSB conditions and Higgs boson masses}
\begin{appendletterA}

In this appendix we provide for completeness the general formulae for
the computation of the radiative corrections to the EWSB conditions
and to the Higgs boson masses in the MSSM, following the notation of
Ref.~\cite{pbmz}. We also explain how to adapt the formulae of
Refs.~\cite{dsz,bdsz,dds,dstad} to the same notation.\s

Once the requirement of correct EWSB is imposed, providing input
values for the $Z$ boson mass and $\tan\beta=v_2/v_1$ [where
$v_i\;(i=1,2)$ are the vacuum expectation values of the neutral Higgs
fields], the minimisation conditions on the MSSM effective potential
translate into conditions for the parameters $\mu^2$ and $B$:
\bea
\label{eqmu}
\mu^2 & = & - \frac{\mz^2}{2} 
- \hlf\,\tan 2\beta\, \left[  
\left( m^2_{H_1} - \frac{t_1}{v_1}\right)\,\cot\beta -
\left( m^2_{H_2} - \frac{t_2}{v_2}\right)\,\tan\beta \right] \,,\\
&\nn\\
\label{eqm3}
B & = & - \frac{\mz^2}{2}\,\sin 2\beta\, 
- \hlf\,\tan 2\beta\, \left( m^2_{H_1} -\frac{t_1}{v_1}  
- m^2_{H_2} + \frac{t_2}{v_2} \right) \, ,
\eea
where $\mz$ is the running mass for the $Z$ boson defined in
eq.~(\ref{mzrun}), $m^2_{H_i}\;(i=1,2)$ are the soft SUSY--breaking
mass terms for the two Higgs doublets and $t_i$ are the tadpole
corrections, corresponding to diagrams with one incoming Higgs
field.
The squared physical masses for the CP--odd Higgs boson $A$ and for
the charged Higgs boson $H^{\pm}$ can be computed as:
\bea
\label{mapole}
\ma^2 & = & 2\,B /\sin 2\beta - {\rm Re}\,\Pi_{AA}(\ma^2) 
+\sin^2\beta \,\frac{t_1}{v_1}+ \cos^2\beta\,\frac{t_2}{v_2}\,,\\
\label{mhpm}
M^2_{H^{\pm}} & = & \ma^2 + M_W^2 + {\rm Re}\,\left[
\Pi_{AA}(\ma^2)-\Pi_{H^+H^-}(M^2_{H^{\pm}}) + \Pi^{T}_{WW}(M^2_W) \right]\,,
\eea
where $M_W$ is the physical $W$ boson mass, and $ \Pi_{AA}(\ma^2)\,,
\Pi_{H^+H^-}(M^2_{H^{\pm}})$ and $\Pi^{T}_{WW}(M^2_W)$ are the
self--energies for the $A\,, H^{\pm}$ and $W$ bosons, respectively,
each computed at a squared external momentum equal to the squared
physical mass of the corresponding particle.\s

Finally, the squared physical masses for the CP--even Higgs bosons,
$\mh^2$ and $\mH^2$, correspond to the real part of the poles of the
$2\times2$ propagator matrix, i.e.~the two solutions of the equation:
\be
\label{detm}
{\rm Det}\left[ \,p_i^2\,{\rm I} - {\cal M}^2(p_i^2)\right] =0\,,
\ee
where the radiatively corrected mass matrix for the CP--even Higgs fields is:
\be
\label{math}
{\cal M}^2(p^2) =
\left(\!\!
\begin{array}{cc}
\mz^2\,\cos^2\beta + m_A^2\,\sin^2\beta - \Pi_{11}(p^2) + \frac{t_1}{v_1} &
\! -(\mz^2+m_A^2)\,\sin\beta\cos\beta - \Pi_{12}(p^2)\\
-(\mz^2+m_A^2)\,\sin\beta\cos\beta - \Pi_{12}(p^2) &
\! \mz^2\,\sin^2\beta + m_A^2\,\cos^2\beta - \Pi_{22}(p^2) + \frac{t_2}{v_2}
\end{array}\!\!
\right)\! .
\ee
In the above formula, $m_A^2 = 2\,B/\sin 2\beta$ is the squared
running mass for the $A$ boson as defined in eq.~(\ref{marun}), and
$\Pi_{ij}(p^2)\; (i,j=1,2)$ are the self--energies of the CP--even
components of the Higgs fields. In \softsusy, \spheno\ and \suspect\
an iterative procedure is employed for solving eq.~(\ref{detm}).  The
Higgs mixing angle $\alpha$, instead, is defined in the codes as the
angle that diagonalises ${\cal M}^2(p^2)$ for $p^2 = \mh^2$.\s

The explicit formulae for the one--loop parts of the self--energies
and tadpole diagrams appearing in eqs.~(\ref{eqmu})--(\ref{math}) can
be found in Ref.~\cite{pbmz}. In addition, \softsusy, \spheno\ and
\suspect\ employ the formulae given in Refs.~\cite{dsz,bdsz,dds,dstad}
for the leading two--loop corrections to the tadpoles and the neutral
Higgs boson masses. \s

The computations in Refs.~\cite{dsz,bdsz,dds,dstad} are performed in
the effective potential approach, which implies that the approximation
of zero external momentum is used in the two--loop
self--energies. Also, the formulae of Refs.~\cite{dsz,bdsz,dds,dstad}
assume that the tree--level mass matrix for the CP--even Higgs bosons
is expressed in terms of the physical $A$--boson mass, $\ma$, whereas
in eq.~(\ref{math}) the mass matrix is expressed in terms of the
running mass $\marun$. It is thus necessary to rearrange the formulae
of Refs.~\cite{dsz,bdsz,dds,dstad} for the corrections to the Higgs
tadpoles, $\Sigma_i$, to the CP--odd Higgs mass, $\Delta m_A^2$, and
to the CP--even Higgs mass matrix, $\left(\Delta {\cal
M}^2_S\right)^{\rm eff}$, in order to adapt them to the notation of
Ref.~\cite{pbmz}.  In particular, the following equalities hold:
\vspace{-3mm}
\bea
\Sigma_i & = & - \frac{t_i}{v_i} \;\;\;(i=1,2)\,, \nn\\
\Delta m_A^2 & = & - {\rm Re}\,\Pi_{AA}(0) 
+\sin^2\beta \,\frac{t_1}{v_1}+ \cos^2\beta\,\frac{t_2}{v_2}\,,\nn\\
\Delta m_A^2\,\sin^2\beta + \left(\Delta {\cal M}^2_S\right)^{\rm eff}_{11} 
& = & -{\rm Re}\,\Pi_{11}(0) + \frac{t_1}{v_1}\,,\nn\\ 
- \Delta m_A^2\,\sin\beta\cos\beta
+\left(\Delta {\cal M}^2_S\right)^{\rm eff}_{12}
& = & -{\rm Re}\,\Pi_{12}(0)\,,\nn\\
\Delta m_A^2\cos^2\beta + \left(\Delta {\cal M}^2_S\right)^{\rm eff}_{22}  
& = & -{\rm Re}\,\Pi_{22}(0) + \frac{t_2}{v_2}\,.\nn
\eea
\end{appendletterA}

\vspace{-5mm}

\section*{Appendix B: ~Formulae for the $\oabatau$ corrections}
\begin{appendletterB}

We present in this appendix the explicit analytical formulae for the
two--loop $\oabatau$ corrections to the MSSM Higgs masses and tadpoles
in the $\dr$ renormalisation scheme. We follow very closely the lines
of Refs.~\cite{dsz,bdsz,bdsz2,dds,dstad}, to which we refer the reader
for further details. The following results complete the computation of
the two--loop corrections to the Higgs masses and tadpoles controlled
by the third--family Yukawa couplings (the inclusion of the $\oatauq$
corrections was discussed in Ref.~\cite{dds}). \s

If we express the MSSM effective potential $V_{\rm eff}$ in terms of
$\dr$--renormalised fields and parameters, the two--loop contribution
to $V_{\rm eff}$ involving both the bottom and tau Yukawa couplings
$\hb$ and $\htau$ reads, in units of $N_c/(16\pi^2)^2$,
\be
\Delta V_{b \tau} = 
\frac{\hb\,\htau}{2}\,\sdb\,\sdtau\,c_{\ptb-\pttau}\,
\left[\hat{J}(\msbu,\mstauu)-\hat{J}(\msbd,\mstauu)-\hat{J}(\msbu,\mstaud)
+ \hat{J}(\msbd,\mstaud) \right]\, ,
\ee
where: $c_{\phi} \equiv \cos \phi$ and $s_{\phi} \equiv \sin \phi$;
$\msqi$ $(q=\tau,b)$ are the field--dependent squark masses; $\thsq$
is the field--dependent squark mixing angle, defined in such a way
that $0\leq \thsq < \pi/2$ (to be contrasted with the usual
field--independent mixing angle $\thq$, such that $-\pi/2 \leq \thq <
\pi/2$); $\widetilde{\varphi}_q$ is the phase in the off--diagonal
element of the squark mass matrix. For the explicit Higgs field
dependence of these parameters, see Refs.~\cite{dsz,bdsz}. Finally, the
$\dr$--subtracted two--loop integral $\hat{J}$,
with $Q^2$ being the renormalisation scale, is defined as
\be
\hat{J}(\msqu,\msqd) = \msqu\,\msqd\,
\left(1-\ln\frac{\msqu}{Q^2}\right)\,\left(1-\ln\frac{\msqd}{Q^2}\right)\, 
,
\ee

Exploiting the field--dependence of the squark masses and mixing
angles, we can express the tau/bottom corrections to the CP--even
Higgs mass matrix, $\left(\Delta {\cal M}^2_S\right)^{\rm eff}$, to
the Higgs tadpoles, $\Sigma_i$, and to the CP--odd Higgs mass, $\Delta
m_A^2$, (see appendix A and Ref.~\cite{dds} for the various
definitions) as
\bea
\label{dms11}
\left(\Delta {\cal M}^2_S\right)^{\rm eff}_{11} & = & 
2 \, \htau^2\, \mtau^2\, F^{\tau}_1
+ 2\, \htau^2\, A_{\tau}\, \mtau\, \Sdtau\, F^{\tau}_2 
+ \hlf \, \htau^2\, A_{\tau}^2\, \Sdtau^2\, F^{\tau}_3 
+ 2 \,\htau\,h_b\, \mb \, A_{\tau} \,  \Sdtau \, F^{\tau}_4\nn\\
&+& 
2 \, h_b^2\, m_b^2\, F^b_1
+ 2\, h_b^2\, A_b\, m_b\, \Sdb\, F^b_2 
+ \hlf \, h_b^2\, A_b^2\, \Sdb^2\, F^b_3 
+ 2 \,\htau\,h_b\, \mtau \, A_b \,  \Sdb \, F^{b}_4 \nn\\
&+&
\htau\,h_b\, A_{\tau} \, A_b \,\Sdtau\,\Sdb\, F_5
+ 4\,\htau\,h_b\,\mtau\,\mb\,F_6\,, \\
&&\nn\\
\label{dms12}
-\left(\Delta {\cal M}^2_S\right)^{\rm eff}_{12} & = & 
\htau^2 \,\mu\, \mtau\, \Sdtau \,  F^{\tau}_2 
+ \hlf \, \htau^2\, A_{\tau} \,\mu \, \Sdtau^2 \, F^{\tau}_3
+ \htau\,h_b\, \mb \, \mu \, \Sdtau \, F^{\tau}_4\nn\\
&+& 
h_b^2 \,\mu\, m_b\, \Sdb \,  F^b_2 
+ \hlf \, h_b^2\, A_b \,\mu \, \Sdb^2 \, F^b_3 
+ \htau\,h_b\, \mtau \, \mu \, \Sdb \, F^b_4\nn\\
&+& \hlf \,\htau\,h_b\,\Sdtau \,\Sdb \,\mu\, (A_b + A_{\tau})\,F_5\,,\\
&&\nn\\
\label{dms22}
\left(\Delta {\cal M}^2_S\right)^{\rm eff}_{22} & = &
\hlf \, \htau^2\, \mu^2 \, \Sdtau^2\, F^{\tau}_3
+ \hlf \, h_b^2\, \mu^2\, \Sdb^2\, F^b_3
+ \htau\,h_b\, \mu^2 \,\Sdtau\,\Sdb\, F_5\,,\\
&&\nn\\
&&\nn\\
\label{sigma1}
v_1^2\,\Sigma_1 
& = & 
\mtau\, A_{\tau}\, \Sdtau\, F^{\tau} + 2\, \mtau^2\, G^{\tau}+ 
\mb\, A_b\, \Sdb\, F^b + 2\, \mb^2\, G^b\,,\\
&&\nn\\
\label{sigma2}
-v_2^2\,\Sigma_2
& = & 
\mtau\, \mu\, \tb\, \Sdtau\, F^{\tau}+ 
\mb\, \mu\, \tb\, \Sdb\, F^b\,,\\
&&\nn\\
\label{deltama}
\Delta m_A^2 & = &
-\frac{1}{\cb\,\sb}\,\left(\, 
2\,h_b\,\htau\,F_A
-\frac{\htau^2 \,\mu\,A_{\tau}}{\difftau}\, F^{\tau}
-\frac{h_b^2 \,\mu\,A_b}{\diffb}\, F^b\,\right)\;.
\eea
In the equations above, $A_{\tau}$ and $A_b$ are the soft
supersymmetry--breaking trilinear couplings of the Higgs fields to the
stau and sbottom squarks, $\mu$ is the Higgs mass term\footnote{~Note
that we employ here the opposite convention for the sign of $\mu$ with
respect to Ref.~\cite{dds}, which explains the minus signs appearing
in eqs.~(\ref{dms12})\,,\,(\ref{sigma2}) and (\ref{deltama}).} in the
superpotential, $\tb = v_2/v_1$ is the ratio of the two Higgs vacuum
expectation values and $\Sdq$ ($q=\tau,b$) refer to the usual
field--independent squark mixing angles. The functions $F_i^{q}\; (i =
1,2,3,4)\,,\; F_5\,,\; F_6\,,\; F^q\,,\; G^q$ and $F_A$ are
combinations of the derivatives of $\Delta V_{b \tau}$ with respect to
the field--dependent parameters, computed at the minimum of the
effective potential; their explicit expressions are, in units of
$N_c/(16\pi^2)^2$:
\bea
\label{f1}
F_1^{\tau} & = & 
- 
\hlf\,\hb\,\htau\,\Sdb\,\Sdtau\,I_{b}\,\frac{\difftau}{\mstauu\mstaud}\,,\\
&&\nn\\
F_2^{\tau} & = & 
\frac{\hb\,\htau\,\Sdb\,I_{b}}{2\,\Sdtau\,\mstauu\mstaud\,(\difftau)}\,
\left[\Sdtau^2\,(\mqstauu-\mqstaud) + 2\,\Cdtau^2\,\mstauu\,\mstaud\,
\ln\frac{\mstauu}{\mstaud}\right]\,,\\
&&\nn\\
F_3^{\tau} & = &
 \frac{\hb\,\htau\,\Sdb\,I_{b}}{2\,\Sdtau\,\mstauu\mstaud\,(\difftau)}\,
\left[\,
2\,(3\,\Sdtau^2-2)\,\mstauu\mstaud\,\frac{\mstauu+\mstaud}{\difftau}\,
\ln\frac{\mstauu}{\mstaud}\right.\nn\\
&&\hspace{4cm} 
-  \Sdtau^2(\mstauu+\mstaud)^2+8\,\Cdtau^2\,\mstauu\mstaud
 \biggr]\,,\\
%
&&\nn\\
F_4^{\tau} & = &
\frac{\hb\,\htau\,\Sdb}{2\,\Sdtau\,(\difftau)}\,
\ln\frac{\msbu}{\msbd}\,
\left[2\,\Cdtau^2\,I_{\tau} + \Sdtau^2\,(\difftau)\,
\ln\frac{\mstauu\mstaud}{Q^4}\right]\,,\\
&&\nn\\
F_5 & = &
\frac{\hb\,\htau}{\Sdb\,\Sdtau}\,\left[
\frac{\Sdb^2\,\Cdtau^2\,I_{\tau}}{\difftau}\,\ln\frac{\msbu\msbd}{Q^4}
+\frac{\Sdtau^2\,\Cdb^2\,I_b}{\diffb}\,\ln\frac{\mstauu\mstaud}{Q^4}
\right.\nn\\
&&
\left.\hspace{1.5cm} 
-\frac{2\,I_{\tau}\,I_b\,(1-\Cdb^2\,\Cdtau^2)}{(\diffb)(\difftau)}
+\hlf\,\Sdb^2\,\Sdtau^2\,\ln\frac{\msbu\msbd}{Q^4}
\,\ln\frac{\mstauu\mstaud}{Q^4}\right]\,,\\
&&\nn\\
F_6 & = &
\hlf\,\hb\,\htau\,\Sdb\,\Sdtau\,\ln\frac{\msbu}{\msbd}\,
\ln\frac{\mstauu}{\mstaud}\,,\\
&&\nn\\
F^{\tau} & = &
\frac{\hb\,\htau\,\Sdb\,I_b}{2\,\Sdtau\,(\difftau)}\,
\left[2\,\Cdtau^2\,I_{\tau} + \Sdtau^2\,(\difftau)\,
\ln\frac{\mstauu\mstaud}{Q^4}\right]\,,\\
&&\nn\\
G^{\tau} & = & 
\hlf\,\hb\,\htau\,\Sdb\,\Sdtau\,I_{b}\,\ln\frac{\mstauu}{\mstaud}\,,\\
&&\nn\\
\label{fa}
F_A & = & 
\frac{2\,\hb\,\htau\,\mb\,\mtau\,\mu^2\,(A_b-A_{\tau})^2
\,\tb\,I_b\,I_{\tau}}{\Sdb\,\Sdtau\,(\difftau)^2\,(\diffb)^2}\,.
\eea
The functions $F_i^b$, $F^b$ and $G^b$ can be obtained from their 
tau counterparts through the replacement $\tau\leftrightarrow b$. The 
function $I_q$ ($q=\tau,b$) entering the above equations is defined as:
\be
I_q = \msqu\,\left(\ln\frac{\msqu}{Q^2} - 1\right)
-\msqd\,\left(\ln\frac{\msqd}{Q^2} - 1\right)\,.
\ee
The results in Eqs.~(\ref{f1})--(\ref{fa}) are valid when the
one--loop $\oatau$ and $\oab$ corrections are written in terms of
$\dr$ parameters computed at the renormalisation scale $Q^2$. In
alternative, an on--shell renormalisation scheme for the tau/stau and
bottom/sbottom sectors that avoids the appearance of
$\tb$--enhanced terms in the two--loop part of the results should
be devised along the lines of Refs.~\cite{bdsz2,dds}.

\end{appendletterB}

\vspace{1cm}


\end{document}